\documentclass[prc,preprint,superscriptaddress,nofootinbib,
  amsmath,amssymb,aps]{revtex4-1}  
\usepackage{graphicx}
\usepackage{dcolumn}
\usepackage[usenames]{color}

\newcolumntype{d}[1]{D{.}{.}{#1}}

\newcommand{\simge}{\hspace*{0.2em}\raisebox{0.5ex}{$>$}
     \hspace{-0.8em}\raisebox{-0.3em}{$\sim$}\hspace*{0.2em}}
\newcommand{\simle}{\hspace*{0.2em}\raisebox{0.5ex}{$<$}
     \hspace{-0.8em}\raisebox{-0.3em}{$\sim$}\hspace*{0.2em}}

\newcommand{\beq}{\begin{equation}}
\newcommand{\eeq}{\end{equation}}
\newcommand{\bea}{\begin{eqnarray}}
\newcommand{\eea}{\end{eqnarray}}

\begin{document}

\title{Renormalization of One-Pion Exchange in\\
Chiral Effective Field Theory 
for Antinucleon-Nucleon Scattering}

\author{Daren Zhou}
\affiliation{School of Electrical and Computer Engineering, Nanfang College, Guangzhou, Guangdong 510970, China}

\author{Bingwei Long}
\affiliation{College of Physics, Sichuan University, Chengdu, Sichuan 610065, China}

\author{R. G. E. Timmermans}
\affiliation{Van Swinderen Institute for Particle Physics and Gravity, Faculty of Science and Engineering, University of Groningen, 9747 AG Groningen, The Netherlands}

\author{U. van Kolck}
\affiliation{Universit\'e Paris-Saclay, CNRS/IN2P3, IJCLab,
91405 Orsay, France}
\affiliation{Department of Physics, University of Arizona, Tucson, Arizona 85721, USA}

\date{\today}
\vspace{3em}

\begin{abstract}
The renormalization of iterated one-pion exchange (OPE) is studied in Chiral Effective Field Theory ($\chi$EFT) for the antinucleon-nucleon ($\overline{N}\!N$) system.
The OPE potential is cut off at a certain distance and contact interactions are represented by a complex spherical well with the same radius.
We investigate the dependence on the cutoff radius of the phase shifts, inelasticities, and mixing angles for the low partial waves in
$\overline{N}\!N$ scattering.
We show that renormalization requires additional contact interactions compared to the expectation based on naive dimensional analysis. 
Results after renormalization are compared with the state-of-the-art energy-dependent partial-wave analysis of $\overline{N}\!N$ data.
We compare our conclusions with applications of $\chi$EFT to the nucleon-nucleon  
system.
\end{abstract}

\maketitle

\section{Introduction} \label{sec:intro}

Chiral Effective Field Theory ($\chi$EFT) is an effective theory of QCD 
for momenta below the nonperturbative scale $M_{\rm QCD}\sim 1$ GeV.
The most general Lagrangian density involving baryons and pions
is constructed based on symmetry arguments,
particularly approximate chiral symmetry,
while the details of short-range dynamics are parametrized 
by interaction strengths or ``low-energy constants'' (LECs).
Observables are calculated in a systematic expansion in powers of
$Q/M_{\rm QCD}$, where $Q$ is the characteristic 
momentum of the process under consideration~\cite{Weinberg:1978kz}.
The model independence of the approach is guaranteed by order-by-order renormalization.
Since Weinberg's seminal articles~\cite{Weinberg:1990rz,Weinberg:1991um},
$\chi$EFT
has been applied 
to a variety of nuclear, hypernuclear, and antinuclear systems \cite{Hammer:2019poc}.
Here we discuss the renormalization of the antinucleon-nucleon
($\overline{N}\!N$) system at low energies.

Weinberg's original approach~\cite{Weinberg:1990rz,Weinberg:1991um} to the nucleon-nucleon ($N\!N$) system was based on the assumption that all LECs satisfy naive dimensional analysis (NDA) \cite{Manohar:1983md}. Weinberg identified the origin of the breakdown in the perturbative expansion of the amplitude in the ``reducible'' diagrams that contain $N\!N$-only intermediate states. He suggested that the potential --- defined as the sum of ``irreducible'' diagrams --- be expanded in $Q/M_{\rm QCD}$, and each successive truncation be solved for exactly in the Lippmann-Schwinger, or equivalently Schr\"odinger, equation.
Following the initial success of this approach for $N\!N$ \cite{Ordonez:1992xp,Ordonez:1995rz}, significant improvements \cite{Machleidt:2011zz,Epelbaum:2012vx} have made chiral potentials the standard input for the ``{\it ab initio}'' methods \cite{Hergert:2020bxy} that nowadays dominate low-energy nuclear physics.
Weinberg's proposal for the $N\!N$ potential was extended to the $\overline{N}\!N$ system in Refs. \cite{Kang:2013uia,Dai:2017ont}.

However, it has been known for a long time that $N\!N$ amplitudes, and thus observables, obtained with Weinberg's prescription are not renormalization-group (RG) invariant. In any EFT, contributions from virtual 
states require a regulator, both for the loops in the potential and the loops in the Lippmann-Schwinger equation (automatically generated by the solution of the Schr\"odinger equation). Observables should be RG invariant, {\it i.e.}, independent of the form of the regulator and the value of the cutoff $\Lambda$, in the sense that they approach finite results as $\Lambda$ becomes much larger than other scales in the problem. In the $N\!N$ case two types of problems have been found. In the $^1S_0$ channel, the exact solution of the Schr\"odinger equation with one-pion exchange (OPE) and the chiral-symmetric contact interaction demanded by NDA cannot be renormalized \cite{Kaplan:1996xu,Beane:2001bc} without a chiral-symmetry-breaking contact interaction that according to NDA should be suppressed by $m_\pi^2/M_{\rm QCD}^2$, where $m_\pi$ is the pion mass. At fixed quark masses, this LEC underestimation has only mild consequences. The second deficiency is more serious, though: in partial waves beyond $S$ waves where the tensor part of OPE is attractive, momentum-dependent LECs are necessary \cite{Nogga:2005hy,PavonValderrama:2005uj}, which are also inappropriately
attributed to higher orders by NDA.

The origin of the failure of NDA is the nonperturbative character of the $N\!N$ amplitude at leading order (LO). NDA results from the assumption of naturalness \cite{tHooft:1979rat,Veltman:1980mj} 
applied to the perturbative series \cite{vanKolck:2020plz}. It is well known that nonperturbative renormalization can be significantly different, however \cite{vanKolck:2020llt}. For example, in the $^3S_1$ channel Weinberg's prescription seems to fail when one examines individual LO diagrams for the amplitude \cite{Savage:1998vh}, and yet it is consistent nonperturbatively \cite{Frederico:1999ps,Beane:2001bc}. An attractive, singular power-law potential, such as the $\sim -r^{-3}$ of the OPE tensor force, requires an LEC in every wave where it is treated exactly \cite{Beane:2000wh,PavonValderrama:2007nu}. In the $^3S_1$ channel NDA happens to prescribe such an LEC; in other waves, it misses them. The resulting cutoff dependence reflects the fact that the LECs are in fact enhanced by the nonperturbative RG running, being suppressed only by powers of the parameter controlling OPE --- the pion decay constant $f_\pi \sim M_{\rm QCD}/4\pi$ --- instead of $M_{\rm QCD}$. The principle of naturalness applied to this case demands the inclusion of the LECs at lower orders than suggested by NDA \cite{Nogga:2005hy,PavonValderrama:2005uj,Birse:2005um}. 

Once LO is corrected, RG invariance can be maintained at higher orders if subleading interactions are treated in distorted-wave perturbation theory \cite{Long:2007vp}. A reasonable description of the $N\!N$ system results \cite{Valderrama:2009ei,PavonValderrama:2011fcz,Long:2011qx,Long:2011xw,Long:2012ve,Long:2013cya,PavonValderrama:2016lqn,SanchezSanchez:2017tws,Wu:2018lai}. Properties of the triton \cite{Song:2016ale,Yang:2020pgi,Peng:2021pvo} and the alpha particle \cite{Yang:2020pgi} also come out well, but an enhancement of few-body forces might take place in heavier nuclei \cite{Yang:2021vxa}.

In this paper, we study the renormalization of the $\overline{N}\!N$ scattering amplitude arising from the iteration of OPE to all orders --- a possible approach to an LO interaction adopted, for example, in Refs. \cite{Kang:2013uia,Dai:2017ont}. We use the results of the Groningen energy-dependent partial-wave analysis (PWA) \cite{Zhou:2012ui,Zhou:2013}, which provide an excellent description of the available data for elastic $\overline{p}p\rightarrow\overline{p}p$ and charge-exchange $\overline{p}p\rightarrow\overline{n}n$ scattering below 925 MeV/$c$ antiproton laboratory momentum.
As for the $N\!N$ system
\cite{Birse:2005um,PavonValderrama:2016lqn,Kaplan:2019znu}, we expect suppression factors of angular momentum to make OPE perturbative in high partial waves. Accordingly, we focus here on the lowest waves: the uncoupled $S$ and $P$, and the coupled $S$-$D$ and $P$-$F$ partial waves. We perform an analysis similar to that of Ref.~\cite{Nogga:2005hy}, except that we use a coordinate-space regulator as in Refs. \cite{Scaldeferri:1996nx,Beane:2000wh,Beane:2001bc}, which resembles the boundary conditions employed in the Groningen PWA. 
We find that, as in the $N\!N$ case, renormalization requires more LECs than suggested by NDA. Lacking further information, we assume that a short-range LEC contains an imaginary part at the same order as the real part,
so as to account for $\overline{N}\!N$ annihilation into mesons at small distances. The LECs are fitted to phase shifts and inelasticities from the PWA at certain energies, with phase shifts and inelasticities at other energies and mixing angles being postdictions of the EFT at LO. Because isospin breaking is a higher-order effect, we work in the isospin basis. In future publications we plan to investigate the extent to which OPE can be treated perturbatively in some of the low $\overline{N}\!N$ partial waves \cite{Chen:2010an,Chen:2011yu} and the ordering of other interactions, such as two-pion exchange (TPE). 

This paper is organized as follows. In Sec.~\ref{sec:OPE}, we describe the OPE potential and our choice of the regulator,
and identify the partial waves that might need counterterms. In Sec.~\ref{sec:Strategy}, we explain our renormalization strategy and  present our results. The conclusions are summarized in Sec.~\ref{sec:Conclusions}, where we also give an outlook.

\section{One-pion exchange and contact interactions} \label{sec:OPE}

At low energies, antinuclear systems are made of nucleon ($N$) and antinucleon ($N^c$) degrees of freedom:
nonrelativistic fermions with the same mass $m_N\simeq 940$ MeV and
long-range
interactions \cite{LongThesis,Oosterhof:2019dlo} related by 
$G$ parity, which is equivalent to charge conjugation 
together with isospin symmetry. 
For processes involving momenta $Q$ on the order of the pion mass $m_\pi\simeq 138$ MeV, the three pions ($\pi^a$, $a=1,2,3$) must also be 
included explicitly as pseudo-Goldstone bosons from the spontaneous breaking of chiral symmetry, with a decay constant $f_\pi\simeq 92.4$ MeV  and interactions constrained by the approximate chiral symmetry of the QCD dynamics.
OPE arises from
the first few terms in the chiral Lagrangian density
\begin{eqnarray}
\mathcal{L}_{\overline{N}\!N} &=&
N^\dagger \left(i \partial_0+\frac{\nabla^2}{2m_N}\right) N
+ {N^c}^\dagger\left(i \partial_0+\frac{\nabla^2}{2m_N}\right)N^c
-\frac{1}{2}\pi^a\left(\partial^2+m_\pi^2\right)\pi^a
\nonumber\\
&&
+\frac{g_A}{2 f_\pi}\left({N^\dagger}\vec{\sigma}\tau^a N
+ {{N^c}^\dagger}{\vec\sigma}\tau^{aT} N^c\right)\cdot\vec{\nabla}\pi^a
+\dots\,, \label{eq:lagran}
\end{eqnarray}
where 
$g_A\simeq 1.29$ is the axial-vector coupling constant
and $\tau^a$ ($\vec{\sigma}$) are the Pauli matrices for isospin (spin).
We focus on the strong interactions and electromagnetic terms are not shown.

After renormalization, the energy transfer between (anti)nucleons 
is of order $Q^2/m_N$ and the pion propagator can be taken as approximately static.
The OPE potential for the $\overline{N}\!N$ system is 
derived from Eq. (\ref{eq:lagran}): at LO, it reads in configuration space
\begin{equation}
V_\pi = -\frac{g_A^2 m_\pi^3}{16\pi f_\pi^2}  \,\boldsymbol{\tau}_1\cdot\boldsymbol{\tau}_2\,
\bigl[v_S(m_\pi r) \,\vec{\sigma}_{1}\cdot\vec{\sigma}_{2} 
+ v_T(m_\pi r)\,S_{12}\bigr] \, ,
\label{OPE}
\end{equation}
where
$S_{12}=3\,\vec{\sigma}_{1}\cdot\hat{r}\,\vec{\sigma}_{2}\cdot\hat{r} -\vec{\sigma}_{1}\cdot\vec{\sigma}_{2}$
is the tensor operator, and
\begin{subequations}
	\begin{eqnarray}
	v_S(x) & = & \frac{e^{-x}}{3x}~, \\
	v_T(x) & = & \left(1+x+\frac{x^2}{3}\right) \frac{e^{-x}}{x^3} \ .
	\end{eqnarray}
\end{subequations}
Because the $G$ parity of the pion is $-1$, Eq.~\eqref{OPE} has the opposite sign as OPE for $N\!N$ ({\it cf.}, for example, Ref. \cite{Nogga:2005hy}). Two orders down in the $Q/M_{\rm QCD}$ expansion, chiral-symmetry-breaking corrections represent the Goldberger-Treiman discrepancy, which increases the strength of static OPE \cite{vanKolck:1996rm,vanKolck:1997fu}, while chiral-symmetric corrections to the pion-(anti)nucleon interaction account for recoil \cite{Ordonez:1992xp}. At this order TPE also appears \cite{Ordonez:1992xp}.

From NDA, one expects the iteration of the OPE potential in Eq.~\eqref{OPE} in low partial waves to be a series in powers of $Q/f_\pi$, and thus to require a full solution of the Schr\"odinger equation for $Q\sim f_\pi \ll M_{\rm QCD}$ \cite{Bedaque:2002mn}. In the absence of fine-tuning, 
$f_\pi$ should set the scale for dimensionful parameters in the amplitude, such as scattering lengths and volumes. Indeed, the majority of these values extracted with a chiral potential in Ref. \cite{Kang:2013uia,Dai:2017ont} have natural sizes.

However, the OPE potential is singular: it diverges as $r^{-3}$ and needs to be regularized. Our choice of regulator is made for ease of comparison with the Groningen PWA~\cite{Zhou:2012ui,Zhou:2013}, which follows the procedure laid out in the Nijmegen PWAs of Refs.~\cite{Timmermans:1990tz,Timmermans:1994pg,Timmermans:1995xb}: the partial-wave Schr\"odinger equation is solved for the coupled $\overline{p}p$ and $\overline{n}n$ channels with a long-range potential outside a radius $r=b=1.2$ fm. The long-range potential is taken to consist of the electromagnetic interaction, OPE, and TPE \cite{Rentmeester:1999vw,Rentmeester:2003mf}. At $r=b$ a boundary condition is chosen that, for convenience, corresponds to a spherical well which is independent of energy, depends on the spin and isospin of the partial wave, and is complex to account for the annihilation into mesons. This strategy allows us to use the code of the PWA to solve the Schr\"odinger equation with only OPE for $r>b$. However, the boundary is now regarded as the regulator: varying the value of $b$ corresponds to varying the ultraviolet 
cutoff $\Lambda\equiv 1/b$. 

The Lagrangian density in Eq.~\eqref{eq:lagran} contains contact interactions among (anti)nucleons with an arbitrary number of derivatives, which represent the short-distance QCD dynamics. For $\overline{N}\!N$, these are complex: annihilation generates mesonic states with energies on the order of $2m_N$ that cannot be accounted for as explicit degrees of freedom in the EFT. 
(One example is the annihilation diagram into one pion, related to OPE by crossing, which would be present in a relativistic theory.)
In addition to imaginary parts, annihilation also generates real contributions to the LECs, making them different from those for $N\!N$.
Like any other EFT LECs, these contact interactions must appear at orders no higher than where they are needed to remove arbitrary regulator dependence. In this first approach, we assume that the real and imaginary parts are equally important in power counting. This is in line with the scattering lengths from Refs. \cite{Kang:2013uia,Dai:2017ont}, where typically real and imaginary parts are of comparable size. Note that there are annihilation states containing soft pions which give rise to long-range effects, but their contributions to elastic scattering are suppressed by powers of $(Q/4\pi f_\pi)^2$, as for other irreducible loops.

With our regulator, the contact interactions, which are (derivatives of) Dirac delta functions in configuration space, are smeared with a spherical well~\cite{Scaldeferri:1996nx,Beane:2000wh,Beane:2001bc}. Schematically, for a channel $c$, 
\begin{equation}
   C_{c} \, \mathcal{O}_{c}\delta^{(3)}(r) 
   \rightarrow 
   \frac{3C_{c}}{4\pi b^3}\,\theta(b-r)\,\mathcal{P}_{c}
   \equiv \left(V_c+iW_c\right)\, \theta(b-r) \, \mathcal{P}_{c} \ ,
\label{eqn:counter}
\end{equation}
where $C_c$ is a complex LEC, $\mathcal{O}_{c}$ is a combination of derivatives, $\mathcal{P}_c$ is the projection operator on channel $c$, and the real short-range parameters $V_c$ and $W_c$ from different channels are independent. 
For a different way of defining the counterterms of the annihilation, see Refs.~\cite{Kang:2013uia, Dai:2017ont}. 

Just like the $N\!N$ case \cite{Nogga:2005hy,PavonValderrama:2005uj}, whether a short-range interaction is needed at LO in a certain channel
hinges on the tensor part of OPE being attractive.
We denote a channel by $^{2I+1\;2S+1}L_J$, where $I$ ($S$) is the total isospin (spin) and $L$ ($J$) is the orbital (total) angular momentum. 
Because the Pauli principle does not apply to the $\overline{N}\!N$ system, the total isospin is independent from other quantum numbers,
unlike $N\!N$. 
Moreover, due to annihilation
there are four times as many phase parameters (phase shifts, inelasticities, and mixing angles) compared to $np$ scattering, $\it viz.$ 8 phase parameters are required for $J=0$ and 20 phase parameters for each value of $J\neq0$ \cite{Timmermans:1995xb,Tim84}.
For uncoupled partial waves, the $S$ matrix is just a complex number, written as
\begin{equation}
S = \eta\,e^{2i\delta} \ ,
\end{equation}
where $\delta$ is the phase shift and $\eta$ ($0\leq\eta\leq1$) is the inelasticity due to 
annihilation. For the coupled spin-triplet partial waves with $L=J\mp1$ ($J\ge1$) the $2\times 2$ $S$ matrix is parametrized 
as~\cite{Timmermans:1994pg,Zhou:2012ui}
\begin{equation}
S^J = \exp(i\bar{\delta}) \,
\exp(i\varepsilon_{J}\sigma_x) \,\,
H^J  \, \exp(i\varepsilon_{J}\sigma_x) \,
\exp(i\bar{\delta})~,
\end{equation}
where $\bar{\delta}$ is a diagonal matrix with real entries
$\delta_{J-1,J}$ and $\delta_{J+1,J}$, and $\varepsilon_{J}$
is the mixing angle.
The matrix $H^J$ parametrizes the inelasticities. It is written as
\begin{equation}
H^J = \exp(-i\omega_J\sigma_y) \,
\left( \begin{array}{cc}
\eta_{J-1,J} &       0        \\
0        &   \eta_{J+1,J}
\end{array} \right) \,
\exp(i\omega_J\sigma_y)~,
\label{Eq:Klarsfeld}               
\end{equation}
where $\eta_{J-1,J}$ and $\eta_{J+1,J}$ are the inelasticities
($0 \leq\eta_{J\mp1,J}\leq 1$) and
$\omega_J$ is the mixing angle for inelasticity.
The $S$ matrix for the coupled partial waves is thus written in terms of six parameters. 

The sign of the singular tensor force depends on the matrix elements of $\boldsymbol{\tau}_1\cdot\boldsymbol{\tau}_2$ and $S_{12}$.
The operator $\boldsymbol{\tau}_1\cdot\boldsymbol{\tau}_2=2I(I+1)-3=-3,+1$ for $I=0,1$, respectively, always makes one channel attractive and its isospin partner repulsive. The matrix elements of the tensor force have the same properties as for $N\!N$. The spin-singlet channels do not have singular long-range forces because the matrix elements of $S_{12}$ between those channels vanish. 
In the coupled channels, the eigenvalues of $S_{12}$ are $-4$ and $2$, 
regardless of $J$: one eigenchannel is attractive, the other repulsive. On account of $\boldsymbol{\tau}_1\cdot\boldsymbol{\tau}_2$, the isoscalar channel is most attractive. Therefore, one short-range interaction is expected at LO for every coupled channel where OPE is iterated to all orders. This leaves us with the uncoupled spin-triplet channels, which are $P$ waves in this paper. We tabulate in Table~\ref{tab:signs} the values of the matrix element of $-\boldsymbol{\tau}_1\cdot\boldsymbol{\tau}_2 \,S_{12}$ in these channels.

\begin{table}[tb]
	\centering
	\caption{Values of $-\boldsymbol{\tau}_1\cdot\boldsymbol{\tau}_2 \,S_{12}$ in the uncoupled spin-triplet $P$ waves. 
	}
	\tabcolsep=1.8em
	\renewcommand{\arraystretch}{0.9}
	\begin{tabular}{ccccccccc}
		\hline
		\hline
		Partial wave & $^{13}P_0$ & $^{33}P_0$ & $^{13}P_1$ &   $^{33}P_1$ \\ 
		\hline
		$-\langle\boldsymbol{\tau}_1\cdot\boldsymbol{\tau}_2 \,S_{12}\rangle$
		& $-12$ & $+4$ & $+6$ & $-2$ \\ 
		\hline
		\hline
	\end{tabular}
	\label{tab:signs}
\end{table}

In summary, as far as $S$ and $P$ waves are concerned, a short-range interaction is expected at LO in:
{\it i}) the lower wave of each of the coupled channels, that is, $^{13}S_1$, $^{33}S_1$, $^{13}P_2$, and $^{33}P_2$;
{\it ii}) the uncoupled channels $^{13}P_0$ and $^{33}P_1$.
In contrast, NDA predicts an LO short-range interaction only in the $S$ waves. Since we see no reason to demote an LEC, we also include short-range interactions in $^{11}S_0$ and $^{31}S_0$. (These interactions break chiral symmetry \cite{Kaplan:1996xu,Beane:2001bc}.) In the next section we confirm numerically the need for the additional $P$-wave LECs and compare the phase shifts they yield with the PWA.

\section{Renormalization and results} \label{sec:Strategy}

Our strategy is the following. The parameters of OPE are known, and we first solve the partial-wave Schr\"odinger equation with $V_c=W_c=0$. 
We check the cutoff dependence of the phase shifts and mixing angles before renormalization.
(Inelasticities are trivial, $\eta_{J\mp 1,J}=1$, and their mixing angles $\omega_J$ are not well defined by Eq.~(\ref{Eq:Klarsfeld}).) 
We show the results at various representative laboratory energies 
($T_{\rm lab} = 10, 50, 100$ MeV).
For each partial wave where significant cutoff dependence is found, we 
adjust the short-range spherical well as function of $b$. Here $V_c$ and $W_c$ play the role of counterterms, whose cutoff dependence ensures that physical observables be cutoff independent within error bars. We determine $V_c$ and $W_c$ from the phase shift and inelasticity of the Groningen PWA~\cite{Zhou:2012ui} at some energy. The PWA results we show here contain more points but are consistent with the values in Tables VIII and IX of Ref.~\cite{Zhou:2012ui}, which assume isospin symmetry. 
The fit results are not significantly sensitive to the choice of fitting energy, which is taken to be $T_{\rm lab}=20$ MeV. 
After verifying that cutoff independence is achieved, we 
take a cutoff value $\Lambda \sim M_{\rm QCD}$ (specifically, $\Lambda= 5$ fm$^{-1}$) and compare phase shifts, inelasticities, and mixing angles as functions of the laboratory momentum $p_{\rm lab}$
with the PWA. 

Since the tensor force is the determining factor for the renormalization of OPE, we split the analysis between spin-singlet and triplet channels.

\subsection{Singlet channels} \label{singlets}

\begin{figure}[tb]
	\centering
	\includegraphics[width=0.45\textwidth]{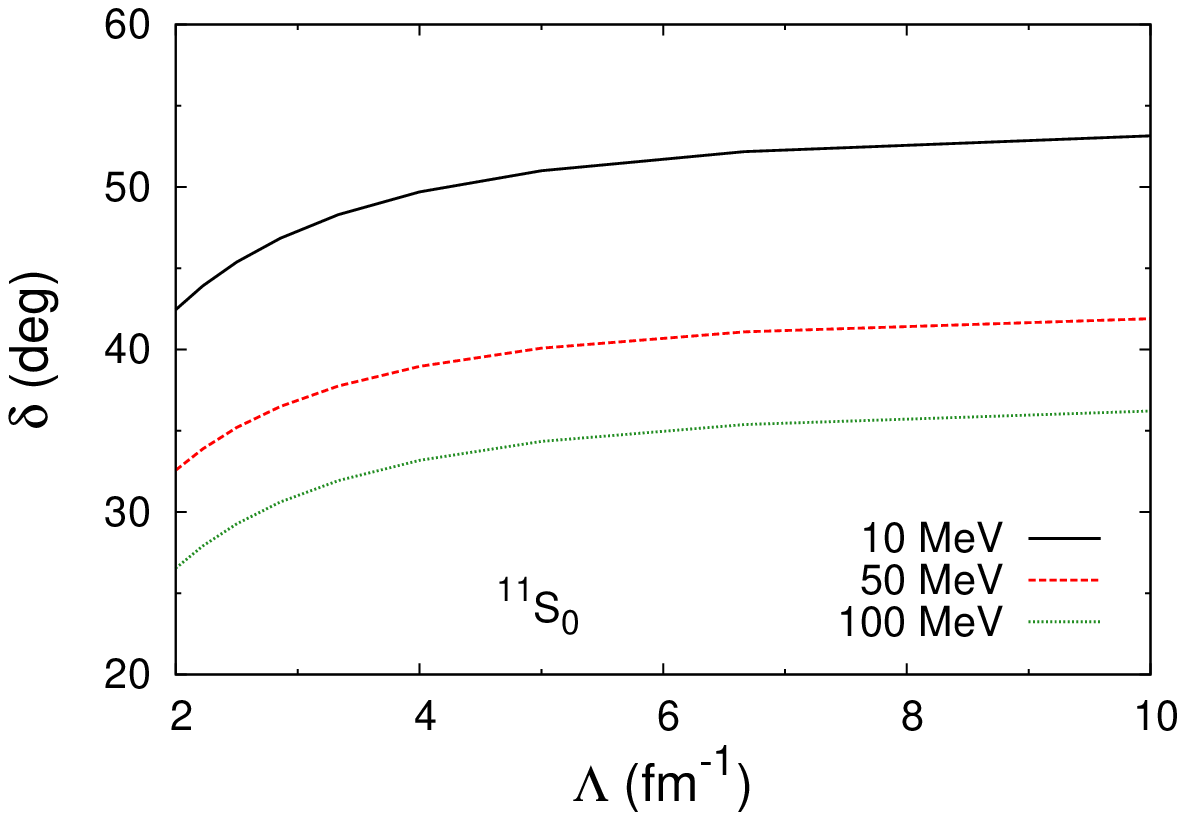} \hspace{2em}
	\includegraphics[width=0.45\textwidth]{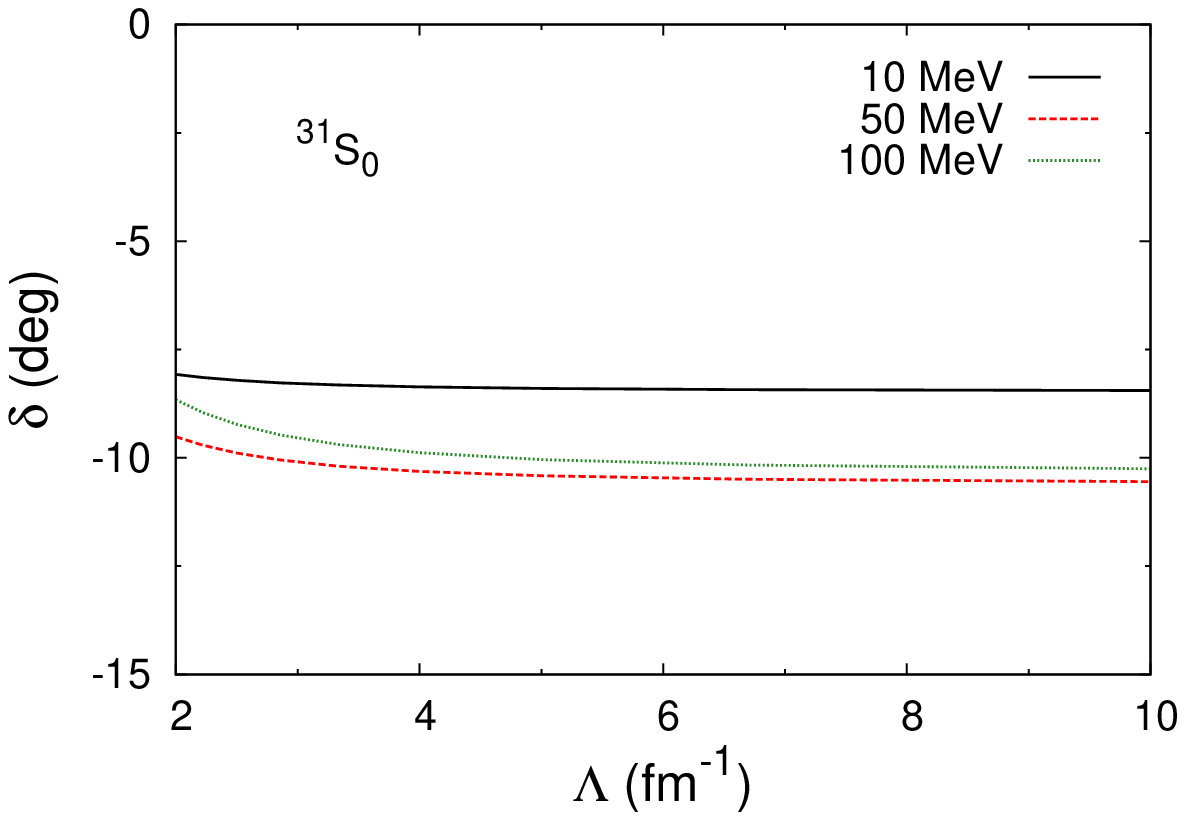}\\	
	\includegraphics[width=0.45\textwidth]{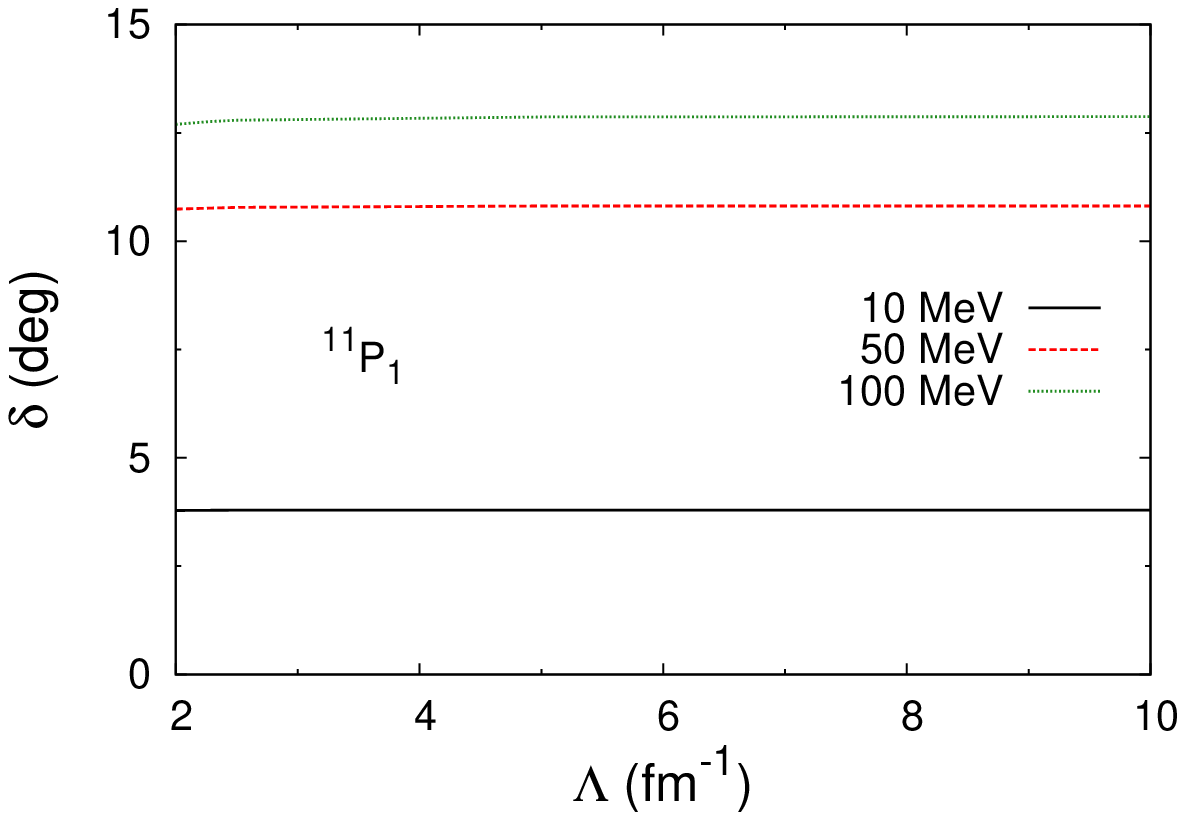} \hspace{2em}
	\includegraphics[width=0.45\textwidth]{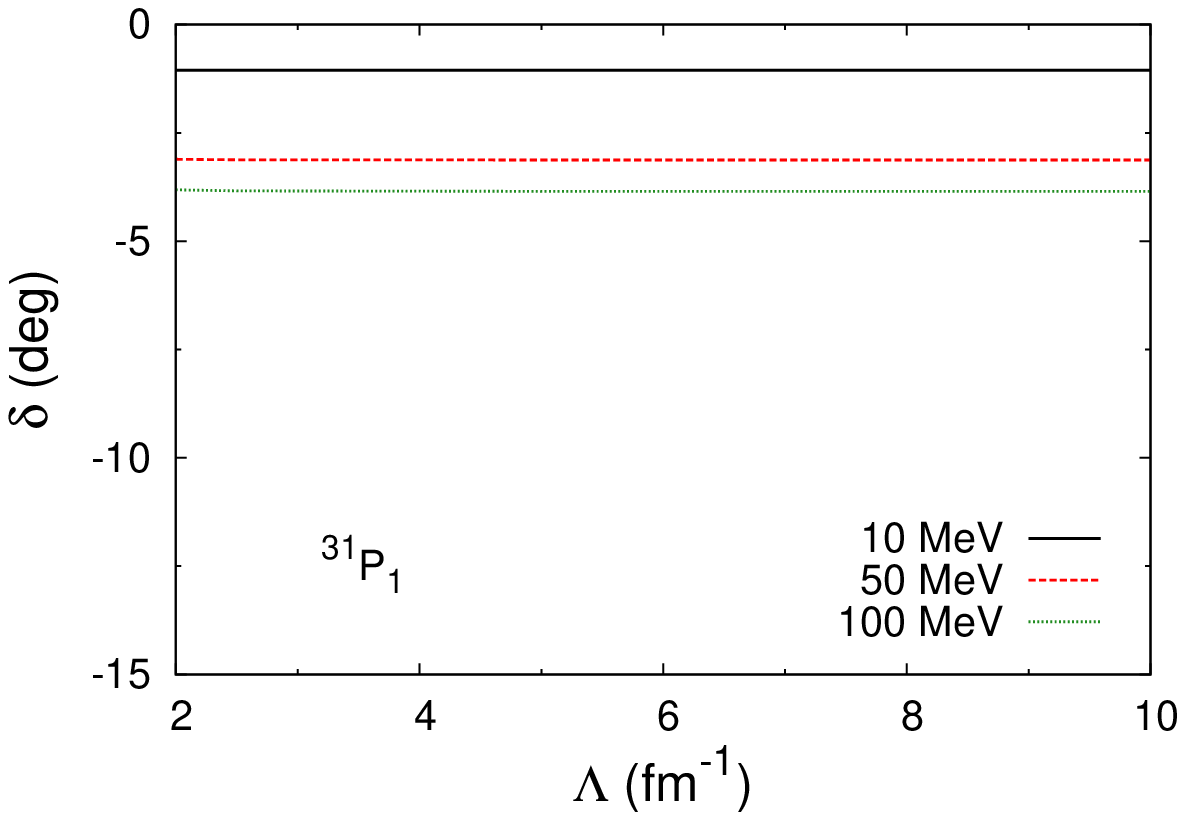}\\
	\caption{\label{SingletPhases_L1}{(Color online) Cutoff dependence of the phase shifts in the spin-singlet $S$ and $P$ waves at the laboratory energies of 10 MeV (black solid line), 50 MeV (red dashed line), and 100 MeV (green dotted line), for $V_c=W_c=0$.}}
\end{figure}

OPE is not singular in the spin-singlet channels and by itself generates no essential cutoff dependence in the solution of the Schr\"odinger equation. Phase shifts in the lowest waves --- $^{11}S_0$, $^{31}S_0$, $^{11}P_1$, and $^{31}P_1$ --- can be seen in Fig.~\ref{SingletPhases_L1} for various laboratory energies. 
All phase shifts approach finite values as the cutoff $\Lambda$ increases, with the fastest variation for $\Lambda\simle 4\pi f_\pi\sim M_{\rm QCD}$. Given that $\vec\sigma_1\cdot\vec\sigma_2=-3$, isospin-singlet phases are attractive and relatively large on account of $\boldsymbol{\tau}_1\cdot\boldsymbol{\tau}_2$ .

\begin{figure}[tb]
	\centering
	\includegraphics[width=0.45\textwidth]{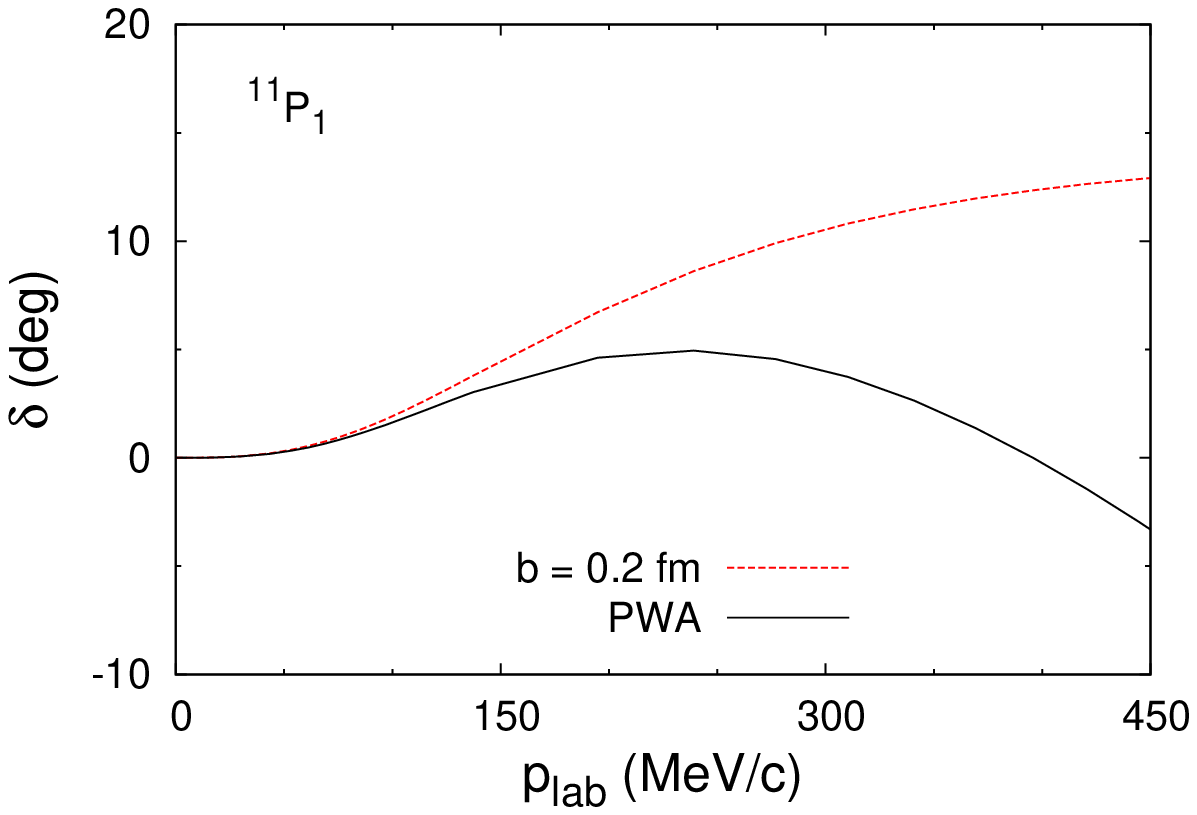} \hspace{2em}	
	\includegraphics[width=0.45\textwidth]{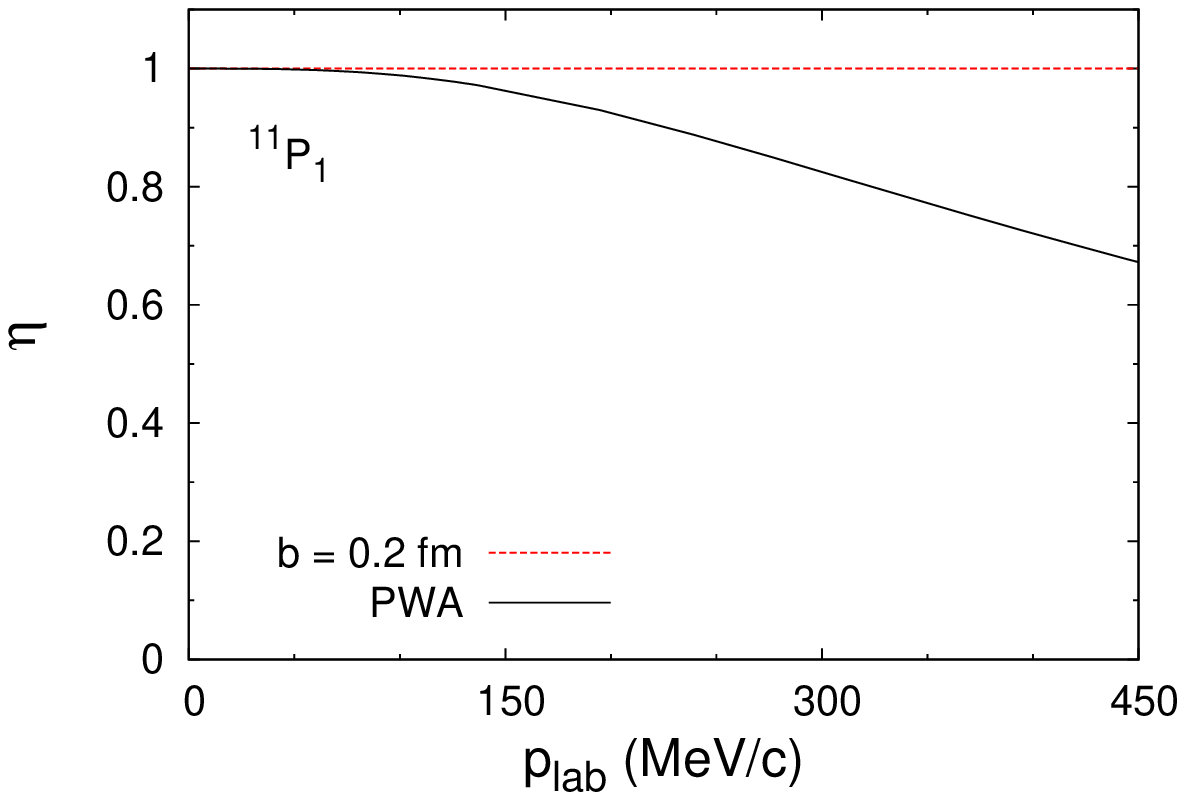} \\   
	\includegraphics[width=0.45\textwidth]{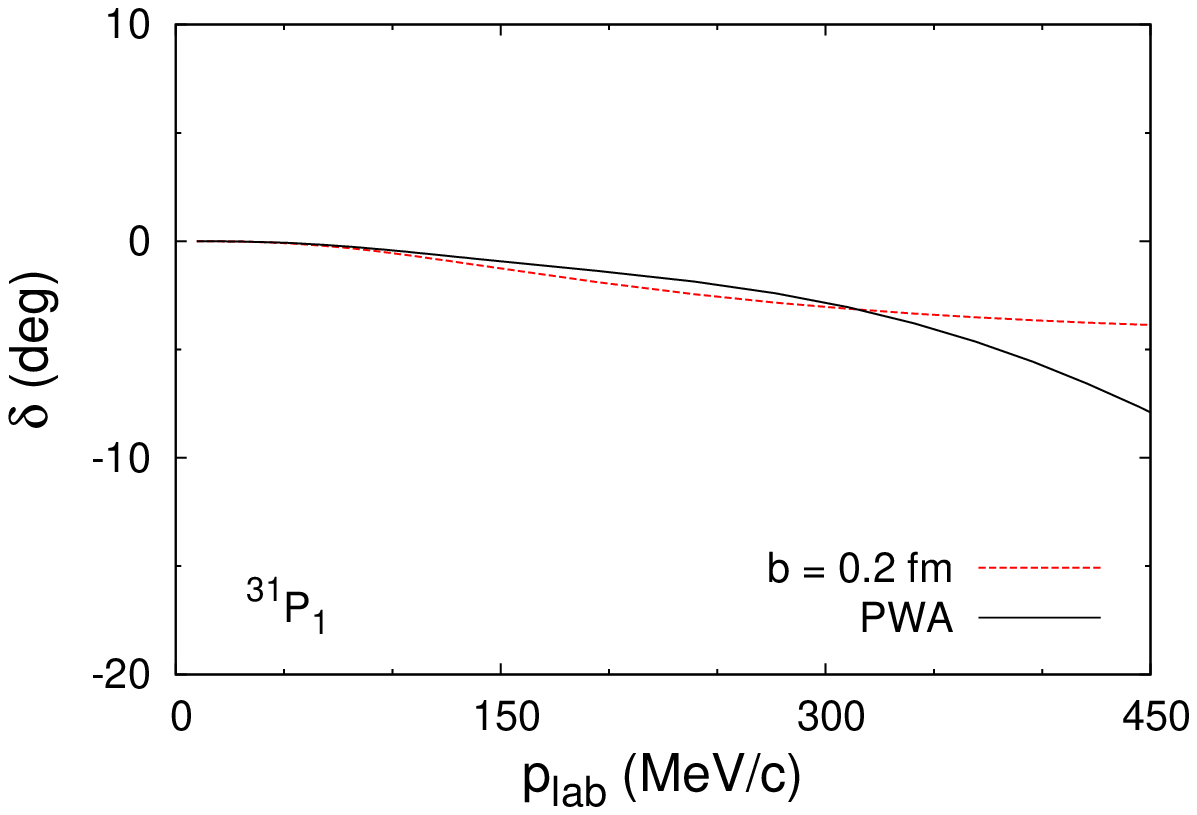} \hspace{2em}	
    \includegraphics[width=0.45\textwidth]{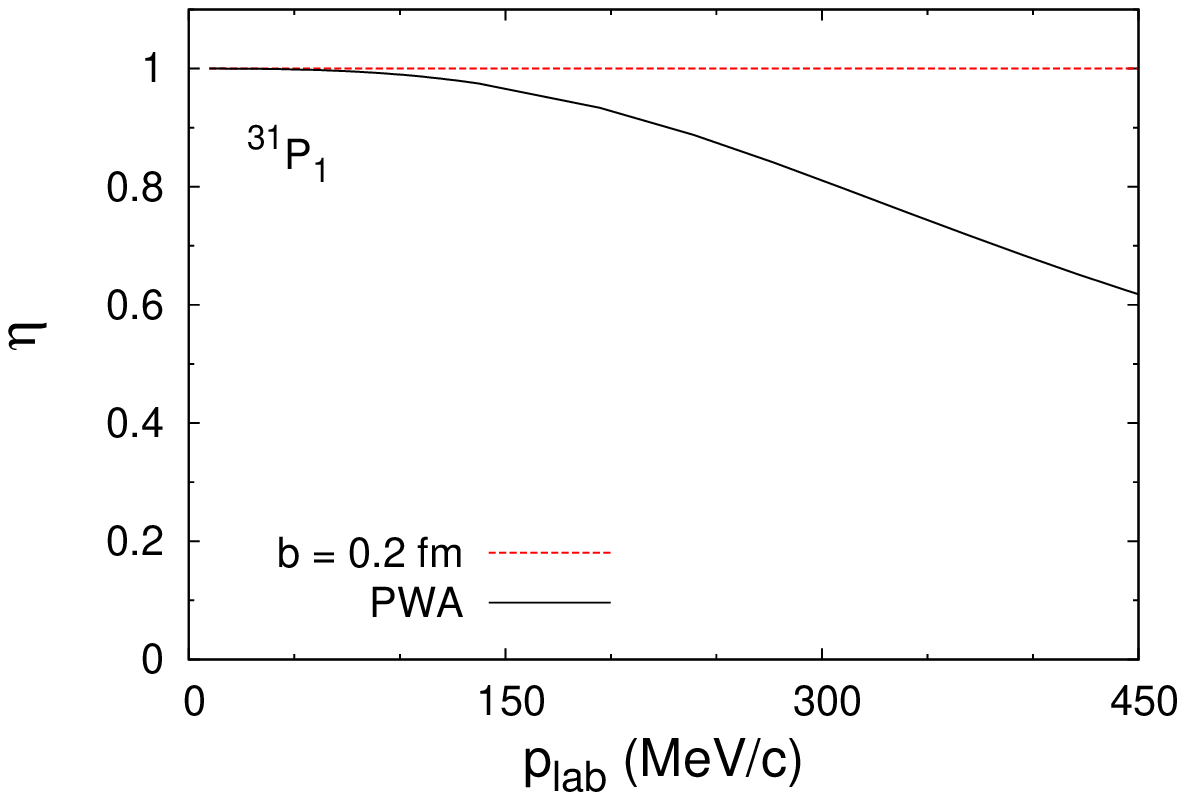} \\  
	\caption{\label{Phase_plab_SingletP}{(Color online) 
	Phase shifts (left panels) and inelasticities (right panels) of the spin-singlet 
	$P$ waves against 
	laboratory momentum.
	The (red) dashed lines are from iterated one-pion exchange for $b=0.2$ fm and $V_c=W_c=0$, while (black) solid lines are the results of the PWA~\cite{Zhou:2012ui,Zhou:2013}.}}
\end{figure}

In the $P$ (and higher) waves NDA suggests LECs only at N$^2$LO (and higher orders). The phase shifts and inelasticities for the $^{11}P_1$ and $^{31}P_1$ channels from iterated OPE are shown as function of the laboratory momentum 
in Fig. \ref{Phase_plab_SingletP} for a cutoff well into the region where very little cutoff dependence is seen in Fig. \ref{SingletPhases_L1}. A good description of the empirical values is found for $p_{\rm lab}\lesssim 250$ MeV/$c$, discrepancies increasing as momentum increases --- as expected in a low-energy EFT. In $^{11}P_1$, however, the phase shift becomes repulsive at large momentum, indicating that another contribution becomes as important as OPE and challenging the convergence of a power counting where OPE is treated alone as LO and iterated. In fact, the magnitudes of $\delta$ and $1-\eta$ are relatively small, suggesting that pions might be perturbative in these waves as in the corresponding $N\!N$ cases \cite{PavonValderrama:2016lqn}.

\begin{figure}[tb]
   \centering
   \includegraphics[width=0.45\textwidth]{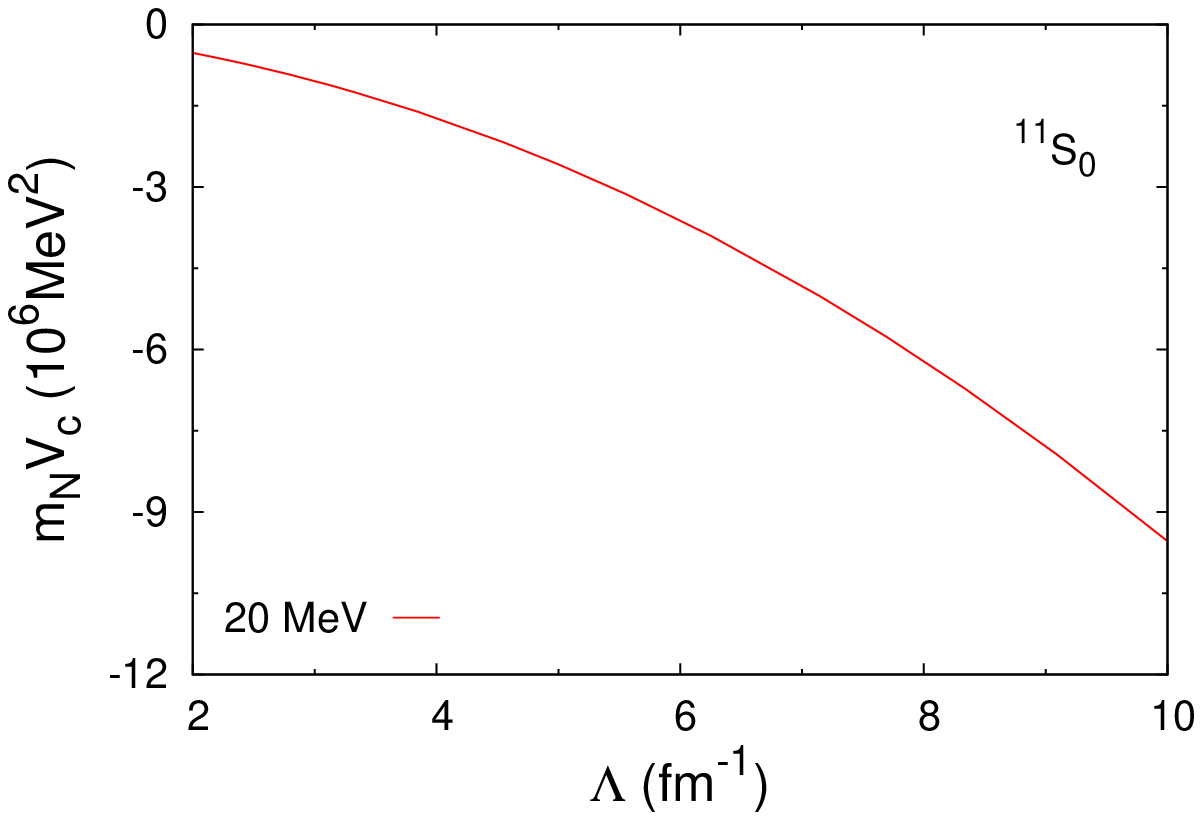} \hspace{2em}
   \includegraphics[width=0.45\textwidth]{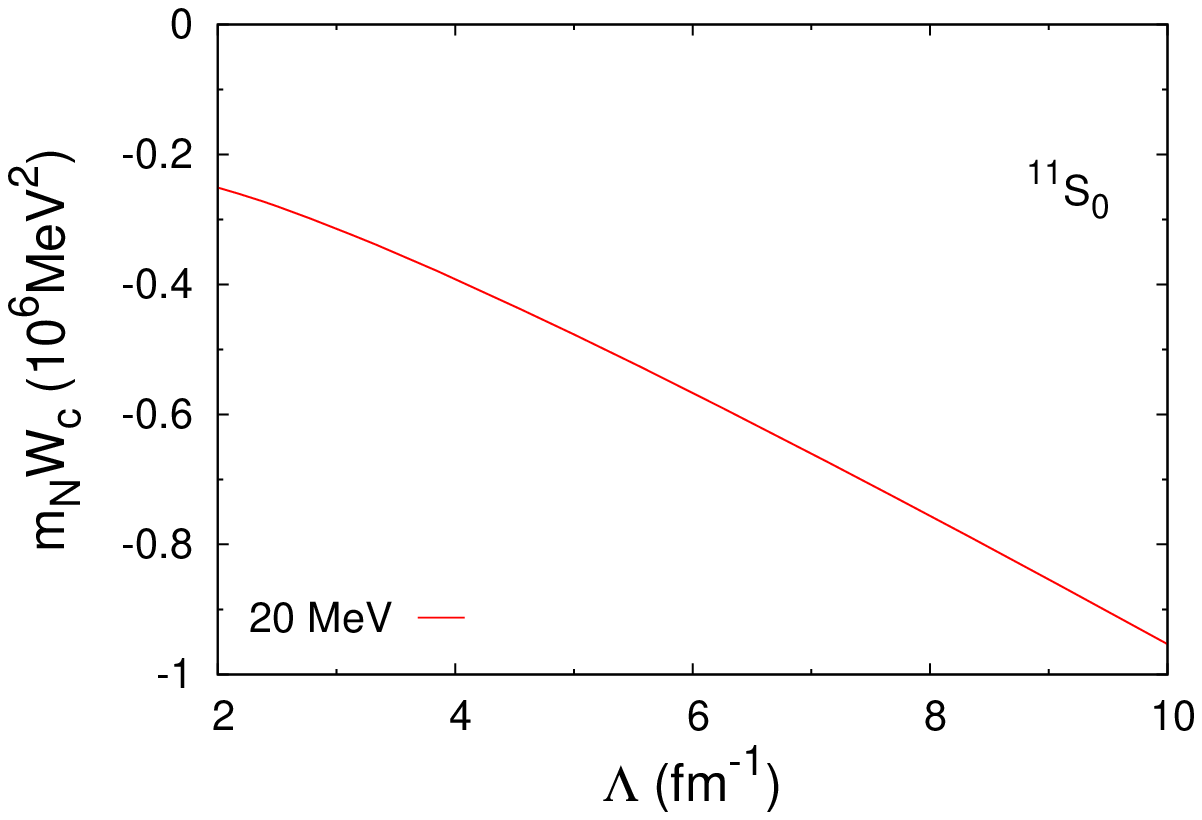} \\   
   \includegraphics[width=0.45\textwidth]{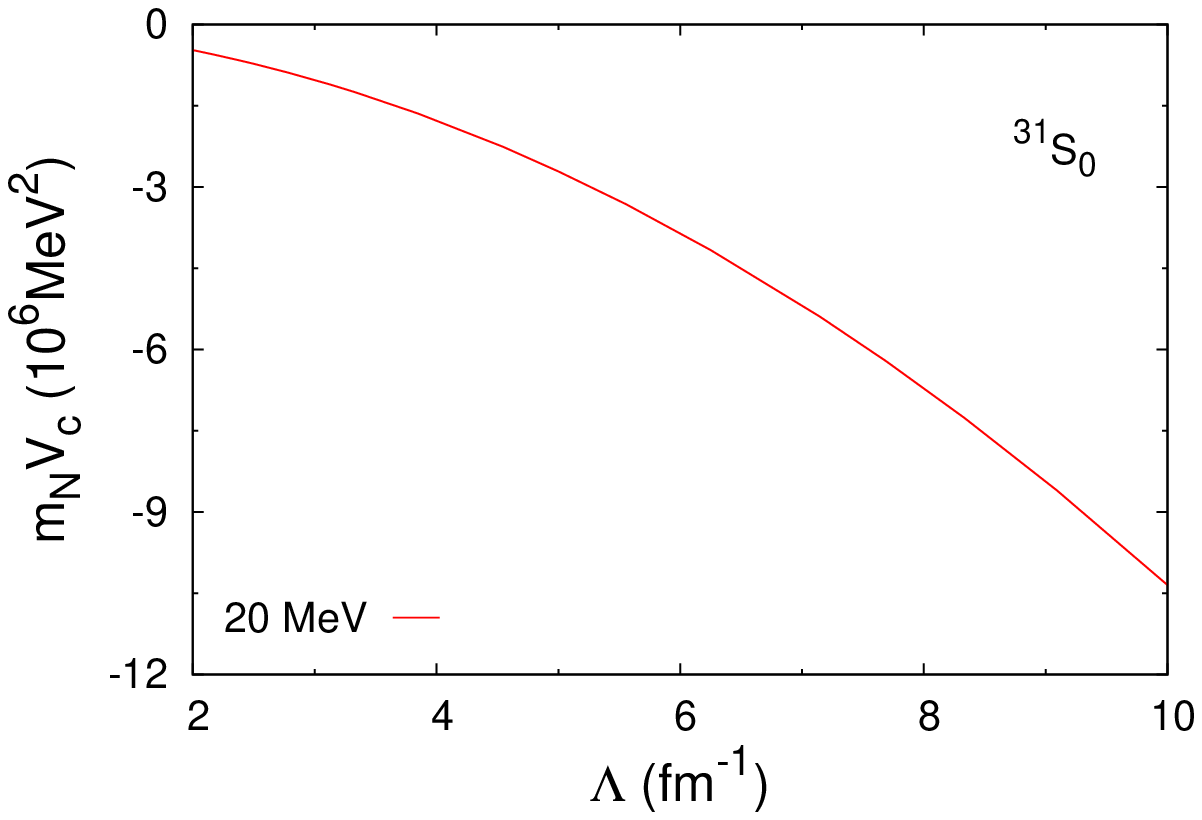} \hspace{2em}
   \includegraphics[width=0.45\textwidth]{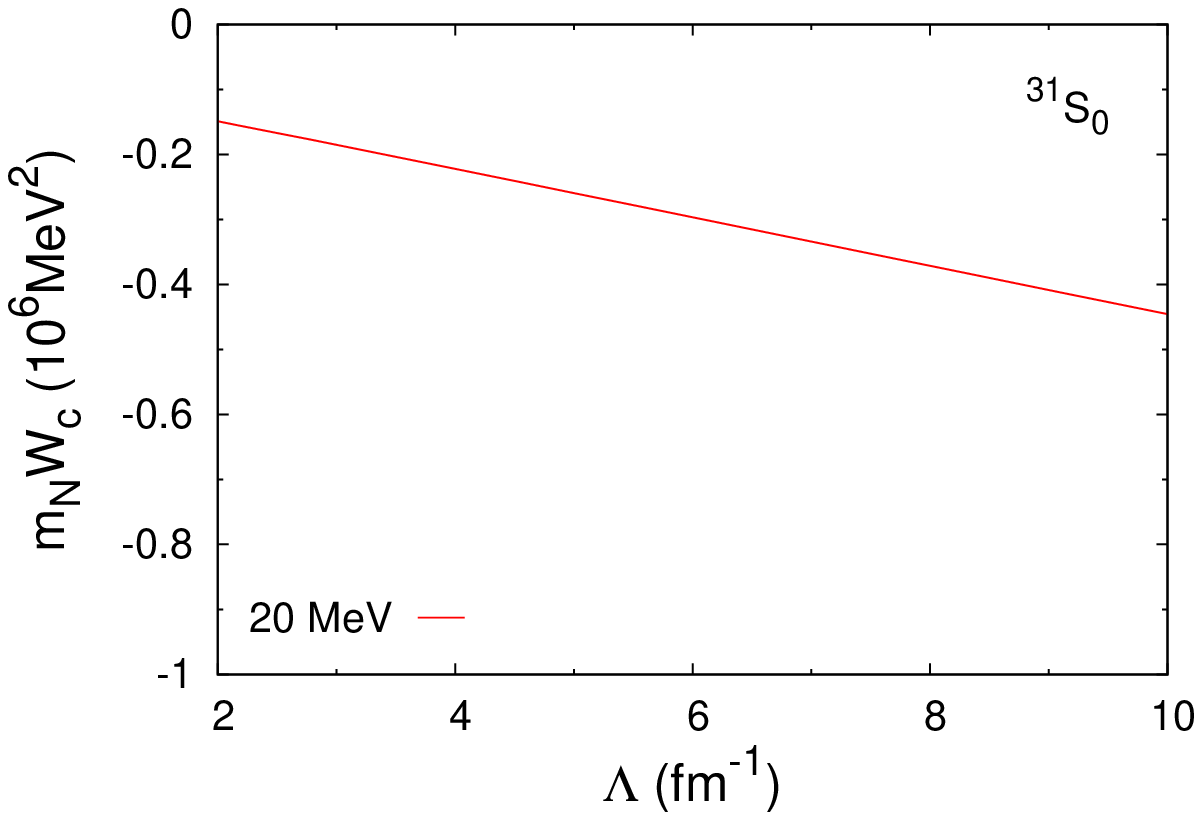} \\
   \caption{\label{Counter_cut_SingletS}{(Color online) 
   Cutoff dependence of $m_N V_c$ (left panels) and $m_N W_c$ (right panels) for the spin-singlet $S$ waves. The PWA phase shifts and inelasticities are fitted at $T_{\rm lab}=20$ MeV.}}
\end{figure}

\begin{figure}[tb]
   \centering
   \includegraphics[width=0.45\textwidth]{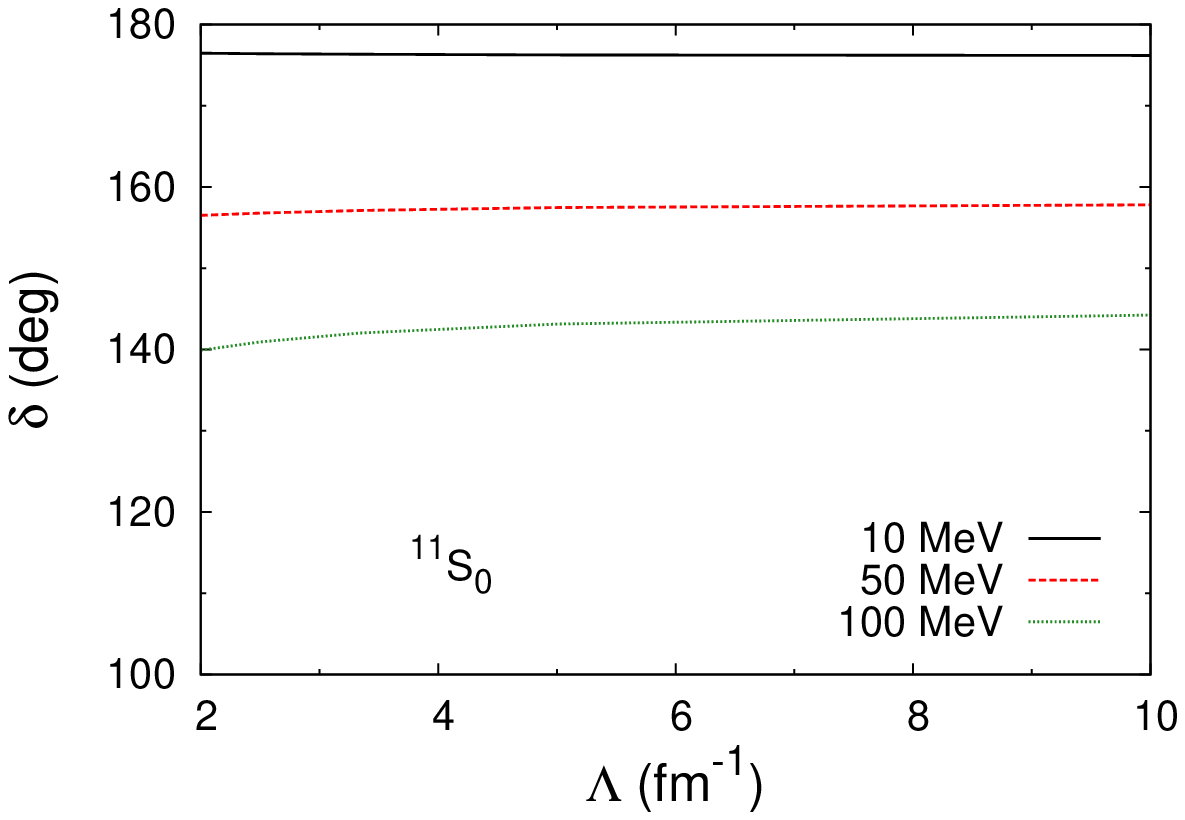} \hspace{2em}
   \includegraphics[width=0.45\textwidth]{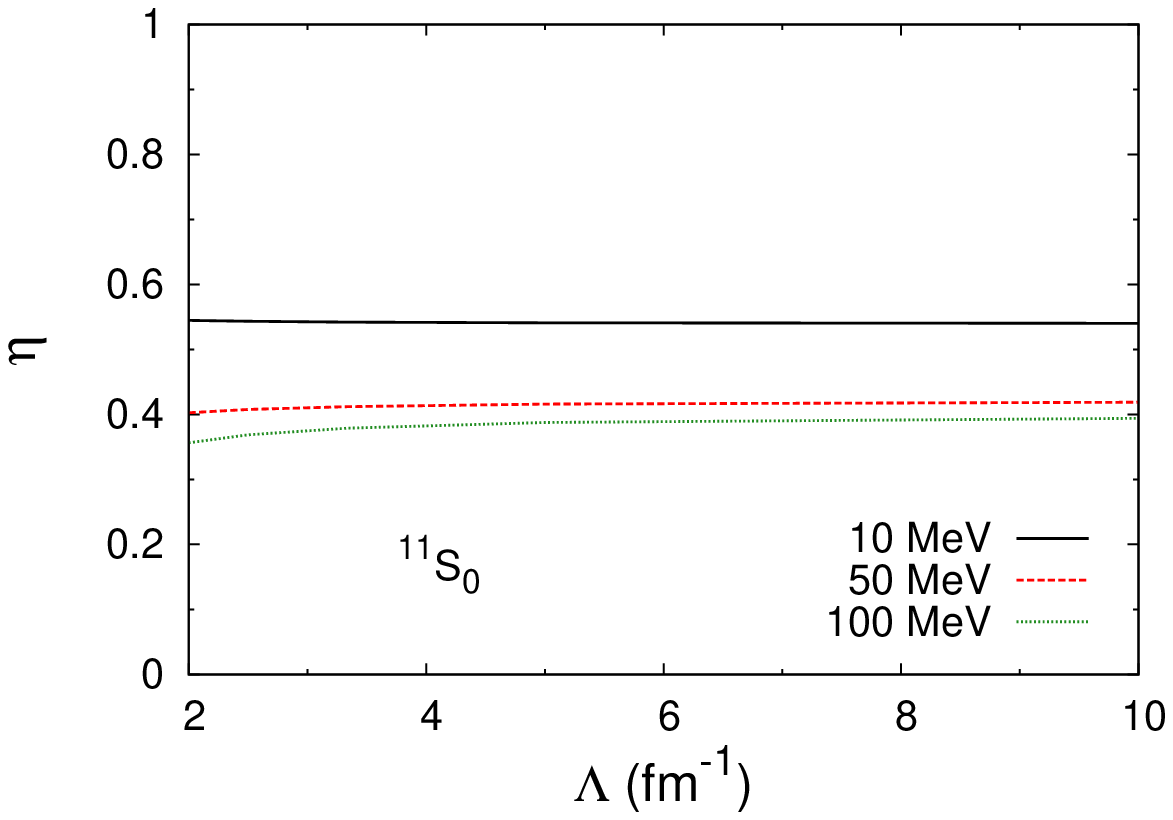} \\   
   \includegraphics[width=0.45\textwidth]{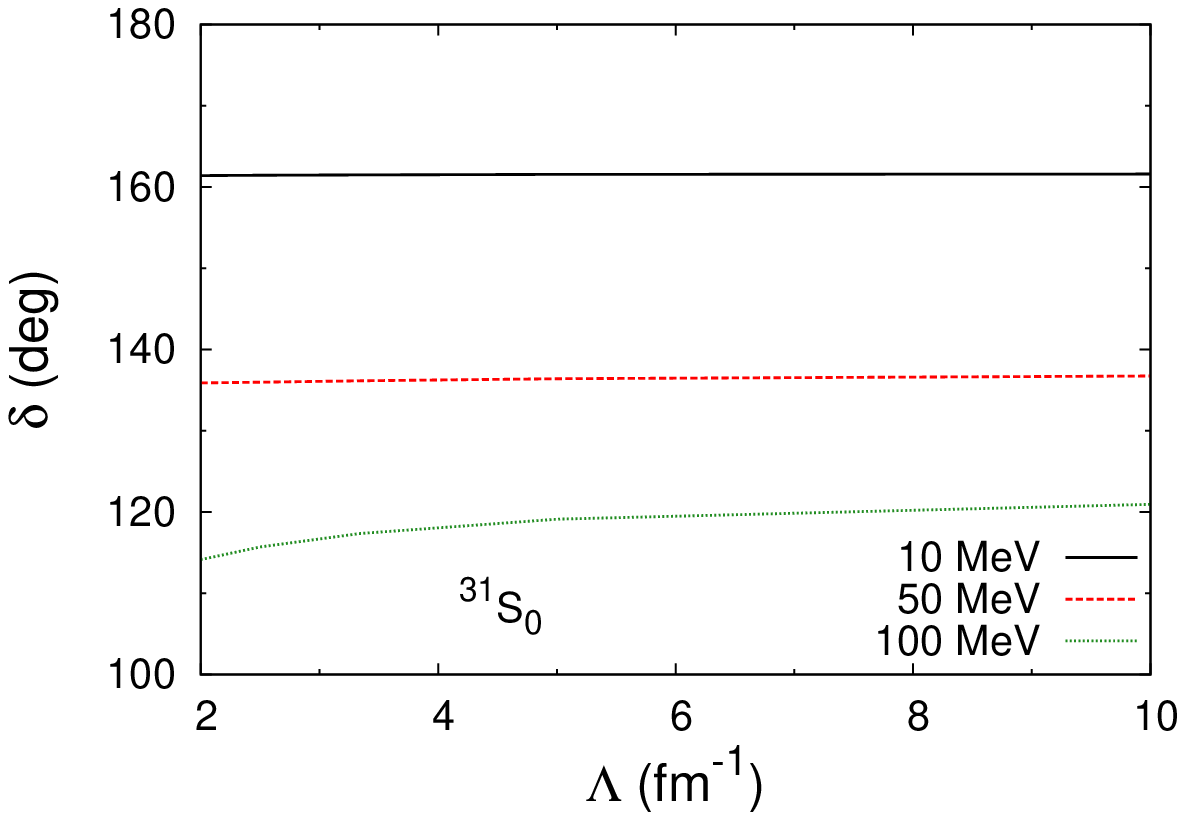} \hspace{2em}
   \includegraphics[width=0.45\textwidth]{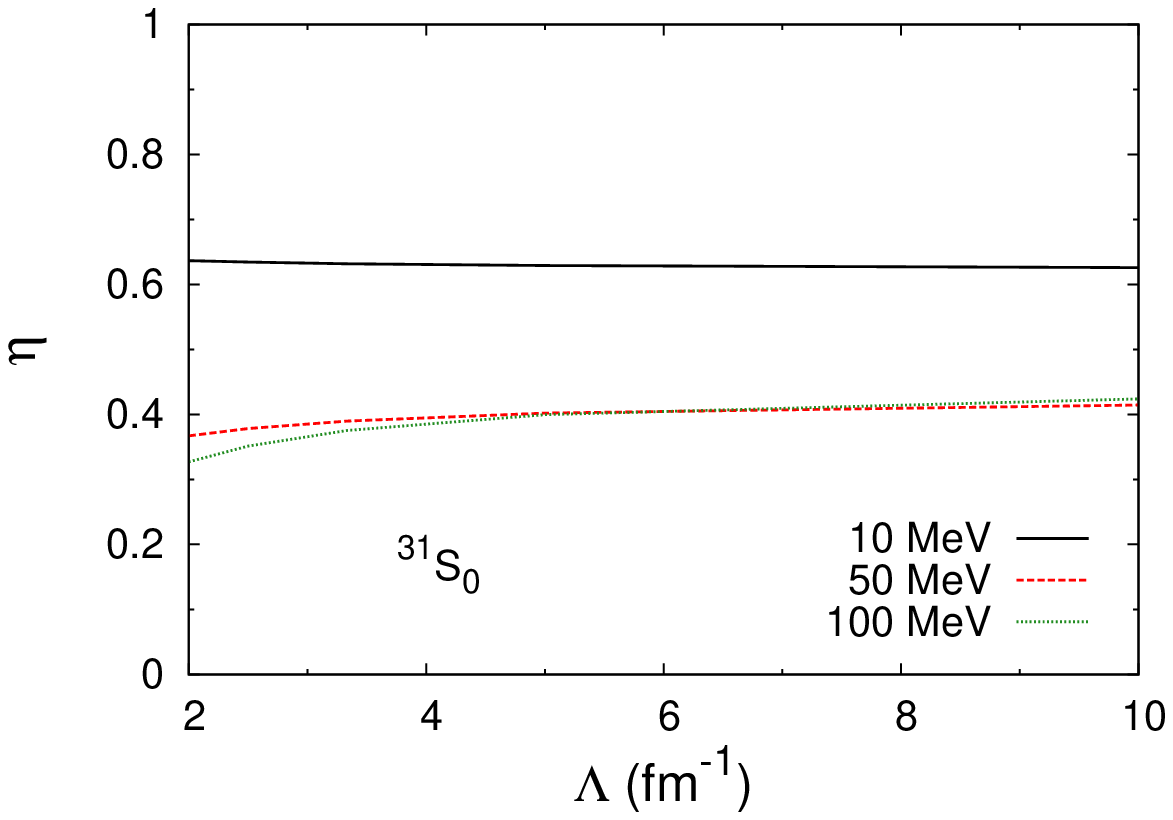}\\
   \caption{\label{SingletPhases_S1}{(Color online) Residual cutoff dependence of the phase shifts and inelasticities in the spin-singlet $S$ waves at the laboratory energies of 10 MeV (black solid line), 50 MeV (red dashed line), and 100 MeV (green dotted line), for $V_c$ and $W_c$ in Fig. \ref{Counter_cut_SingletS}.}}
\end{figure}

\begin{figure}[tb]
	\centering
	\includegraphics[width=0.45\textwidth]{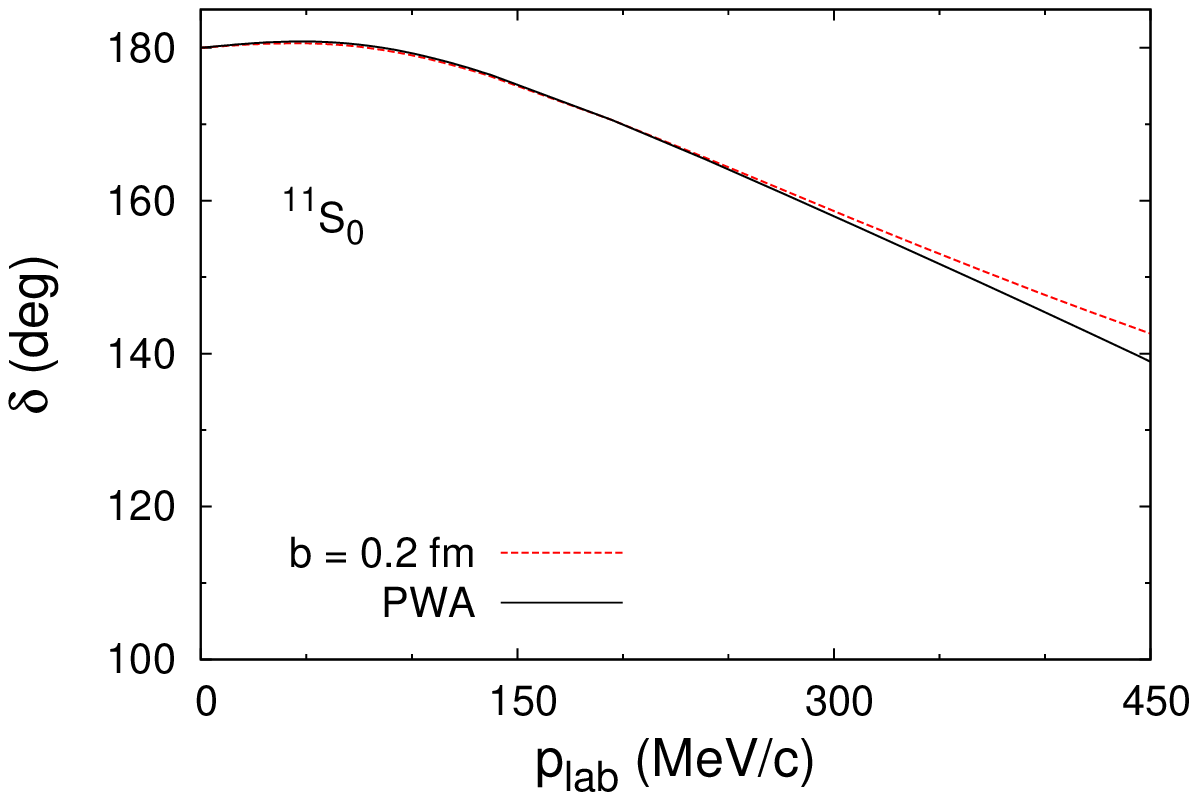} \hspace{2em}
	\includegraphics[width=0.45\textwidth]{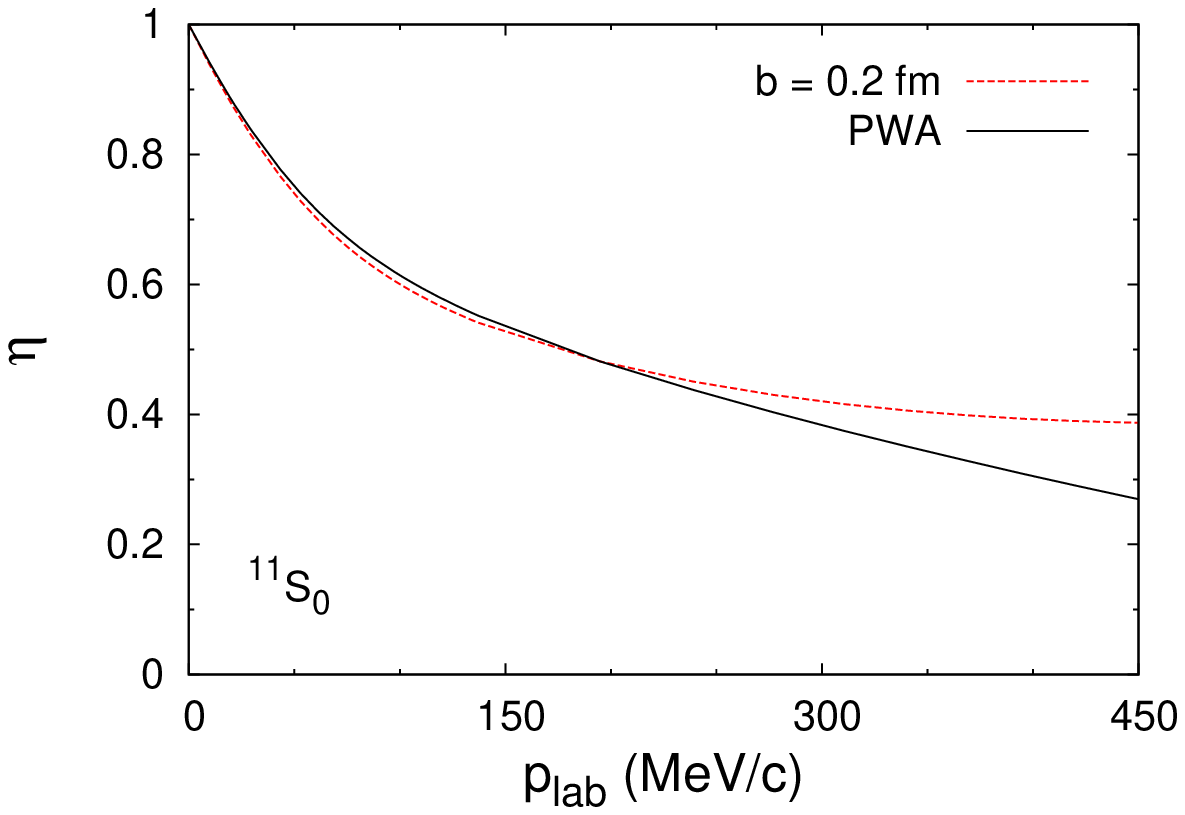} \\   
	\includegraphics[width=0.45\textwidth]{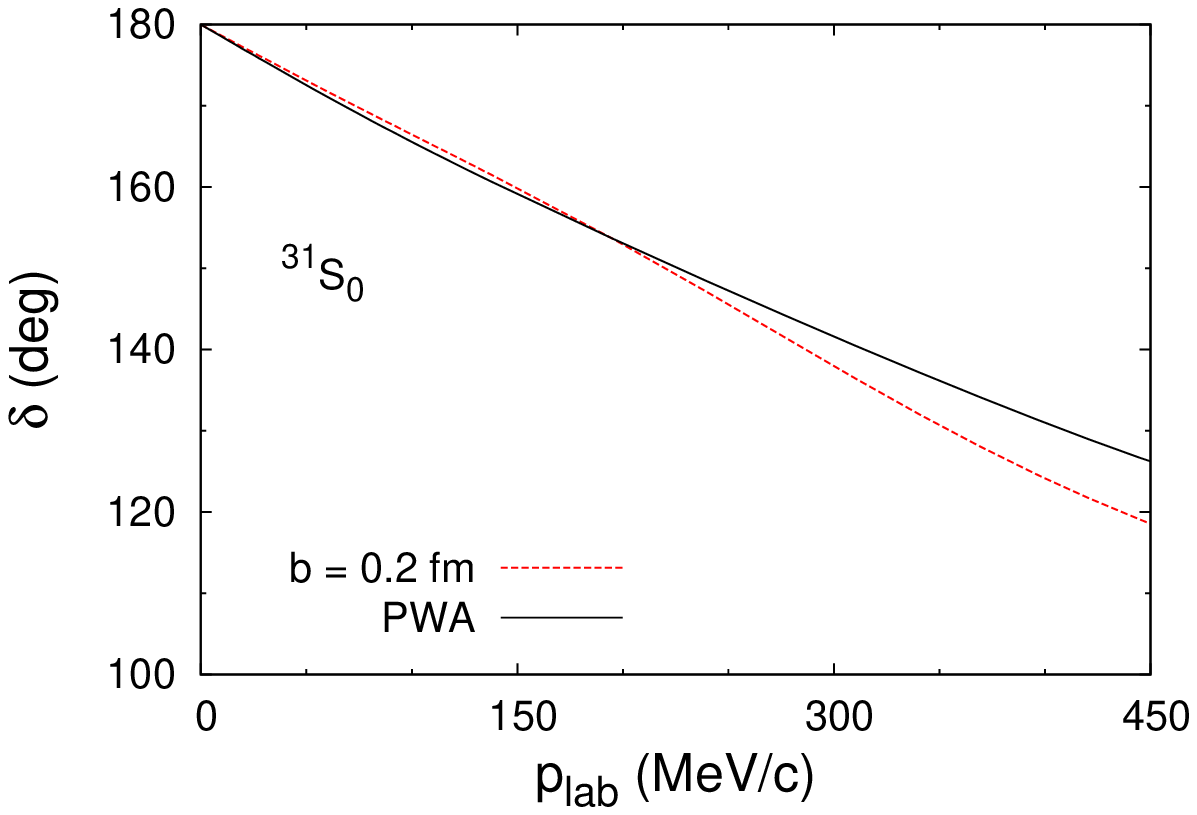} \hspace{2em}
	\includegraphics[width=0.45\textwidth]{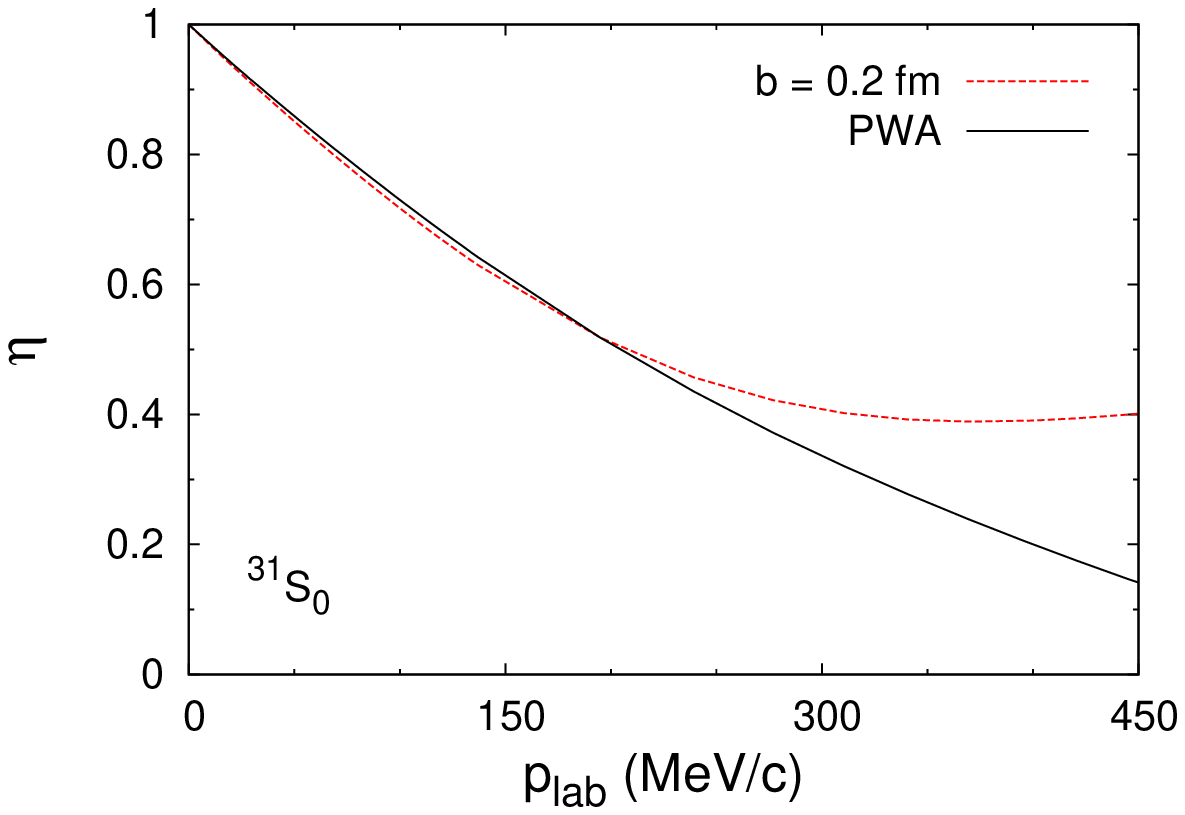}\\
	\caption{\label{Phase_plab_SingletS}{(Color online) 
	Phase shifts (left panels) and inelasticities (right panels) of the spin-singlet 
	$S$ waves against 
	laboratory momentum.
	The (red) dashed lines are from iterated one-pion exchange for $b=0.2$ fm and $V_c$, $W_c$ from Table \ref{tab:potentials1}, while (black) solid lines are the results of the PWA~\cite{Zhou:2012ui,Zhou:2013}.}}
\end{figure}

In contrast, NDA prescribes LECs in the $S$ waves even though they are not needed for renormalization of iterated OPE. Since there is no obvious reason to demote the short-range interactions, we include them and fit the corresponding $V_c$, $W_c$ to the phase shifts and inelasticities of the PWA at the chosen low energy. The resulting LECs, all attractive, are shown in Fig. \ref{Counter_cut_SingletS}. The magnitudes of $V_c$ are about $10$ to $20$ times larger than those of $W_c$ for large $\Lambda$. The simultaneous iteration of OPE and short-range interactions induces new cutoff dependence, shown in Fig. \ref{SingletPhases_S1}. Still, the amplitude is renormalized (at least at a fixed pion mass), and phase shifts and inelasticities approach new values as the cutoff increases. 
The LECs increase the attraction of the $^{11}S_0$ phase shift,  
overcome the $^{31}S_0$ OPE repulsion, and introduce a non-zero inelasticity. The resulting phase shifts and inelasticities at 
a large cutoff, where $V_c$ and $W_c$ take the values given in Table \ref{tab:potentials1}, are shown as functions of the laboratory momentum in Fig. \ref{Phase_plab_SingletS}. A very good description of the empirical values is obtained through most of the displayed momentum range. 
(The non-monotonic behavior of the $^{11}S_0$ phase shift at low momentum has no significance because it is within the errors we expect for both the PWA and the EFT.)
Neither the $^{11}S_0$ 
strong attraction nor the significant inelasticities could be reproduced without the LECs, in line with the NDA expectation.

Thus, a satisfactory, renormalized description of the spin-singlet channels is obtained at low energies with a LO that consists of iterated OPE and $S$-wave short-range interactions. However, the non-monotonic behavior of the $^{11}P_1$ phase shift in the PWA reflects a short-range repulsion, which at $p_{\rm lab}\simeq 400$ MeV/$c$ entirely cancels the OPE attraction. The expansion of the amplitude for momenta in that region will not converge unless the two effects are accounted for at the same order. The extent to which pions can be treated perturbatively in these waves should be investigated in future work. While this is the case in higher $N\!N$ partial waves \cite{PavonValderrama:2016lqn}, the effects of the additional OPE strength in $I=0$ $\overline{N}\!N$ waves need to be studied. We turn now to the spin-triplet channels where renormalization of OPE is more dramatic.

\begin{table}[h]
	\centering
	\caption{Values of the real ($V_c$) and imaginary ($W_c$) components of the short-range potential at $b=0.2$ fm for the uncoupled $S$ and $P$ partial waves, obtained by fitting to the PWA ``data" ~\cite{Zhou:2012ui,Zhou:2013} at $T_{\rm lab}=20$ MeV.} 
	\tabcolsep=2.4em
	\renewcommand{\arraystretch}{0.9}
	\begin{tabular}{cd{4.2}d{4.2}d{4.2}d{4.2}}
		\hline
		\hline
		Partial wave &\multicolumn{1}{c}{$^{11}S_0$}&\multicolumn{1}{c}{$^{31}S_0$}&\multicolumn{1}{c}{$^{13}P_0$} & \multicolumn{1}{c}{$^{33}P_1$} \\ 
		\hline
		$V_c$(fm$^{-1}$)     & -13.9 & -14.6 & -147.4 & -3.0 \\
		$W_c$(fm$^{-1}$)     &  -2.6 & -1.4 & -25.5 & -22.2 \\
		\hline
		\hline
	\end{tabular}
	\label{tab:potentials1}
\end{table}

\subsection{Triplet channels} \label{triplets}

\begin{figure}[tb]
	\centering
    \includegraphics[width=0.45\textwidth]{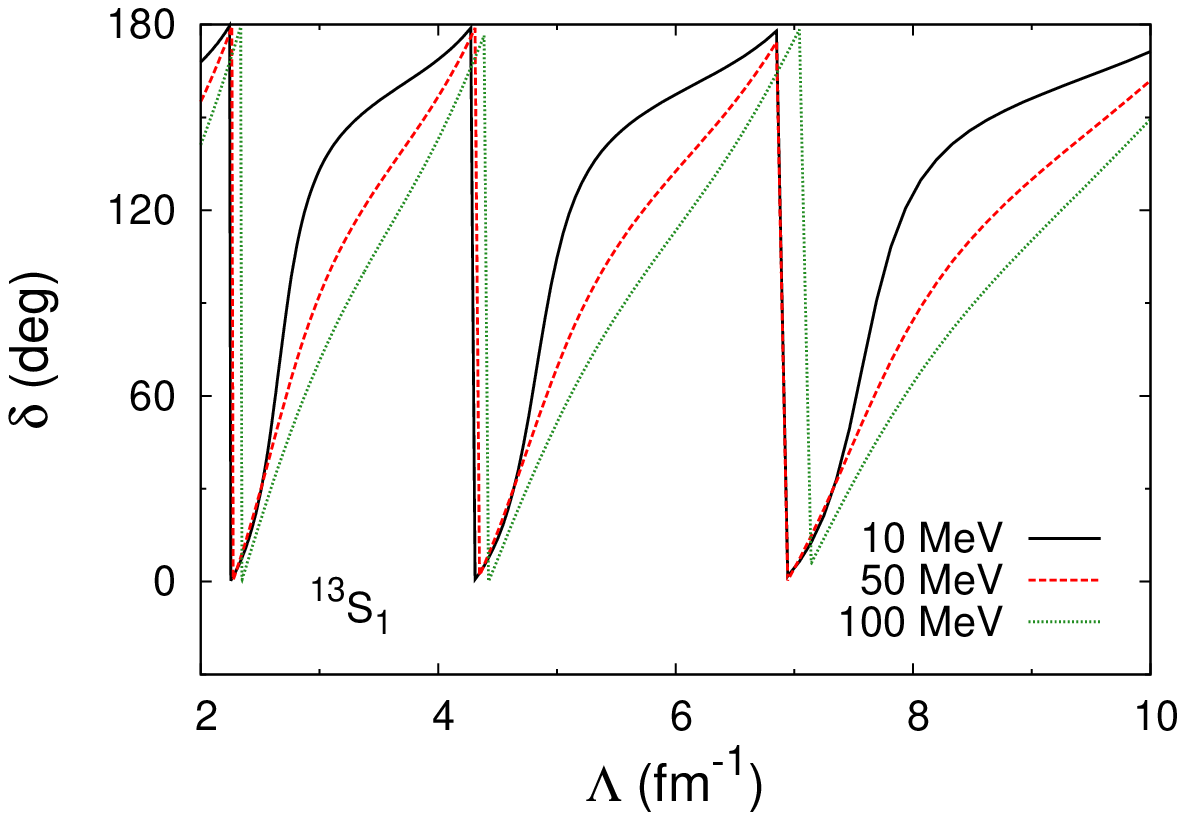} \hspace{2em}
    \includegraphics[width=0.45\textwidth]{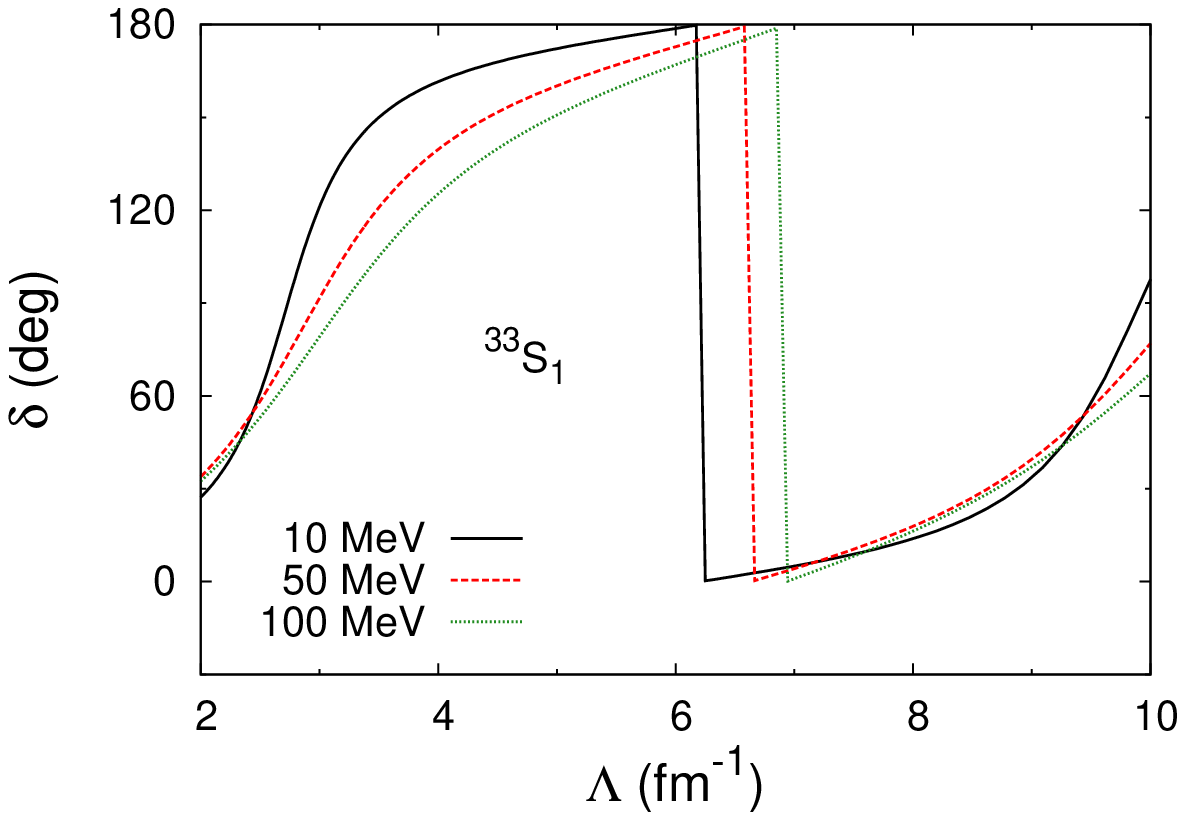} \\
    \includegraphics[width=0.45\textwidth]{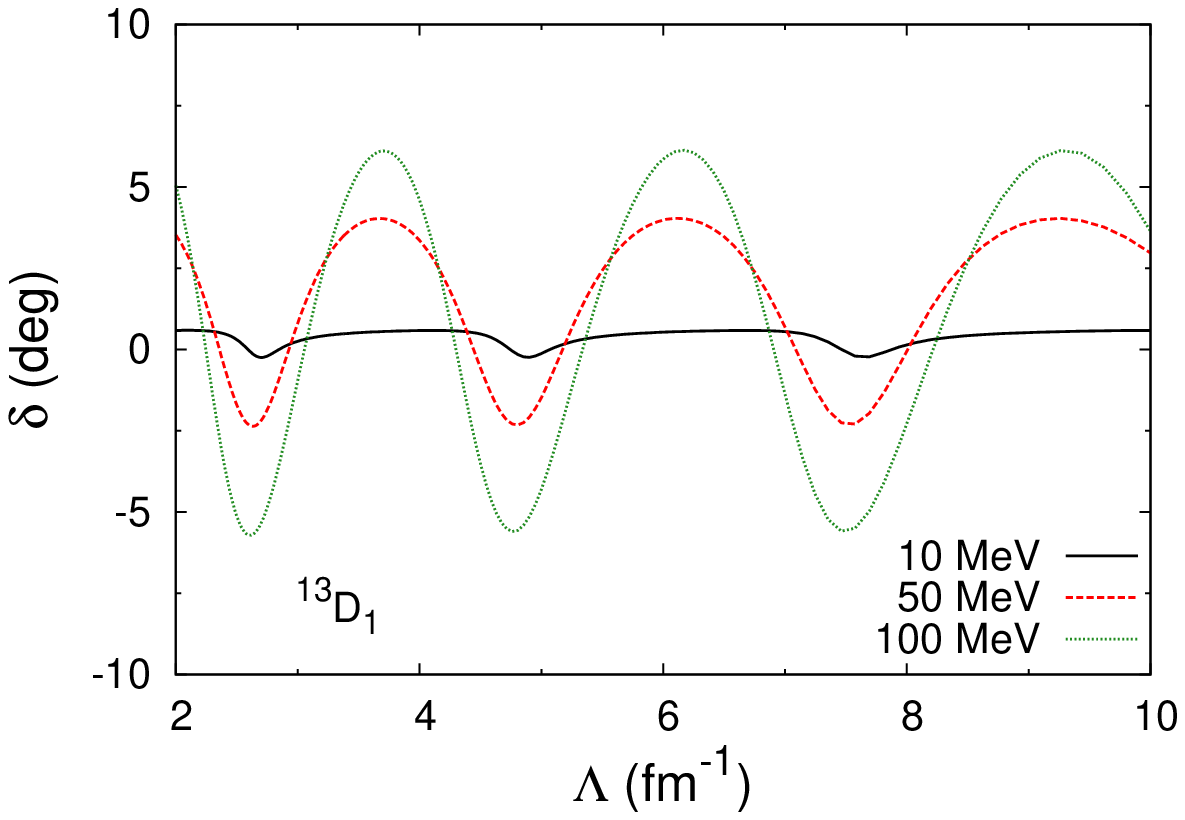}\hspace{2em}
    \includegraphics[width=0.45\textwidth]{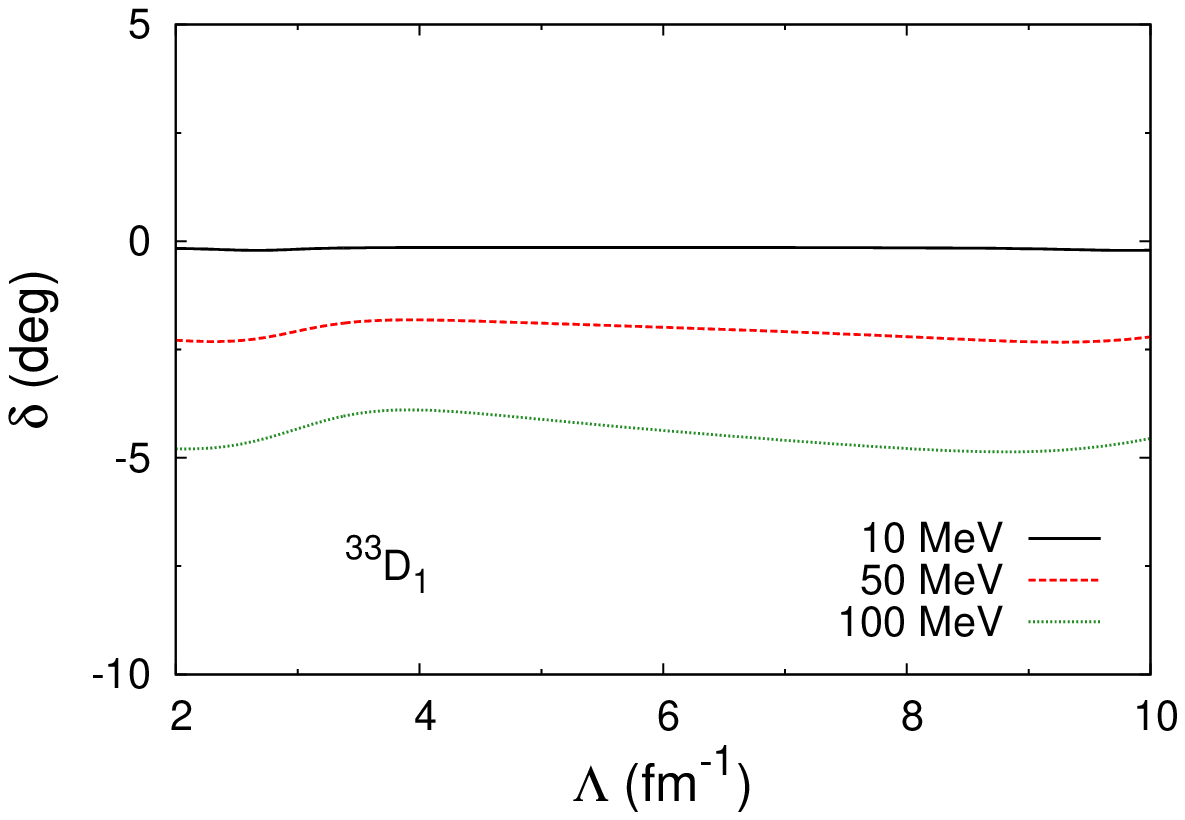}\\
    \includegraphics[width=0.45\textwidth]{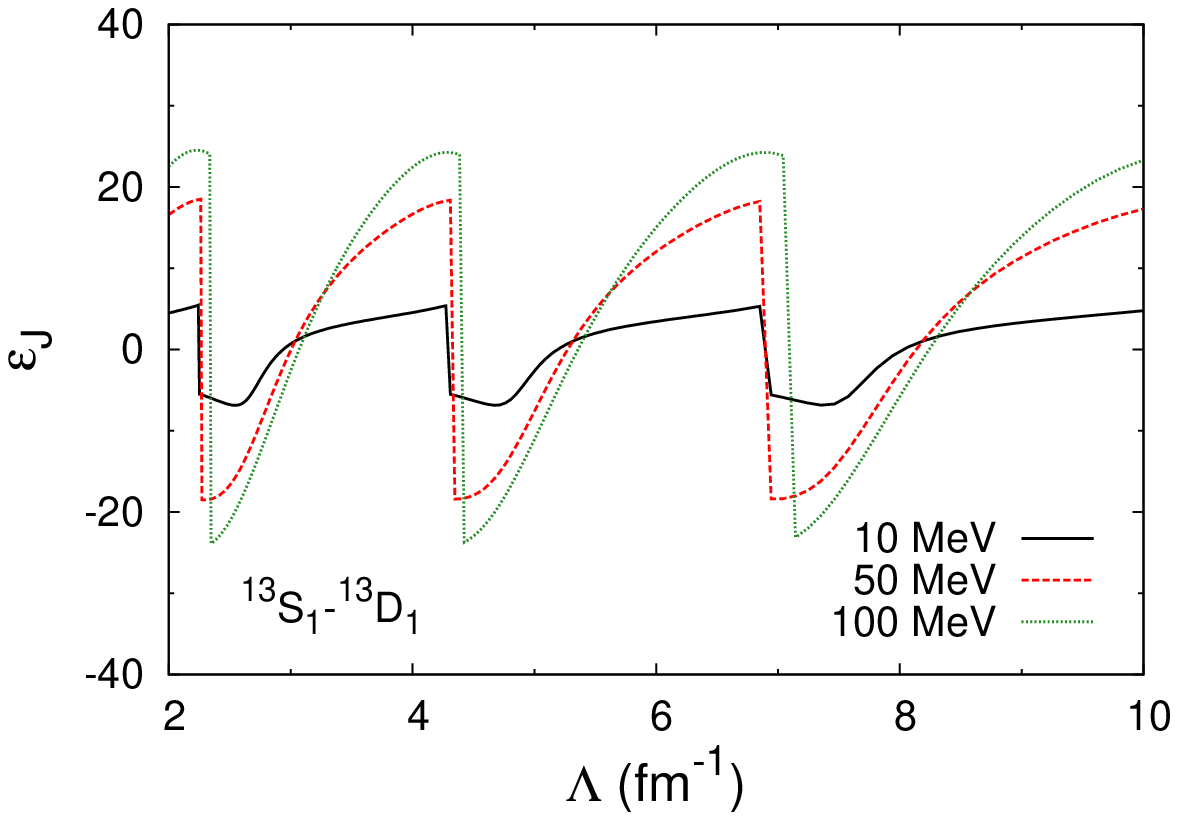} \hspace{2em}
    \includegraphics[width=0.45\textwidth]{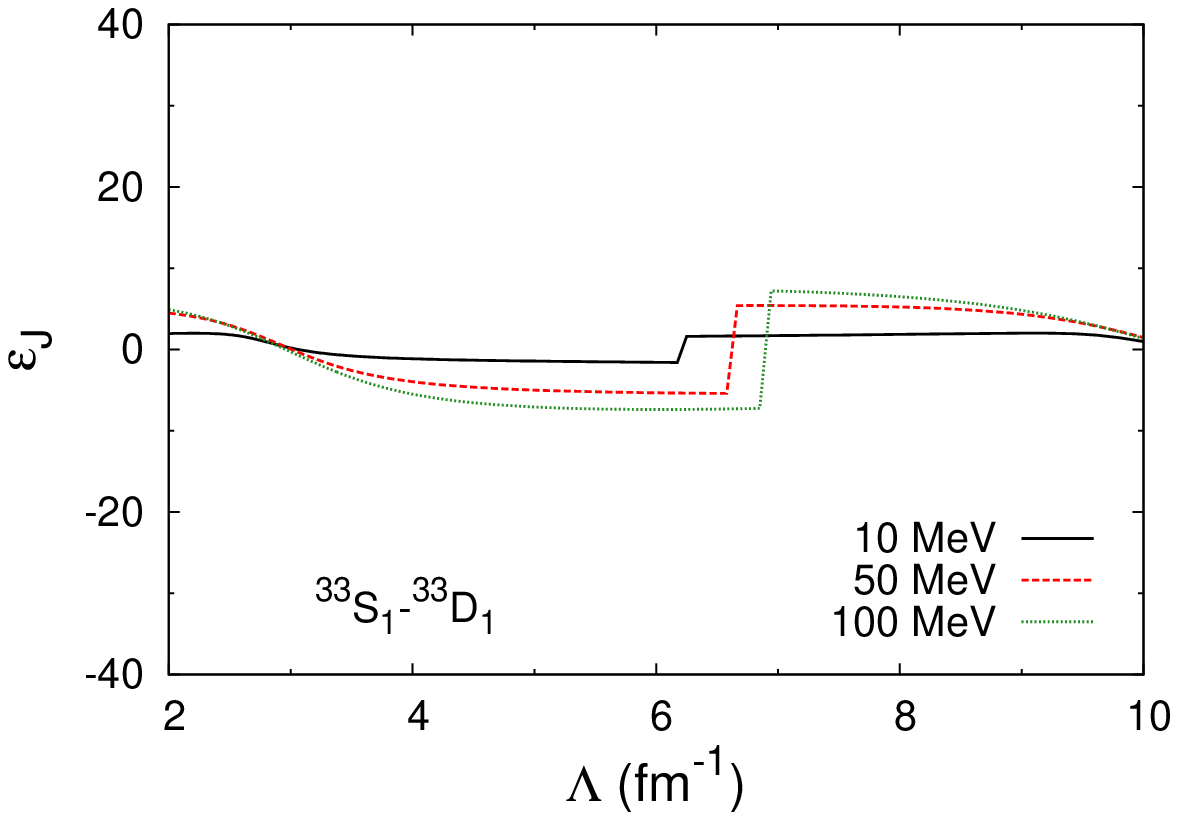}\\
 \caption{\label{TripletSDPhases_L1}{(Color online) Cutoff dependence of the phase shifts and mixing angles $\varepsilon_{J}$ in the spin-triplet coupled $S$-$D$ waves at the laboratory energies of 10 MeV (black solid line), 50 MeV (red dashed line), and 100 MeV (green dotted line), for $V_c=W_c=0$.}}
\end{figure}

\begin{figure}[tb]
	\centering
	\includegraphics[width=0.45\textwidth]{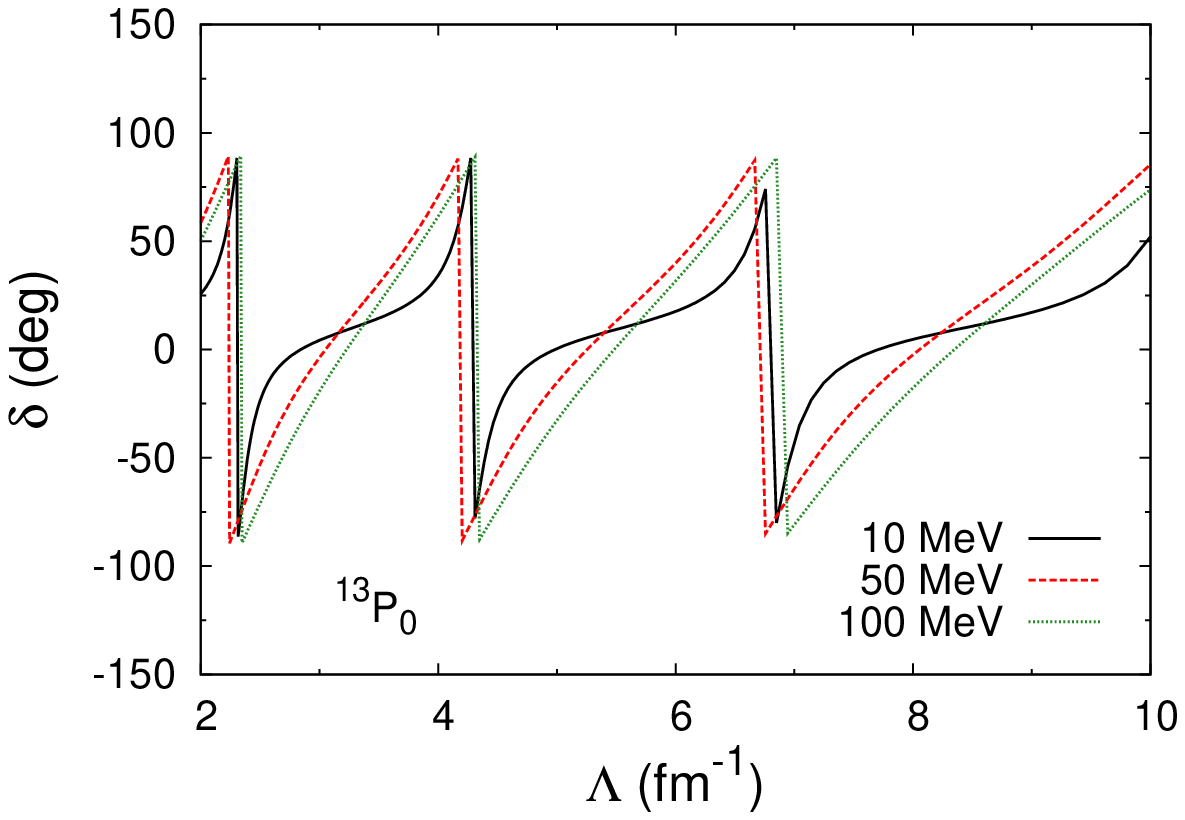} \hspace{2em}	
	\includegraphics[width=0.45\textwidth]{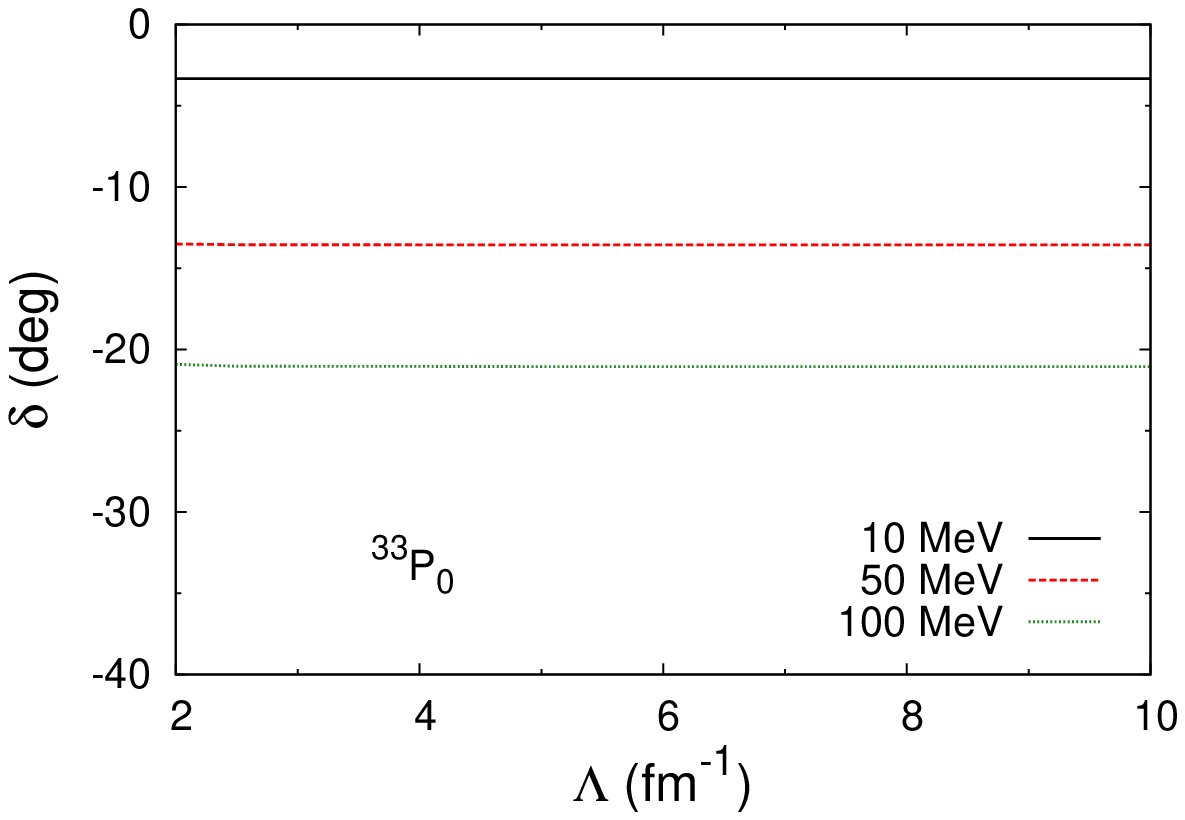} \\
	\includegraphics[width=0.45\textwidth]{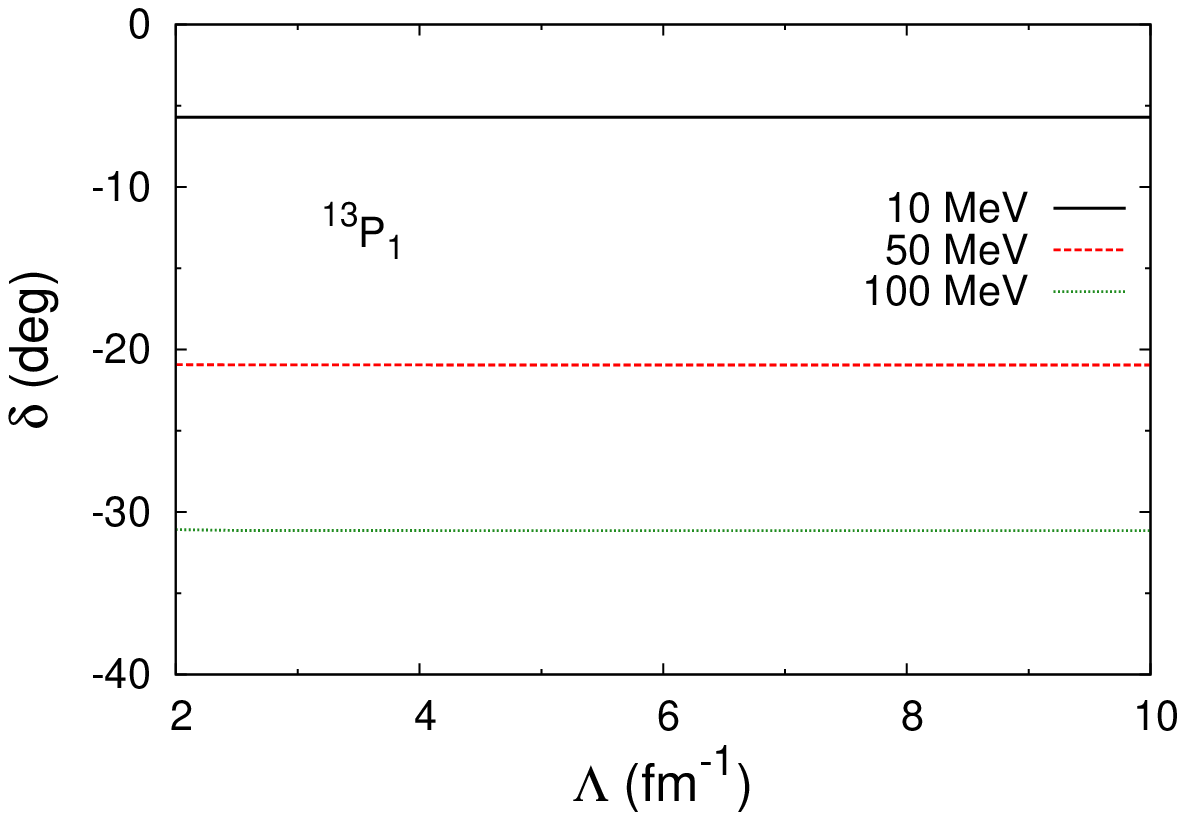} \hspace{2em}
	\includegraphics[width=0.45\textwidth]{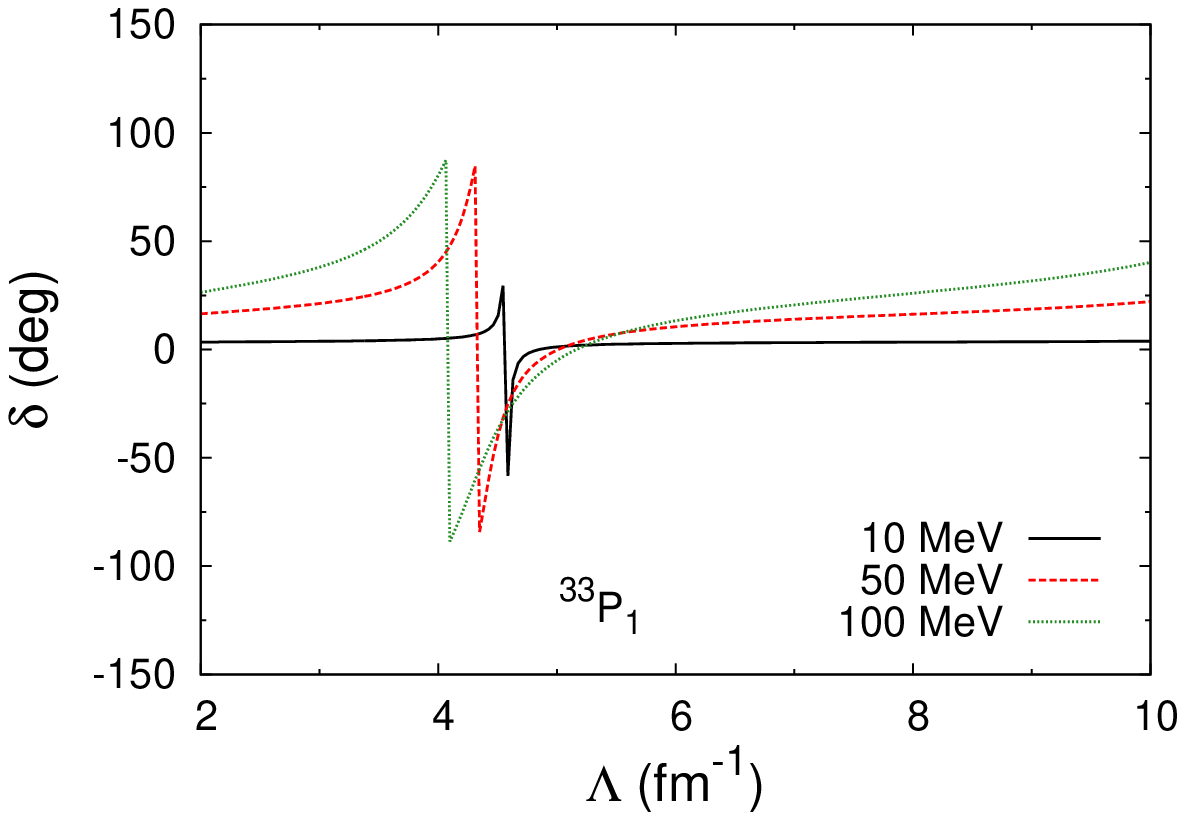} \\
	\caption{\label{TripletUncPPhases_L1}{(Color online) Cutoff dependence of the phase shifts in the uncoupled spin-triplet $P$ waves at the laboratory energies of 10 MeV (black solid line), 50 MeV (red dashed line), and 100 MeV (green dotted line), for $V_c=W_c=0$.}}
\end{figure}

\begin{figure}[tb]
	\centering
	\includegraphics[width=0.45\textwidth]{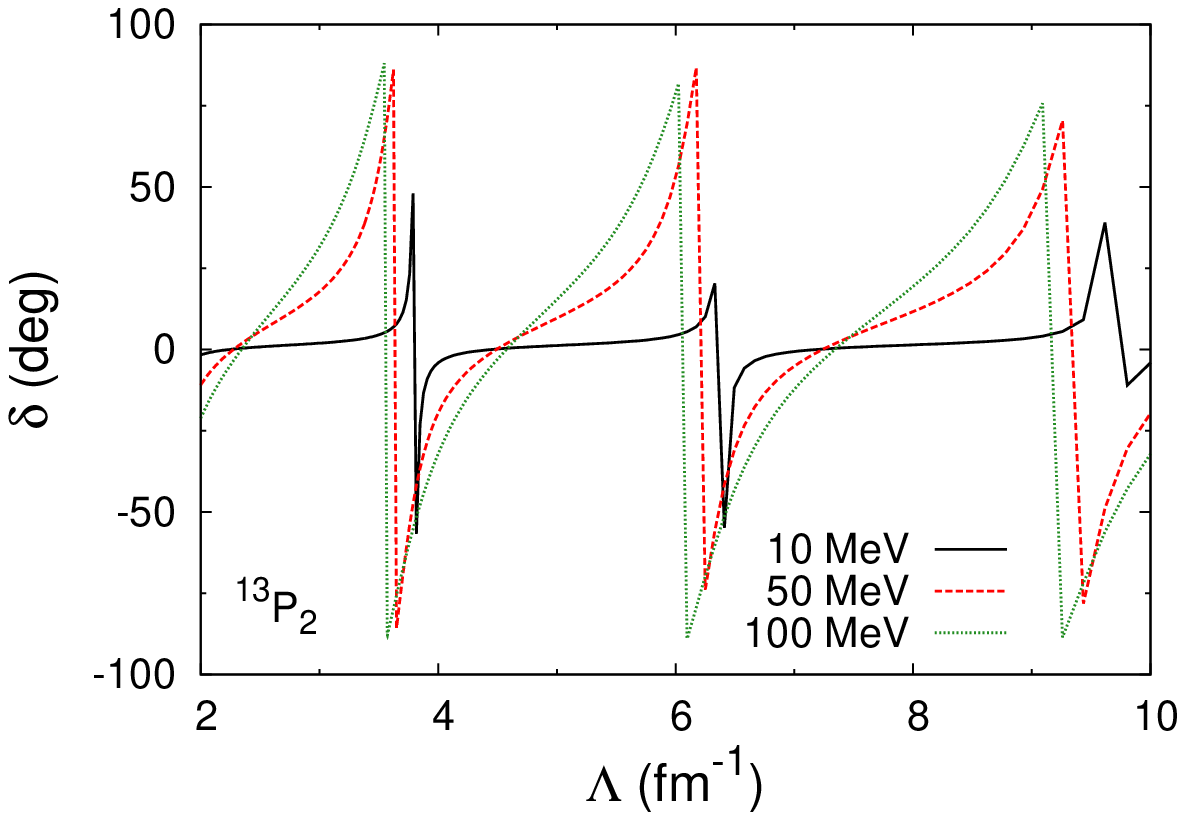} \hspace{2em}
	\includegraphics[width=0.45\textwidth]{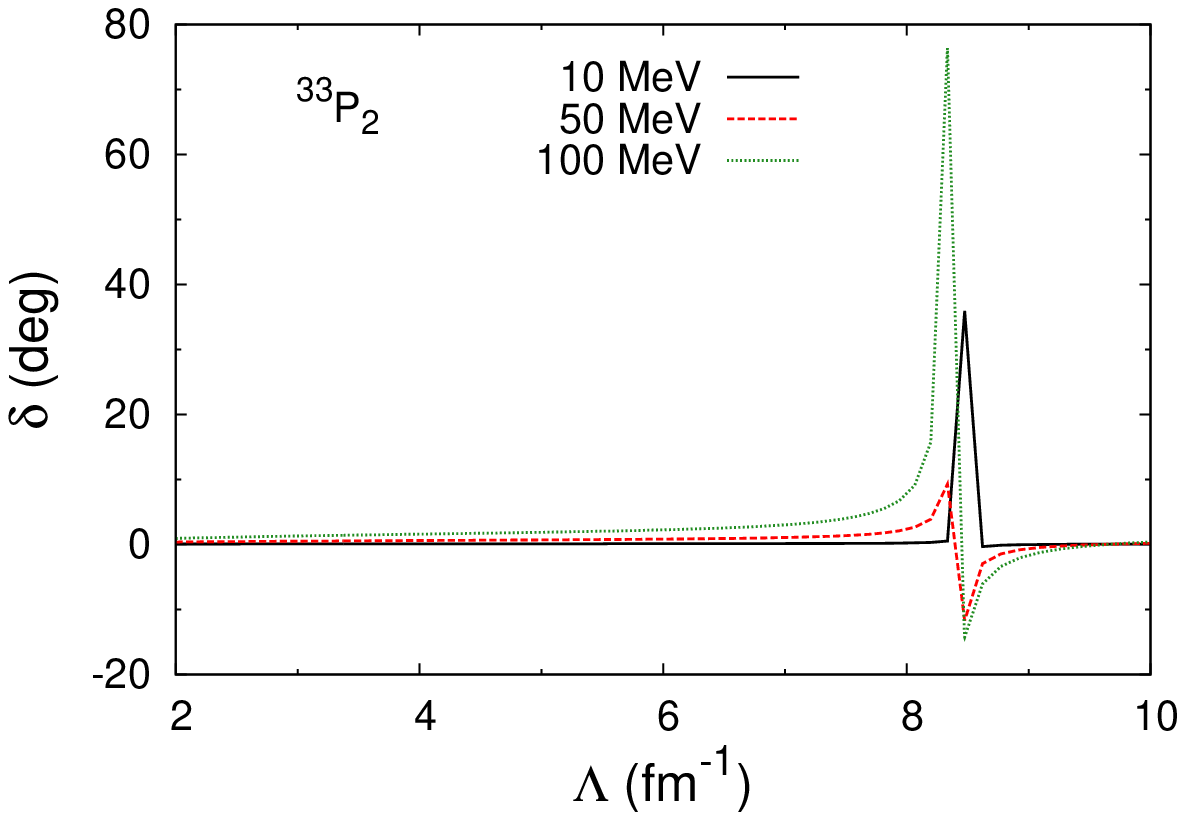} \\	
	\includegraphics[width=0.45\textwidth]{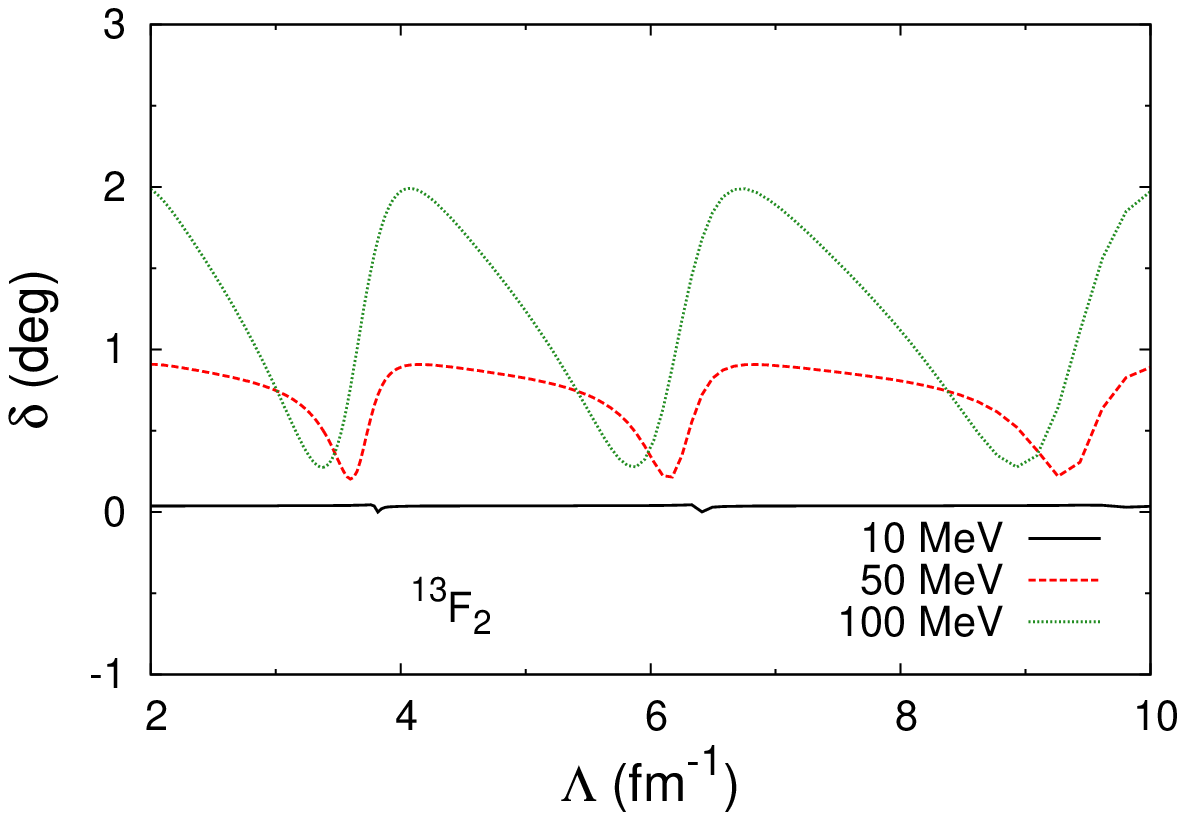} \hspace{2em}
	\includegraphics[width=0.45\textwidth]{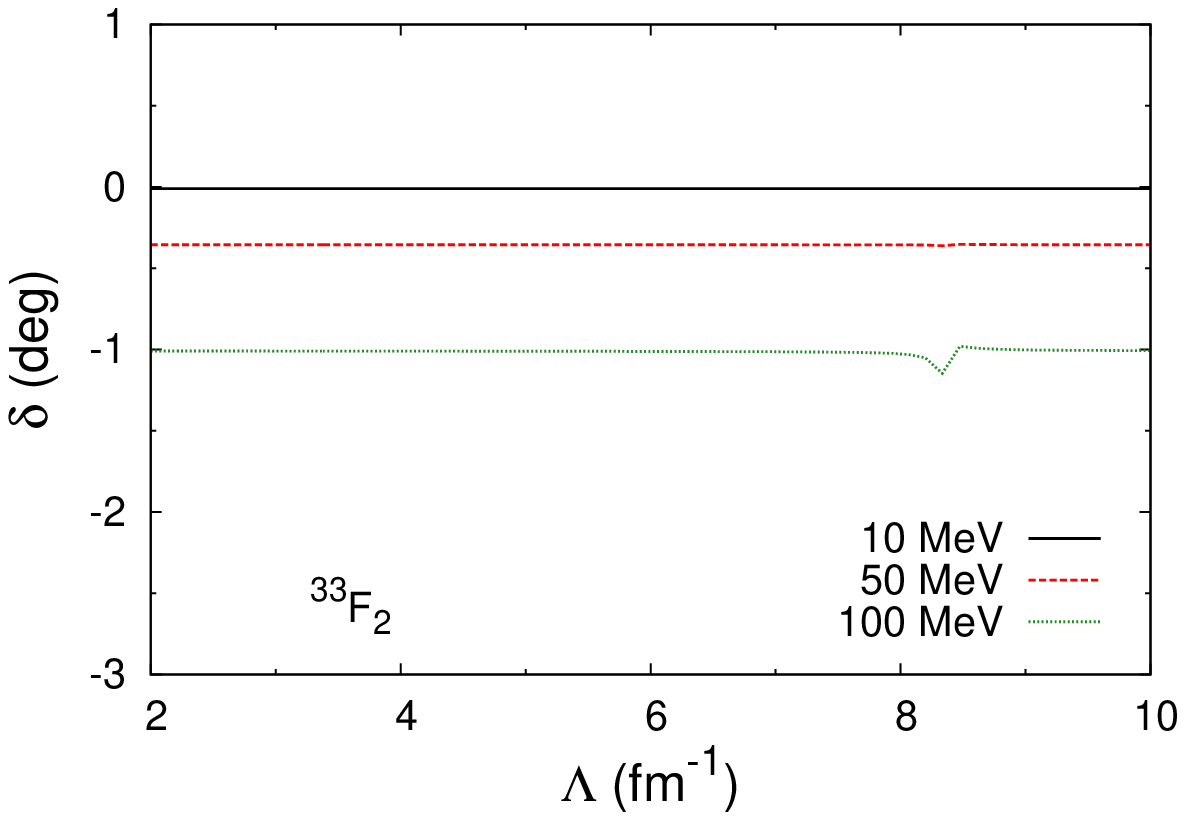}\\	
	\includegraphics[width=0.45\textwidth]{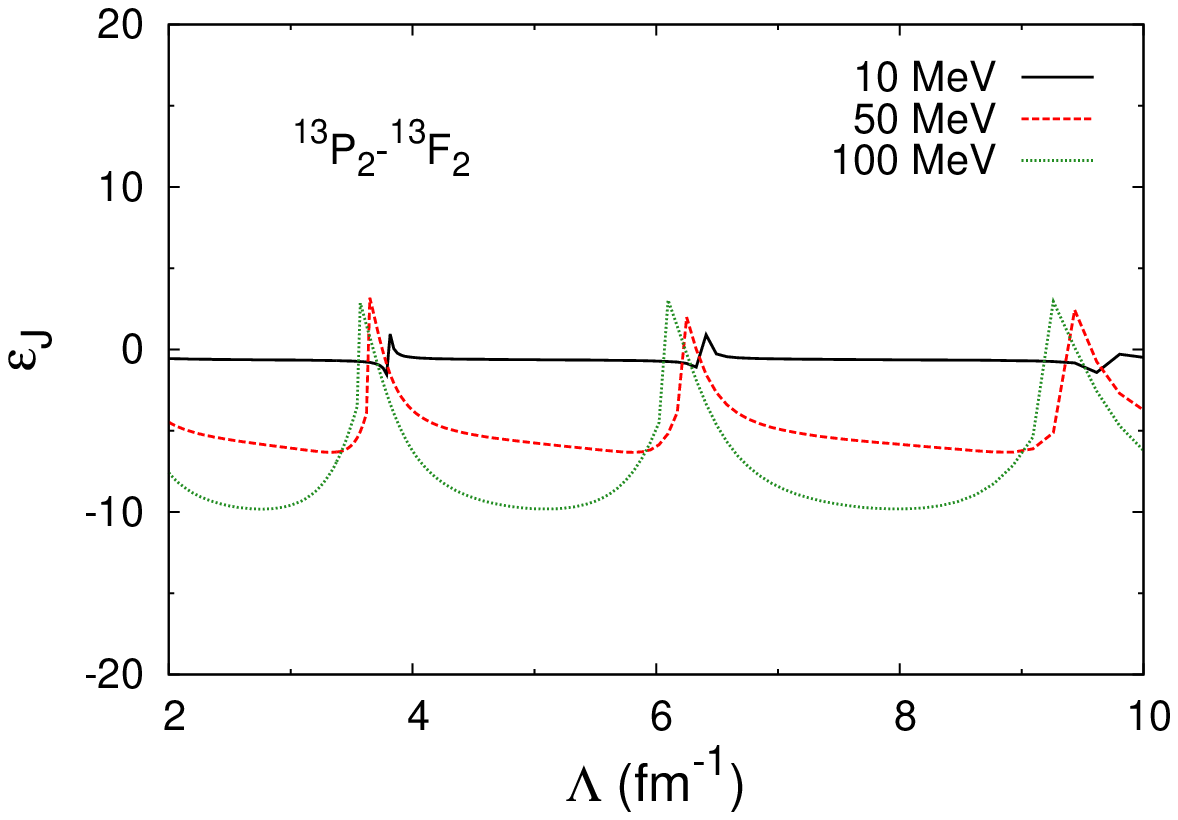} \hspace{2em}
	\includegraphics[width=0.45\textwidth]{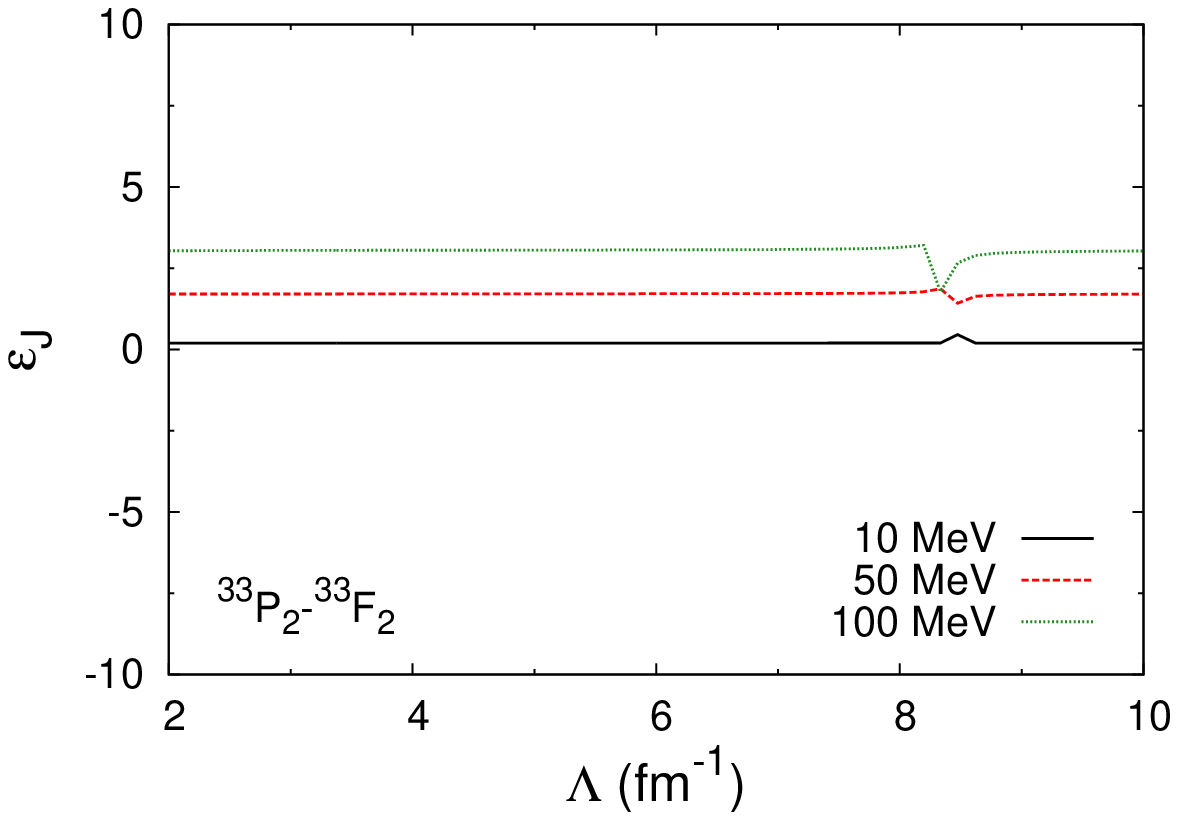}   
 \caption{\label{TripletPFPhases_L1}{(Color online) Cutoff dependence of the phase shifts and mixing angles $\varepsilon_{J}$ in the spin-triplet coupled $P$-$F$ waves at the laboratory energies of 10 MeV (black solid line), 50 MeV (red dashed line), and 100 MeV (green dotted line), for $V_c=W_c=0$.}}
\end{figure}

The tensor force makes OPE strong in some $N\!N$ spin-triplet channels already at relatively low momenta \cite{Fleming:1999ee,Nogga:2005hy,
Birse:2005um,Kaplan:2019znu}. The remarkable cutoff dependence of iterated OPE in some $\overline{N}\!N$ channels is evident in Figs. \ref{TripletSDPhases_L1}, \ref{TripletUncPPhases_L1}, and \ref{TripletPFPhases_L1}, which display the coupled $S$-$D$, uncoupled $P$, and coupled $P$-$F$ waves, respectively. As expected, there is strong cutoff dependence in the coupled channels and attractive uncoupled $P$ waves.

\begin{figure}[tb]
	\centering
	\includegraphics[width=0.45\textwidth]{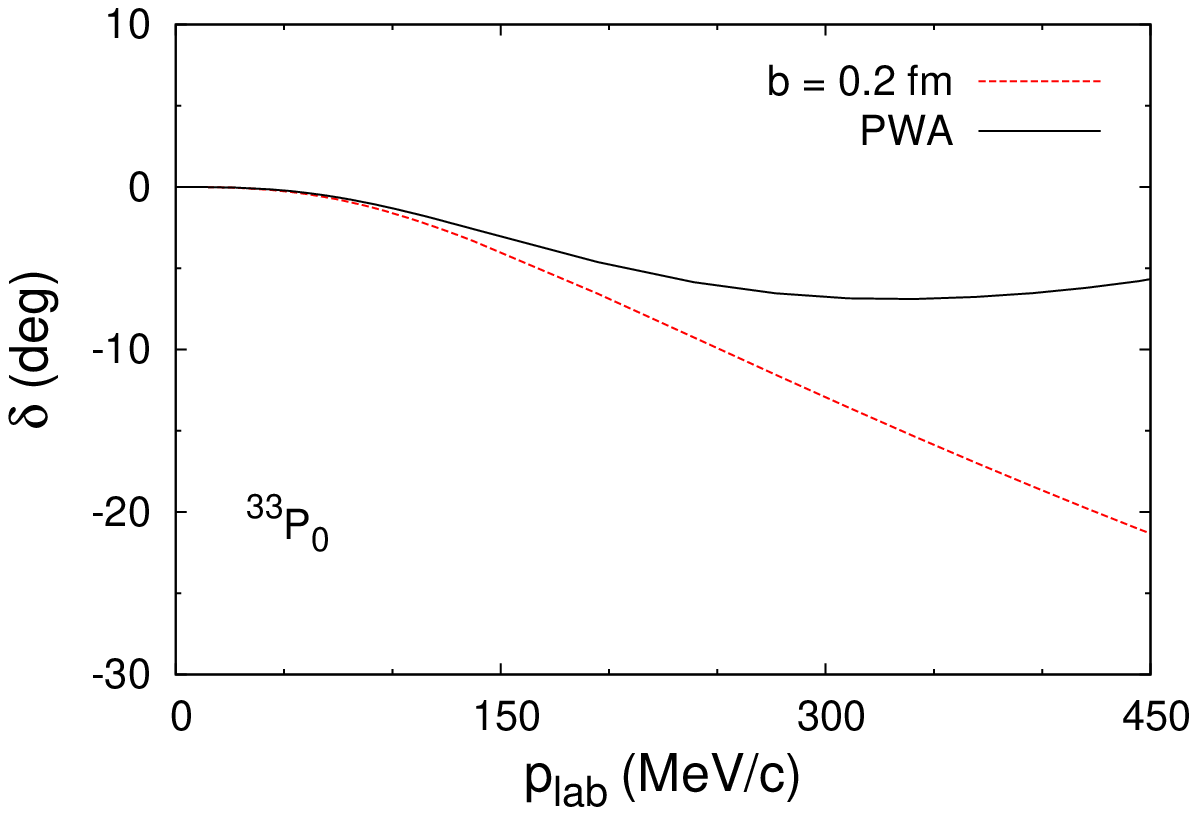} \hspace{2em}	
    \includegraphics[width=0.45\textwidth]{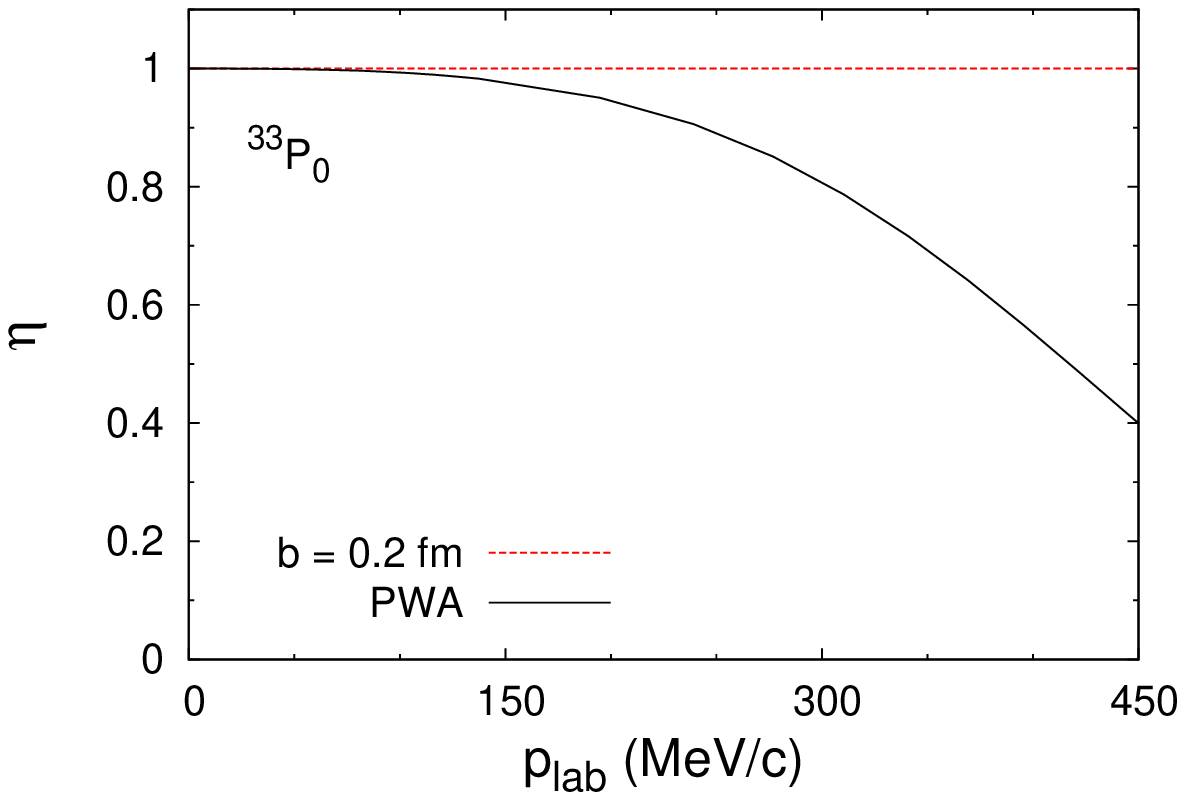} \\    
	\includegraphics[width=0.45\textwidth]{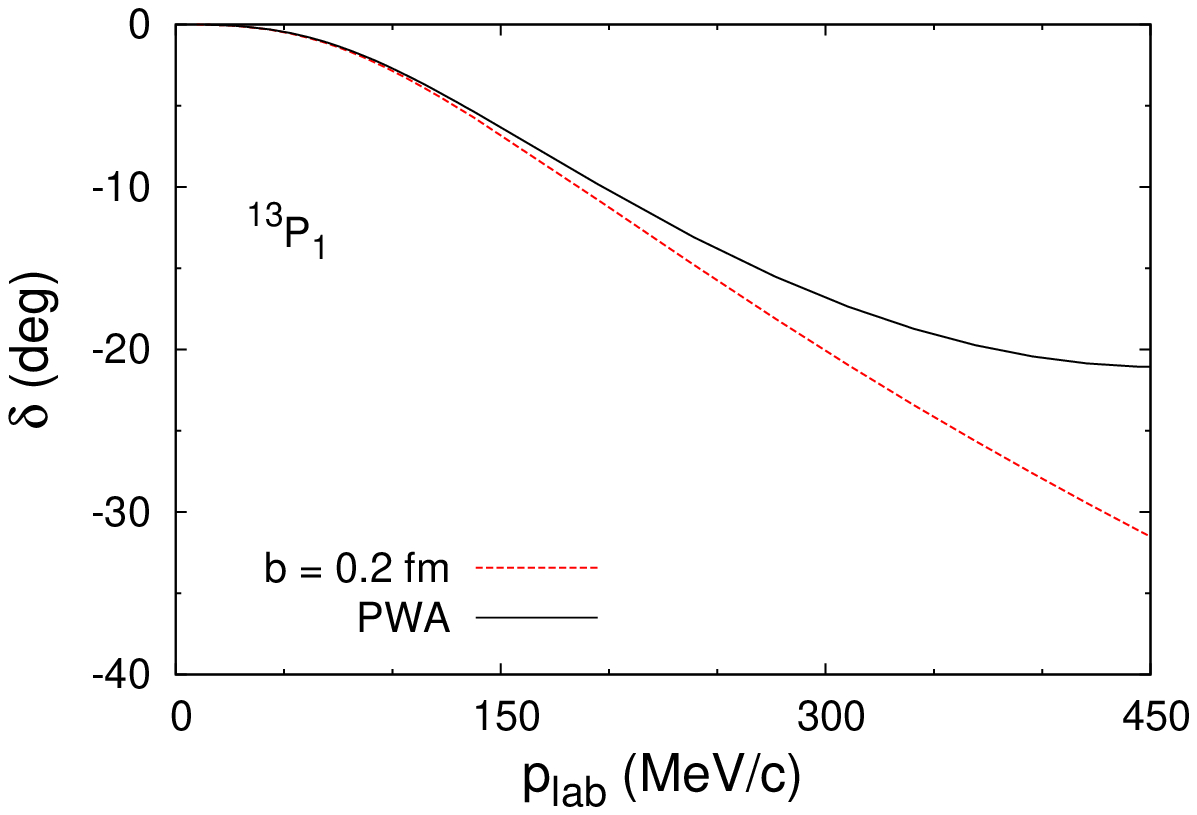} \hspace{2em}	
    \includegraphics[width=0.45\textwidth]{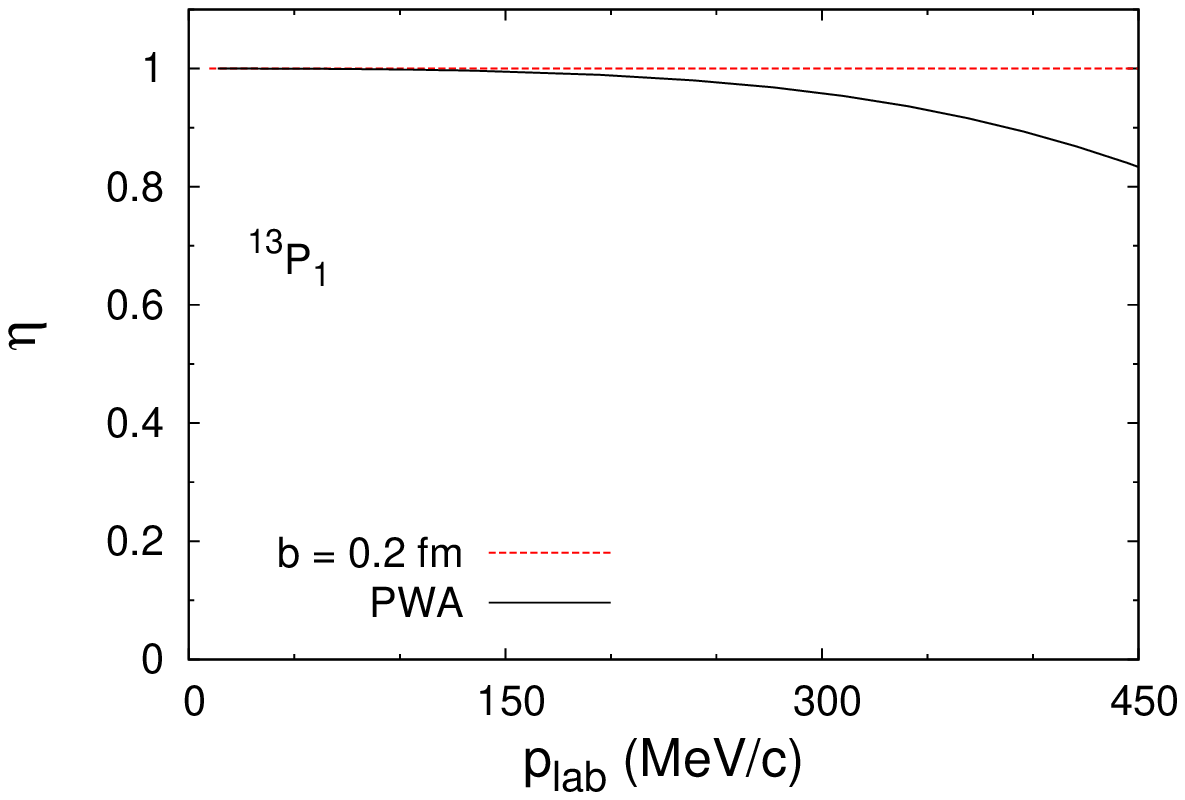} \\   
	\caption{\label{Phase_plab_uncoup0}{(Color online) 
	Phase shifts (left panels) and inelasticities (right panels) of the spin-triplet 
	uncoupled $P$ waves that do not require counterterms against 
	laboratory momentum.
	The (red) dashed lines are from iterated one-pion exchange for $b=0.2$ fm and $V_c=W_c=0$, while (black) solid lines are the results of the PWA~\cite{Zhou:2012ui,Zhou:2013}.}}
\end{figure}

OPE is repulsive in the uncoupled $^{33}P_0$ and $^{13}P_1$ waves and we see that they indeed require no LECs for renormalization. NDA also does not prescribe LECs at LO. In Fig. \ref{Phase_plab_uncoup0} we compare the iterated-OPE phase shifts and inelasticities for $\Lambda = 5$ fm$^{-1}$ with the PWA results. Except for the larger inelasticity gap in $^{33}P_0$, discrepancies are comparable to those for the spin-singlet $P$ waves in Fig. \ref{Phase_plab_SingletP}, which also do not require counterterms.

\begin{figure}[tb]
   \centering
   \includegraphics[width=0.45\textwidth]{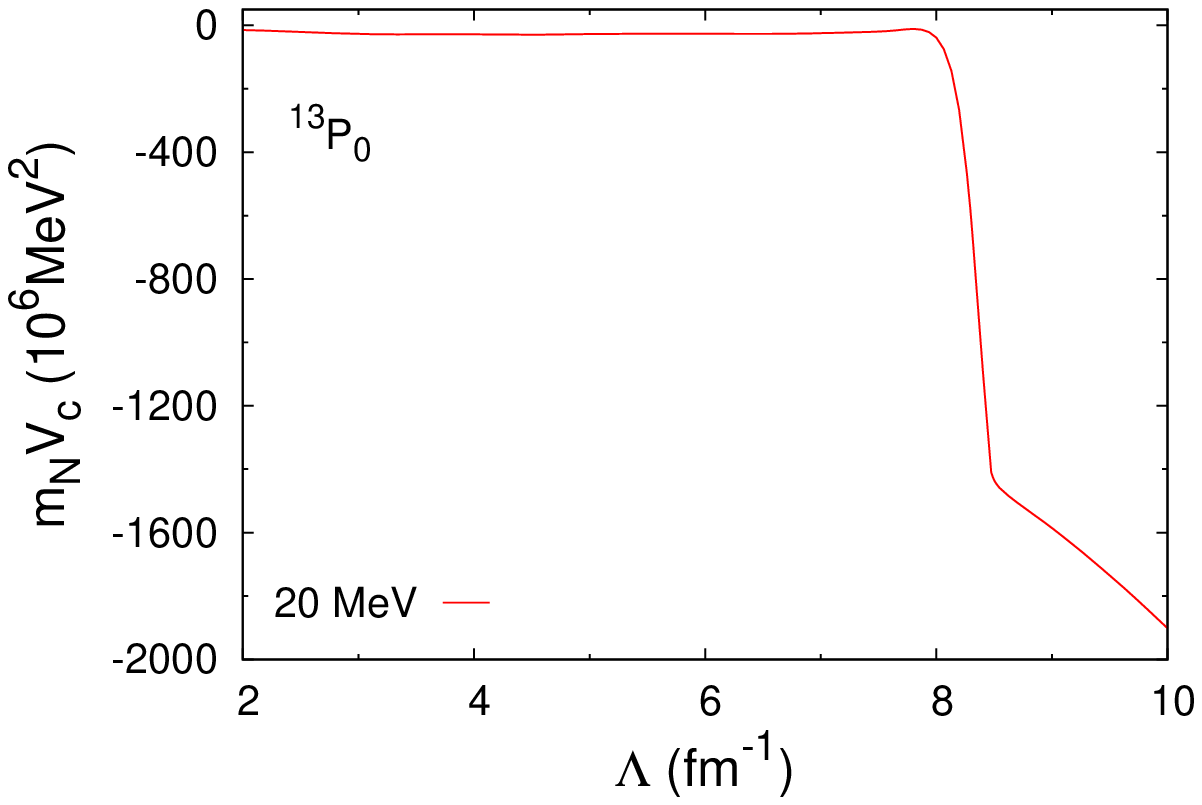} \hspace{2em}
   \includegraphics[width=0.45\textwidth]{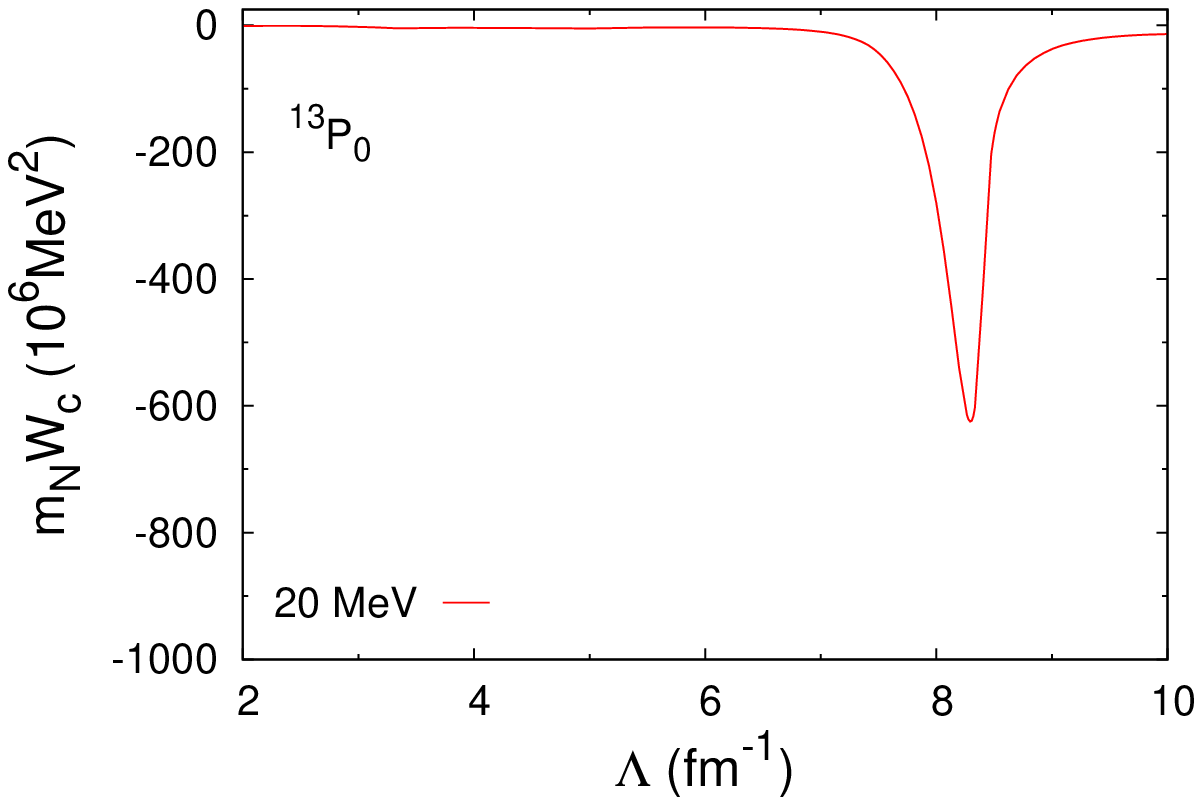} \\   
   \includegraphics[width=0.45\textwidth]{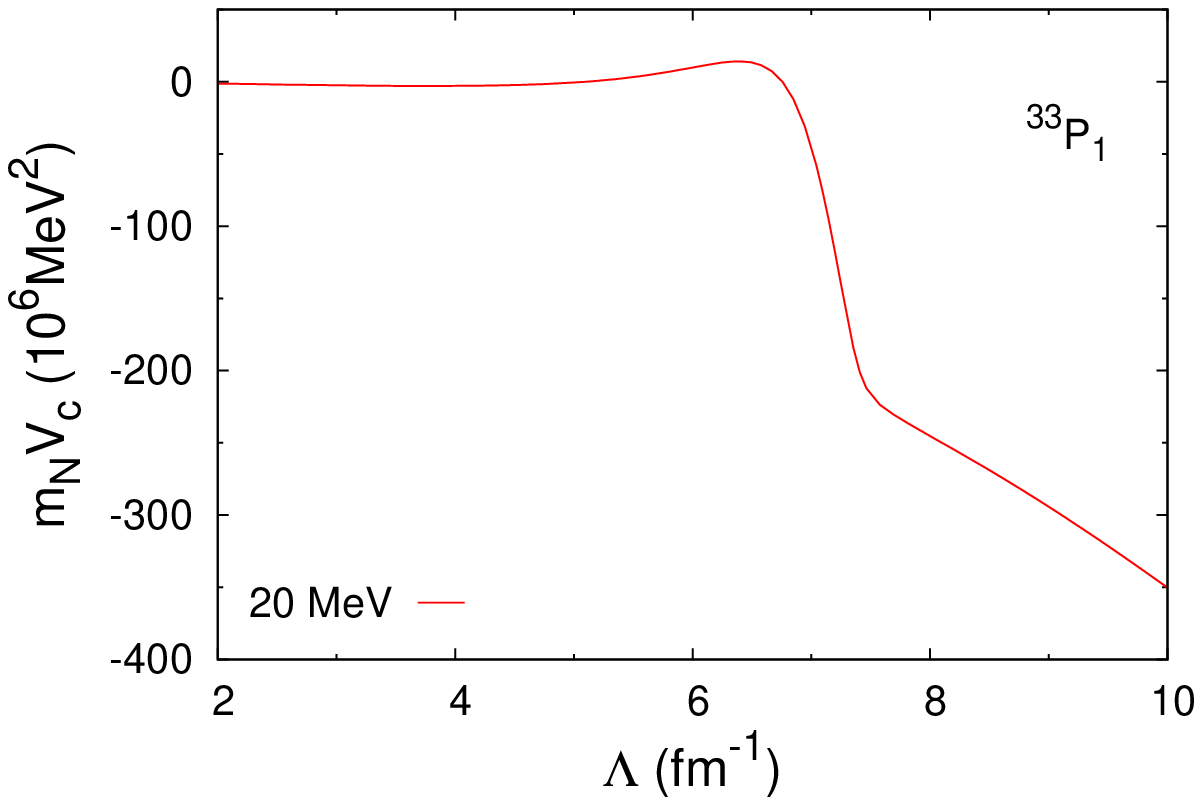} \hspace{2em}
   \includegraphics[width=0.45\textwidth]{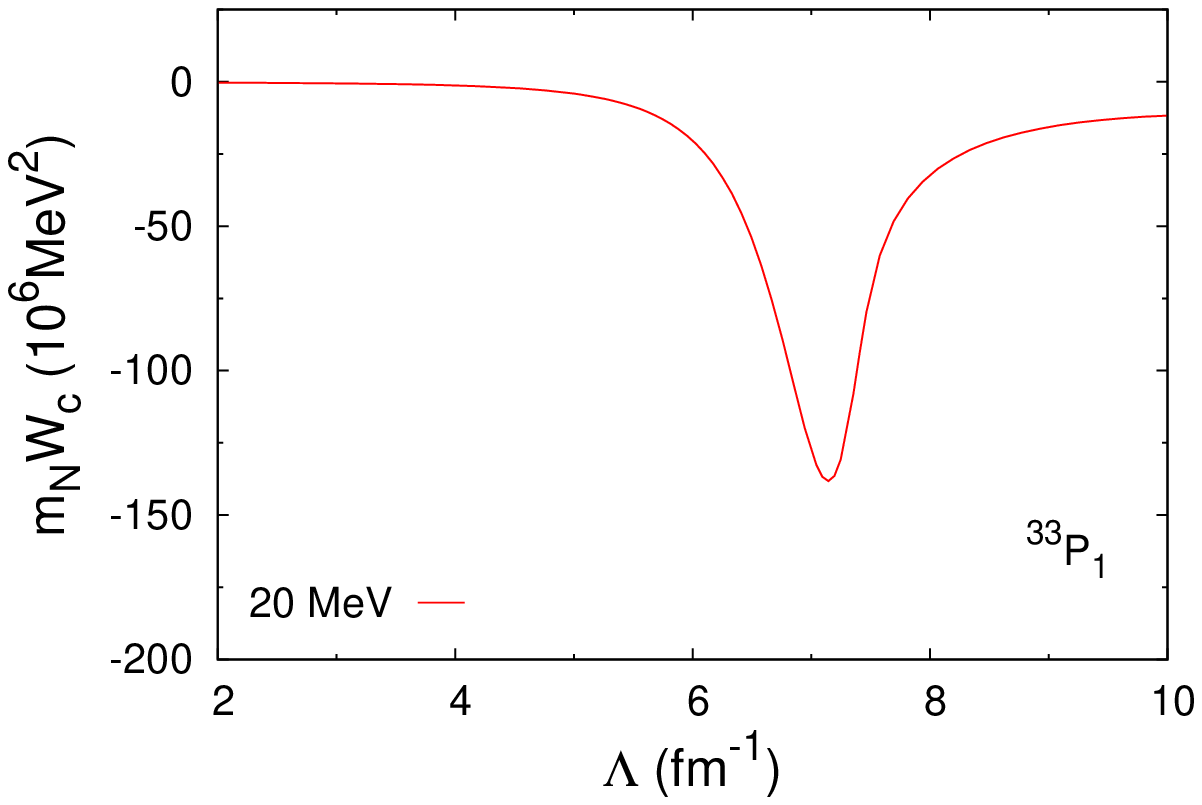}  
   \caption{\label{Counter_cut1}{(Color online) 
   Cutoff dependence of $m_N V_c$ (left panels) and $m_N W_c$ (right panels) for the spin-triplet uncoupled $P$ waves where the OPE tensor force is attractive. The PWA phase shifts and inelasticities are fitted at $T_{\rm lab}=20$ MeV.}}
\end{figure}

\begin{figure}[thbp]
   \centering
   \includegraphics[width=0.45\textwidth]{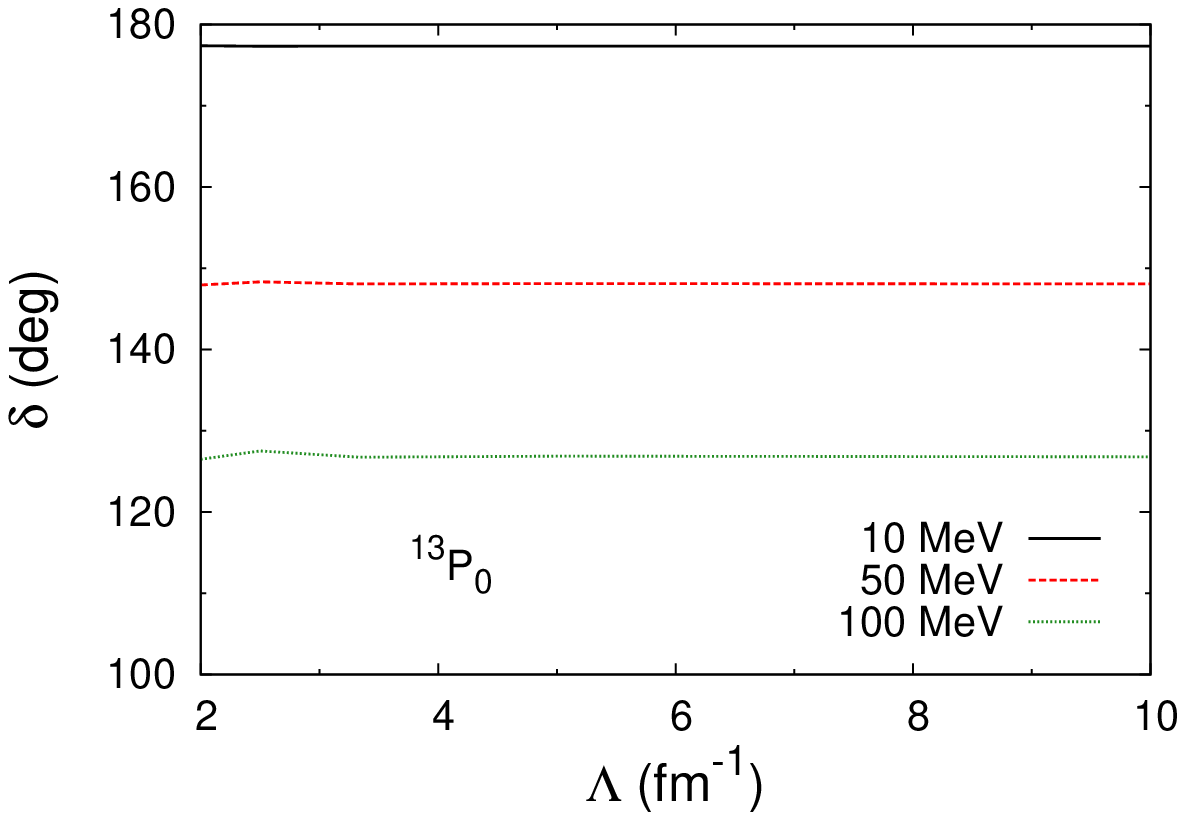} \hspace{2em}
   \includegraphics[width=0.45\textwidth]{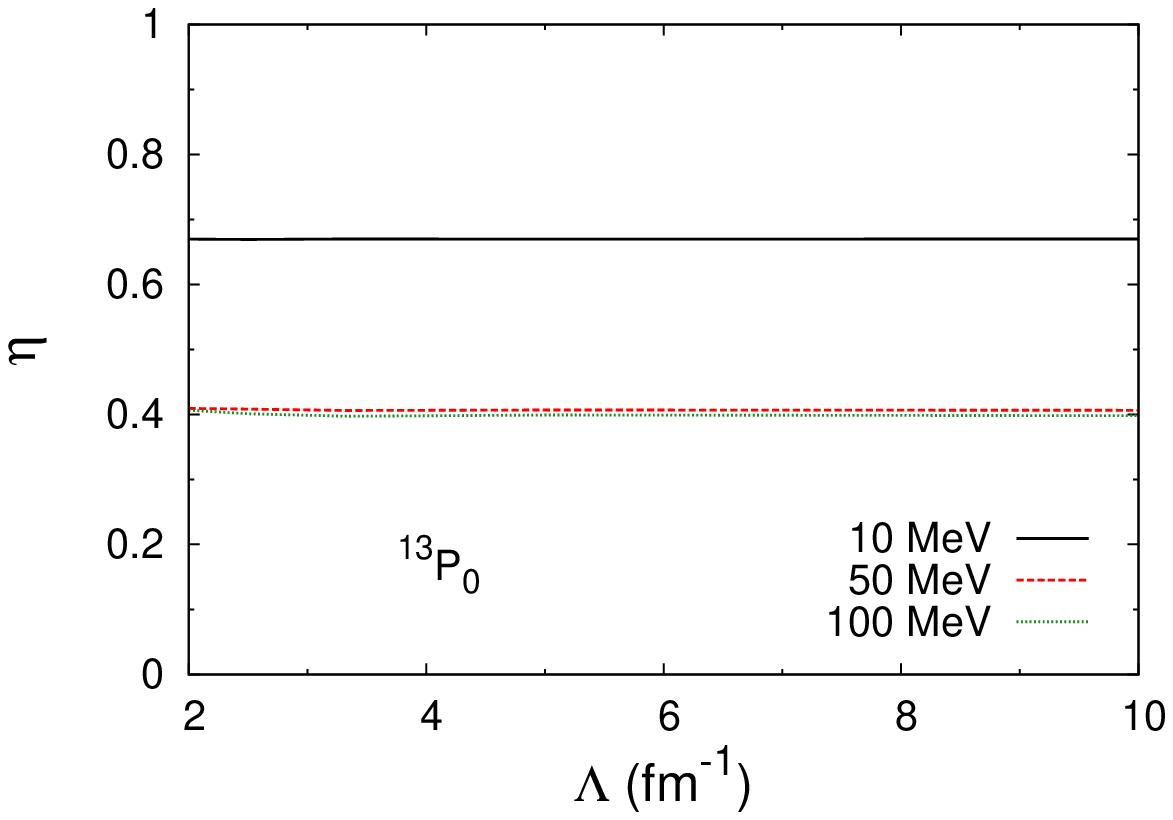} \\   
   \includegraphics[width=0.45\textwidth]{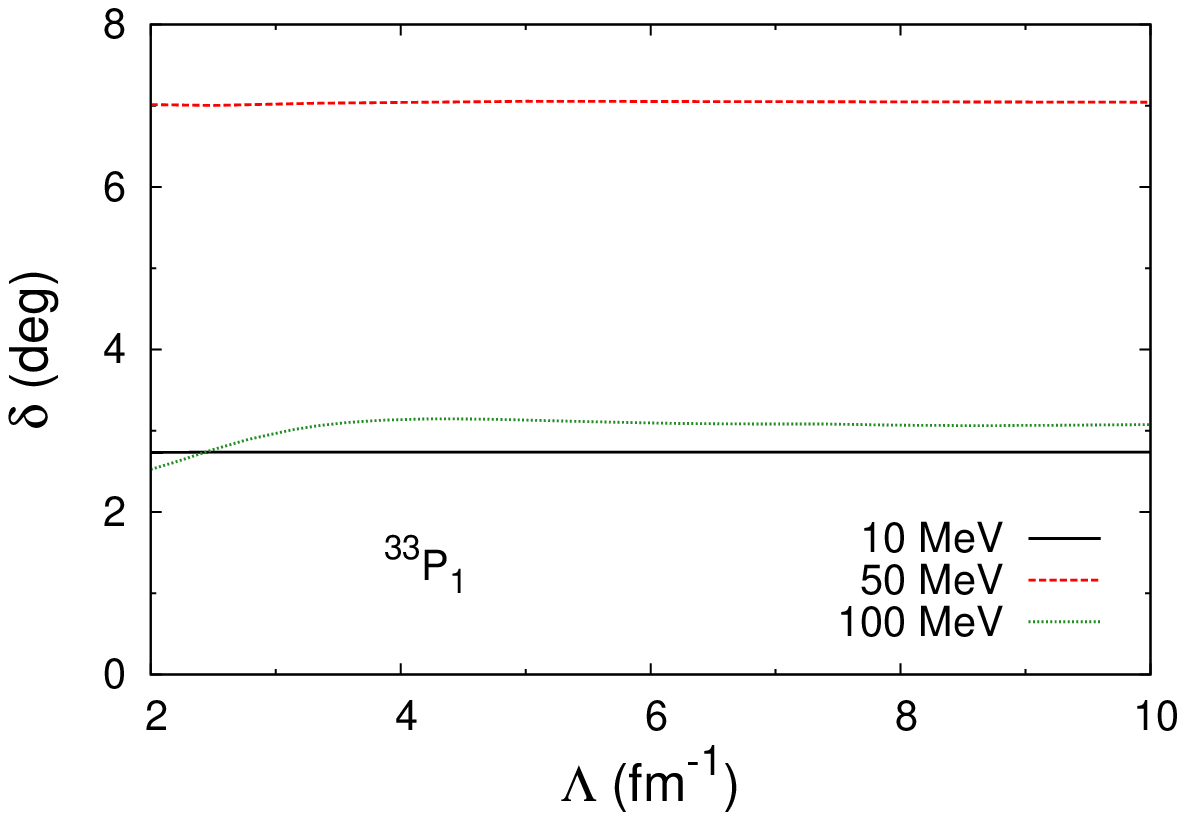} \hspace{2em}
   \includegraphics[width=0.45\textwidth]{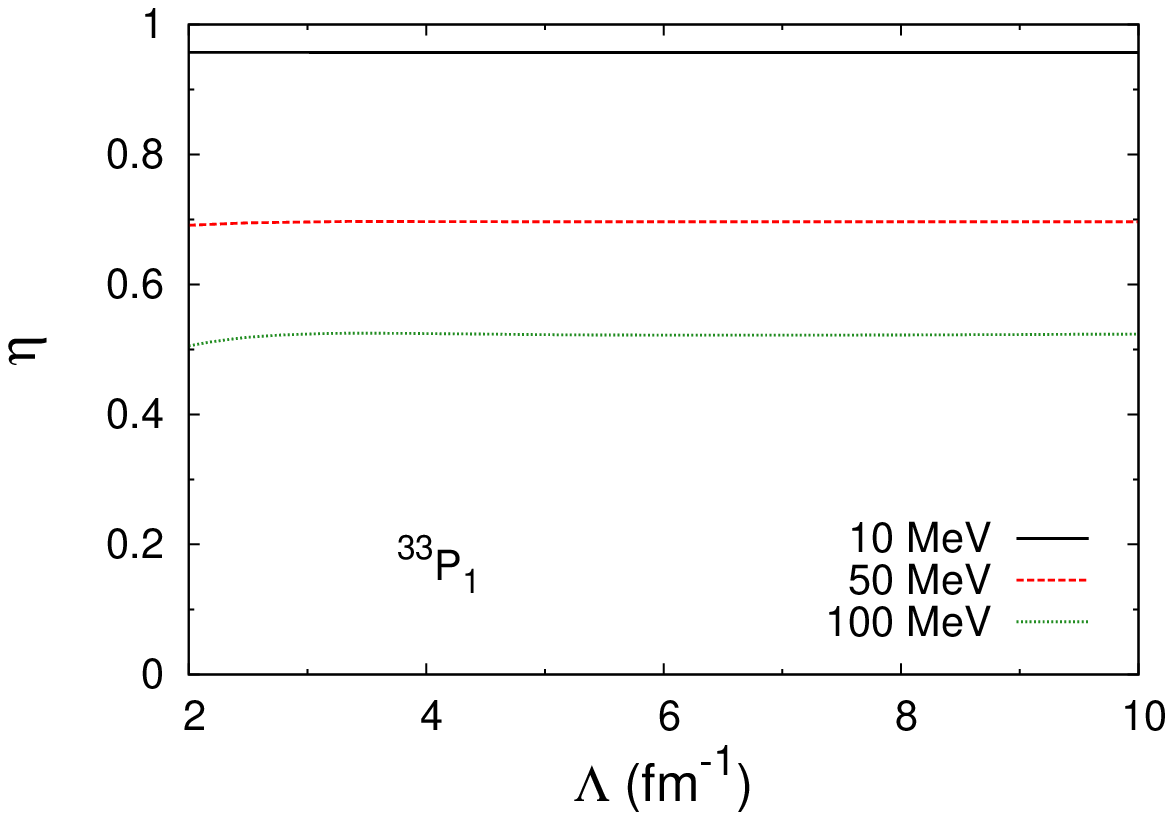}
   \caption{\label{Phase_LC1}{(Color online) Residual cutoff dependence of the phase shifts and inelasticities in the spin-triplet uncoupled $P$ waves that require counterterms at the laboratory energies of 10 MeV (black solid line), 50 MeV (red dashed line), and 100 MeV (green dotted line), for $V_c$ and $W_c$ in Fig. \ref{Counter_cut1}.}}
\end{figure}

\begin{figure}[thbp]
	\centering
	\includegraphics[width=0.45\textwidth]{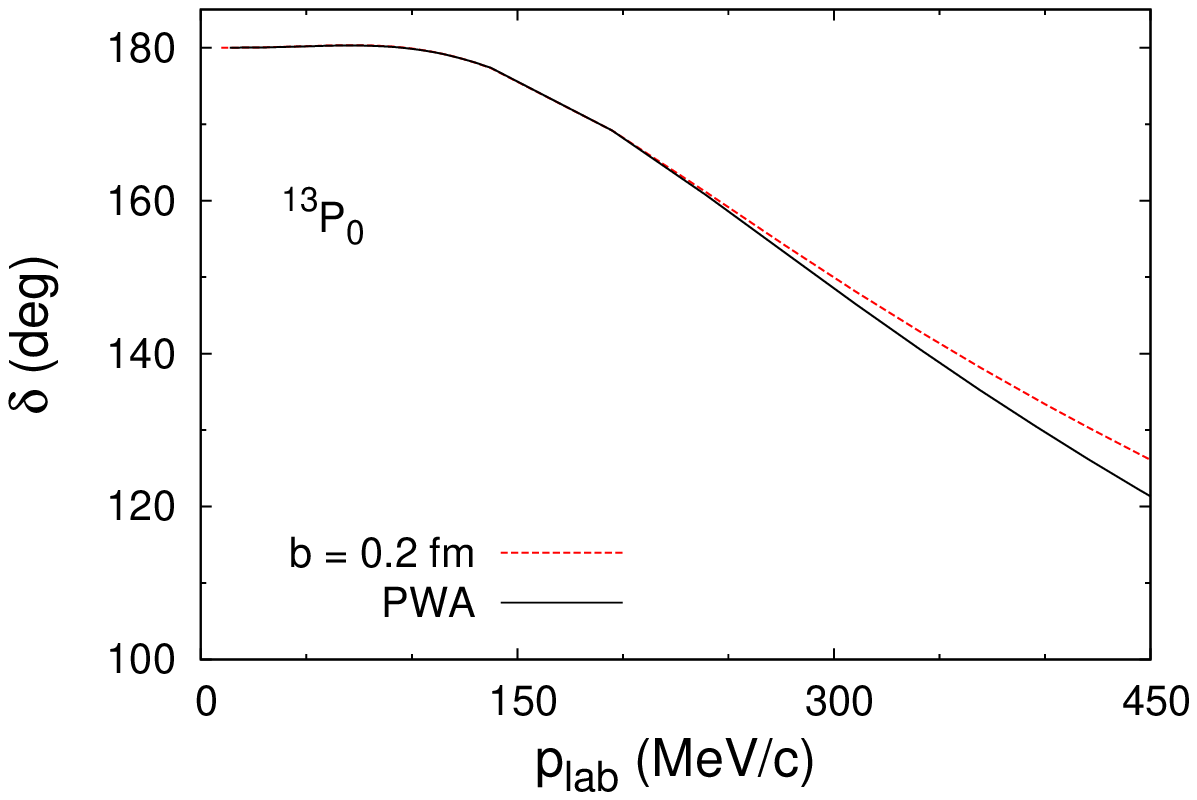} \hspace{2em}
	\includegraphics[width=0.45\textwidth]{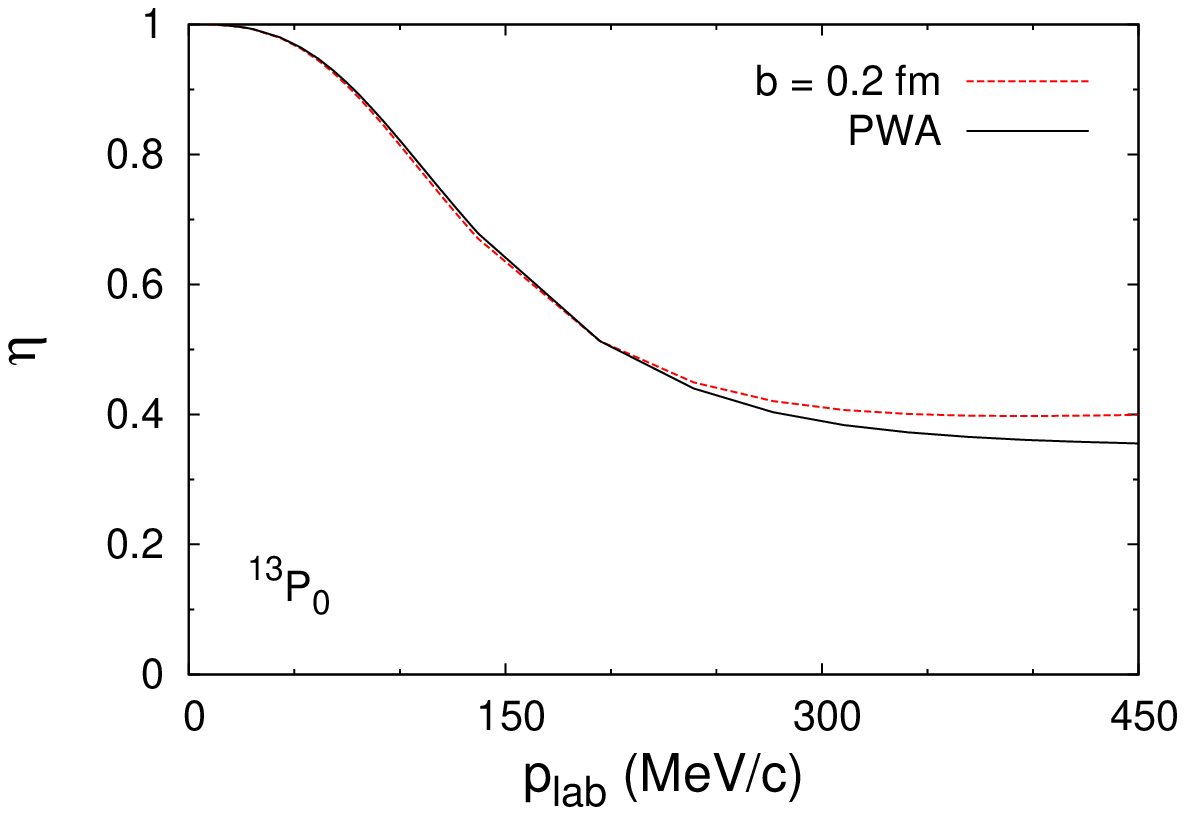} \\   
	\includegraphics[width=0.45\textwidth]{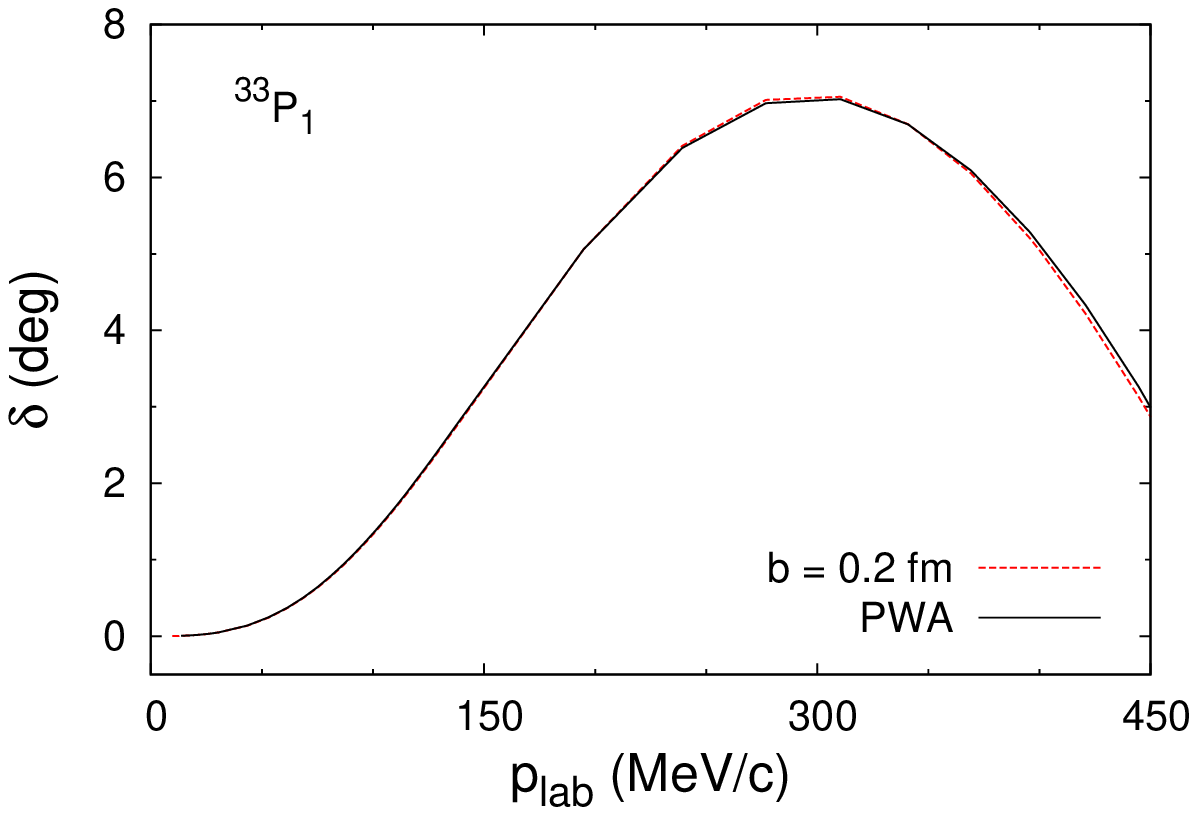} \hspace{2em}
	\includegraphics[width=0.45\textwidth]{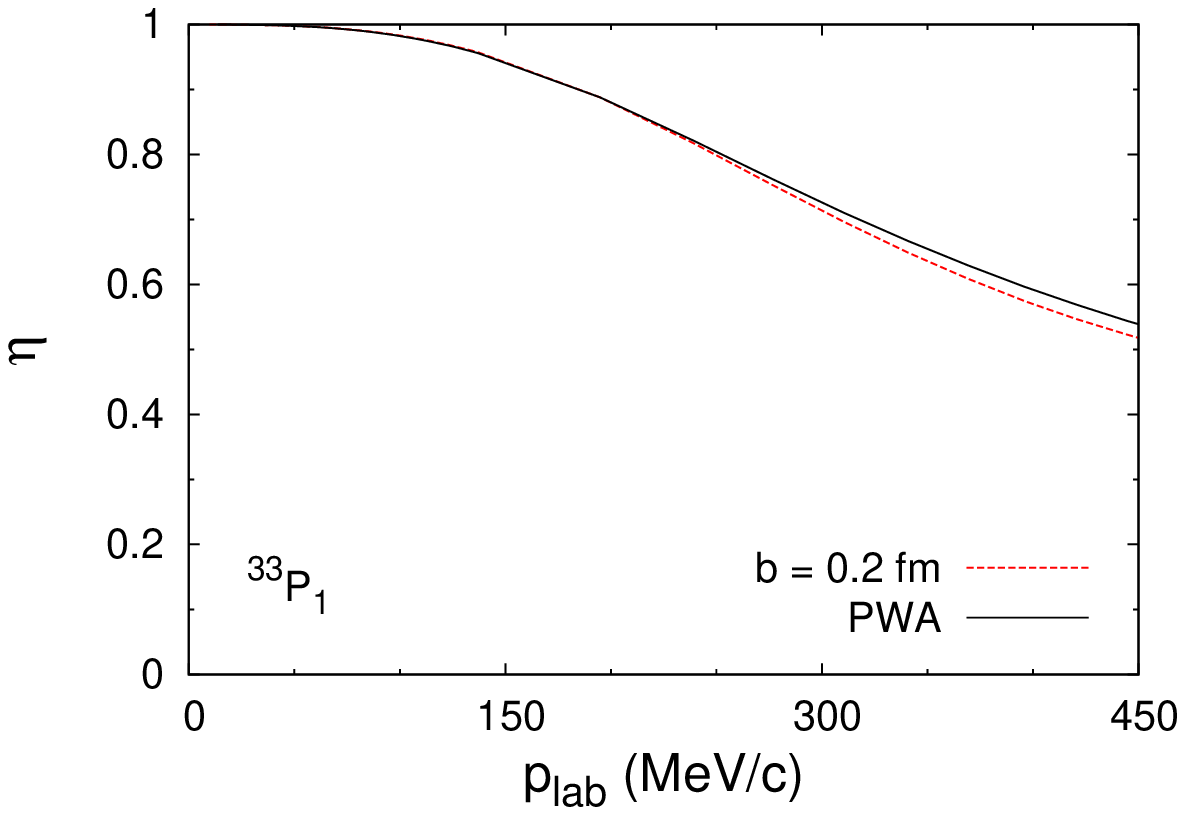}  
	\caption{\label{Phase_plab1}{(Color online) 
	Phase shifts (left panels) and inelasticities (right panels) of the spin-triplet uncoupled $P$ waves that require counterterms against laboratory momentum.
	The (red) dashed lines are from iterated one-pion exchange for $b=0.2$ fm and $V_c$, $W_c$ from Table \ref{tab:potentials1}, while (black) solid lines are the results of the PWA~\cite{Zhou:2012ui,Zhou:2013}.}}
\end{figure}

In the other two uncoupled spin-triplet waves, $^{13}P_{0}$ and $^{33}P_{1}$, where the OPE tensor force is singular and attractive, the cutoff dependence of the phase shifts is very obvious. The oscillation structure seen in the $N\!N$ $^{33}P_0$ channel \cite{Nogga:2005hy} is evident, especially in the analogous $\overline{N}\!N$ $^{13}P_{0}$. It represents the repeated appearance of shallow bound states as $\Lambda$ increases. In between the regions of fast variation there are plateaus that become less visible as the energy increases. These two $\overline{N}\!N$ channels require counterterms, which again we fit to the PWA at $T_{\rm lab}=20$ MeV. The running of the corresponding LECs $V_c$ and $W_c$ is shown in Fig. \ref{Counter_cut1}. In each of the $V_c$ there is a striking ``ankle'' structure associated with a dip in the corresponding $W_c$, beyond which $V_c$ is much larger in magnitude than $W_c$. These 
short-range interactions --- attractive but not as much as the singular OPE they replace at short distances --- prevent the appearance of shallow bound states and guarantee the cutoff independence of the phase shifts and inelasticities, see Fig. \ref{Phase_LC1}. This generalizes for $\overline{N}\!N$ the result found for $^{33}P_0$ in $N\!N$ \cite{Nogga:2005hy,PavonValderrama:2005uj}. We plot the observables as functions of the laboratory momentum in Fig. \ref{Phase_plab1} for $b=0.2$ fm, where the LECs take the values given in Table \ref{tab:potentials1}. The quality of the reproduction of the PWA values is even higher than for the spin-singlet $S$ waves in Fig. \ref{Phase_plab_SingletS}, where counterterms are also present.

\begin{figure}[tb]
   \centering
   \includegraphics[width=0.45\textwidth]{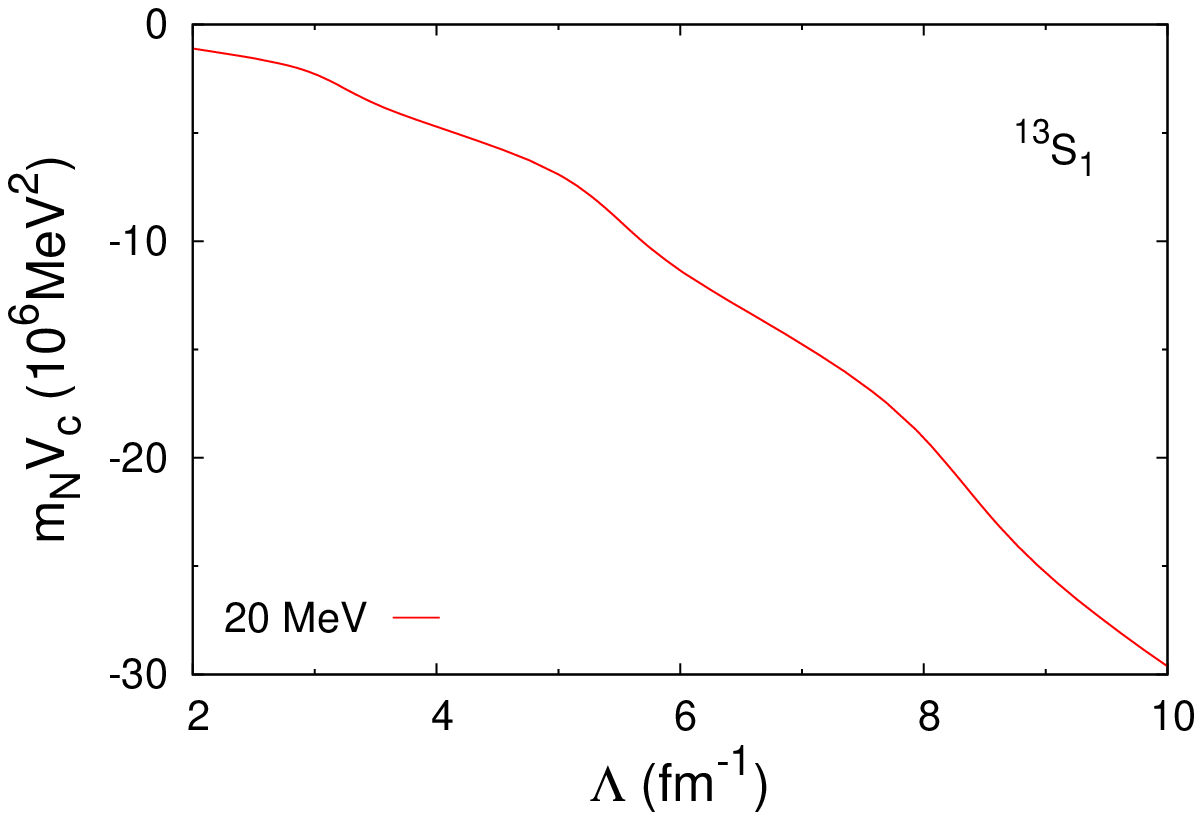} \hspace{2em}
   \includegraphics[width=0.45\textwidth]{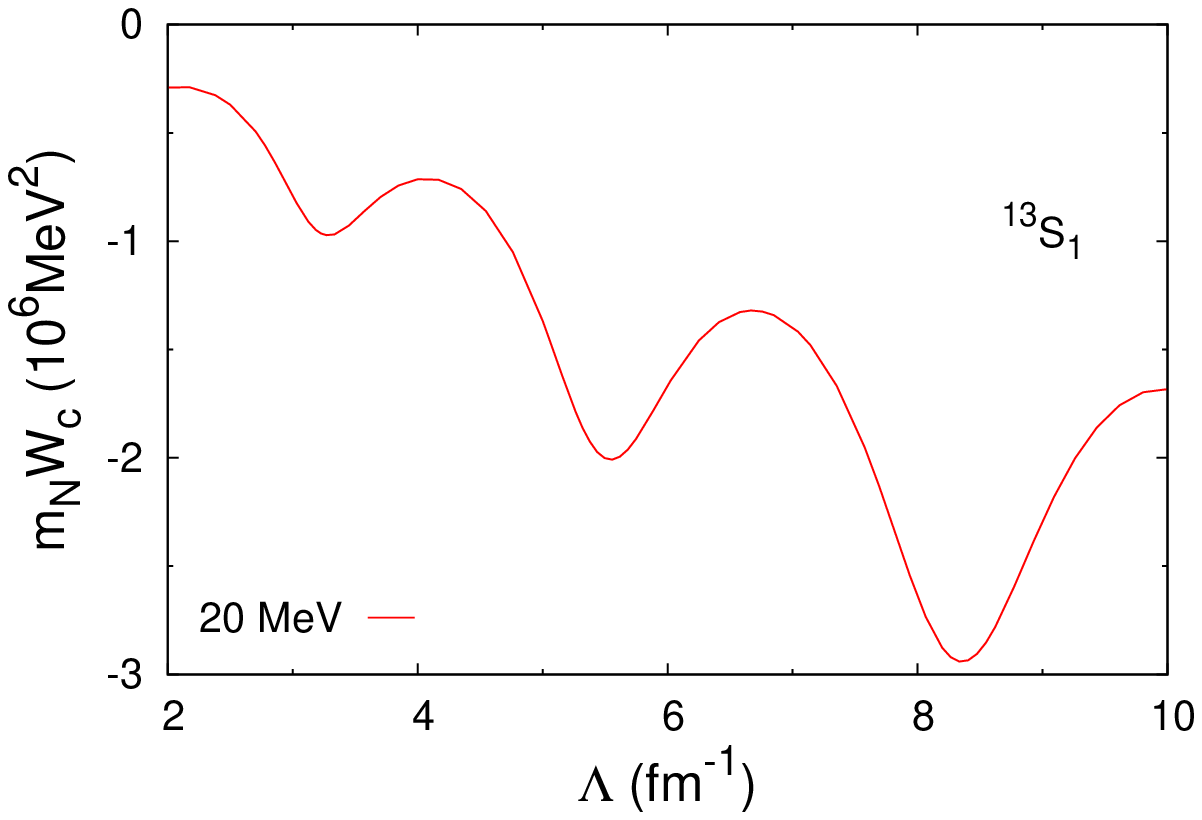} \\   
   \includegraphics[width=0.45\textwidth]{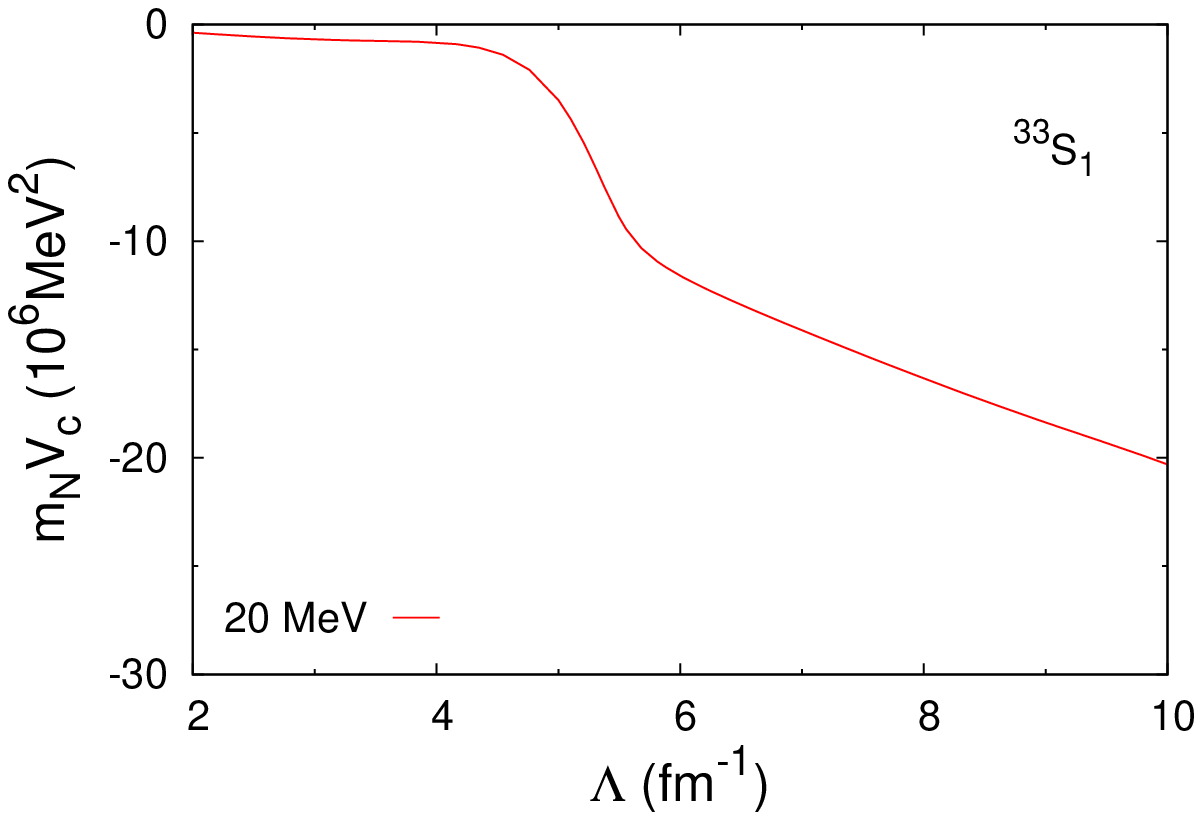} \hspace{2em}
   \includegraphics[width=0.45\textwidth]{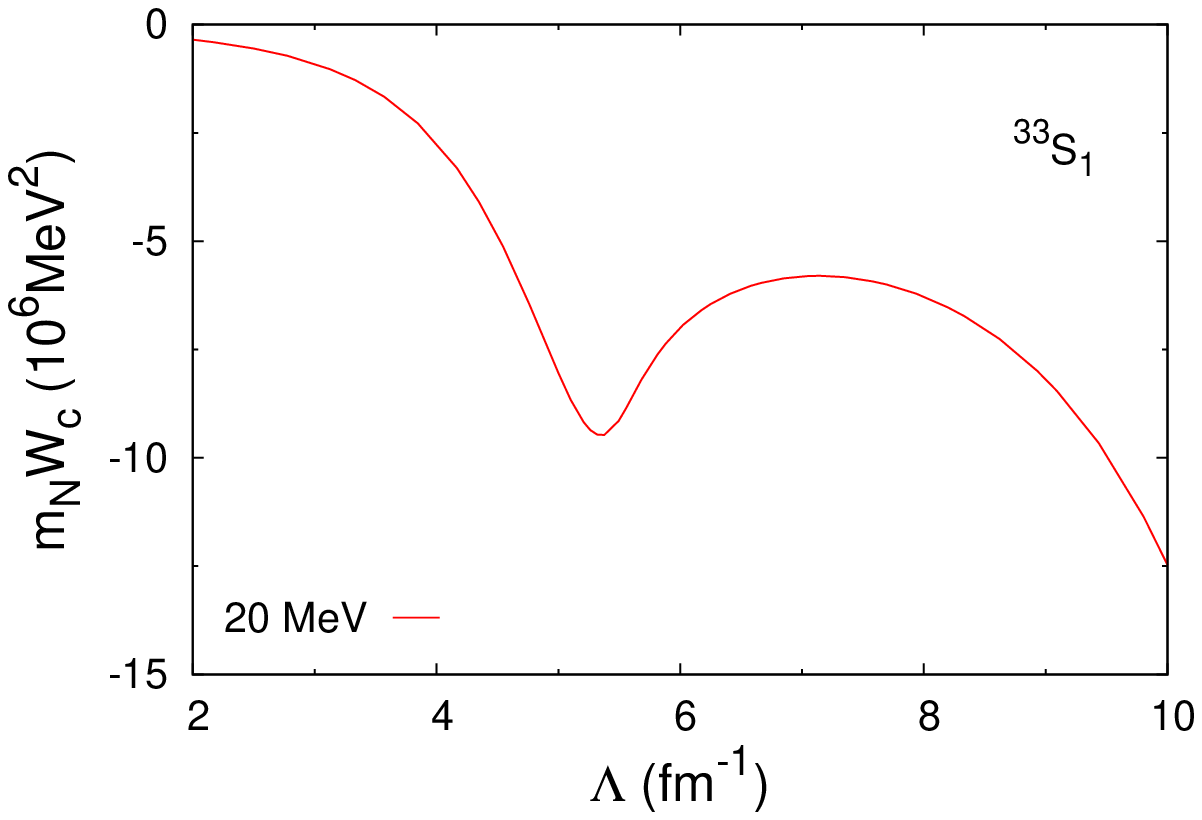}  
   \caption{\label{Counter_cutTripletS}{(Color online) 
   Cutoff dependence of $m_N V_c$ (left panels) and $m_N W_c$ (right panels) for the spin-triplet $S$ waves. The PWA phase shifts and inelasticities are fitted at $T_{\rm lab}=20$ MeV.}}
\end{figure}

\begin{figure}[tb]
	\centering
	\includegraphics[width=0.45\textwidth]{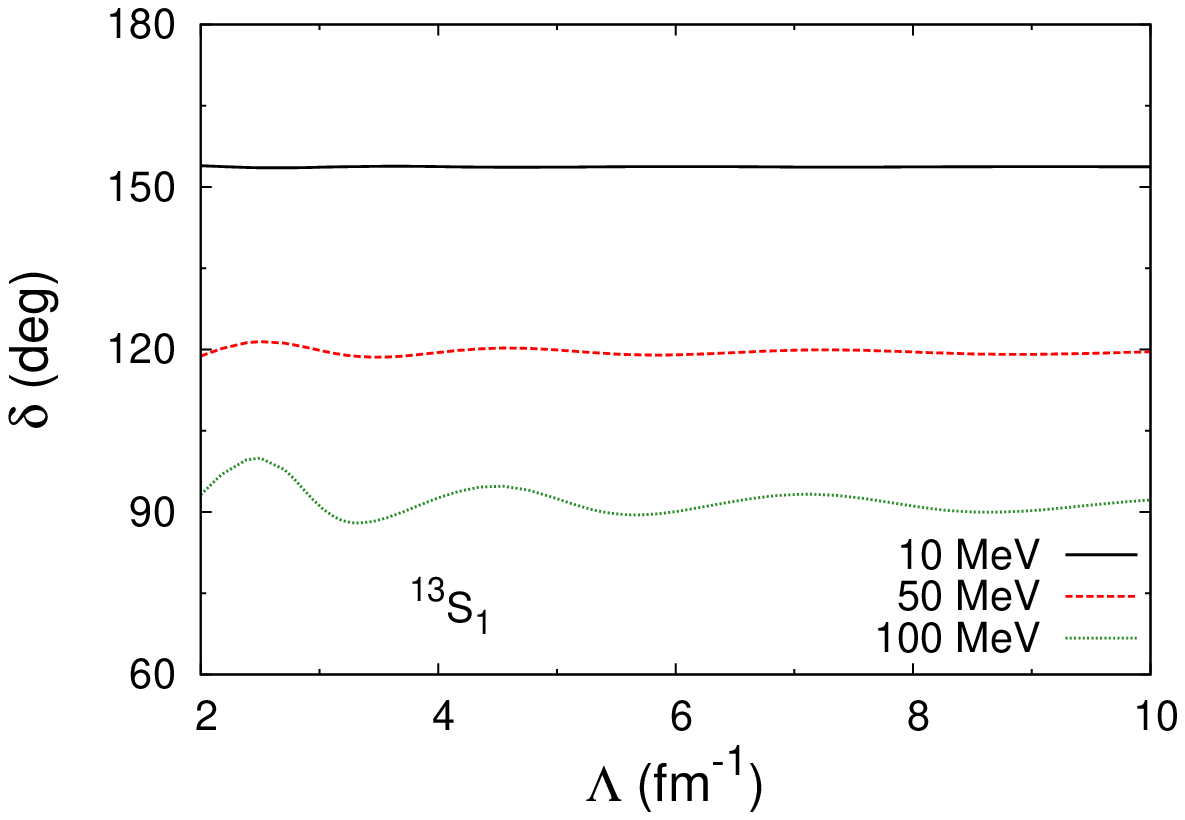} \hspace{2em}
	\includegraphics[width=0.45\textwidth]{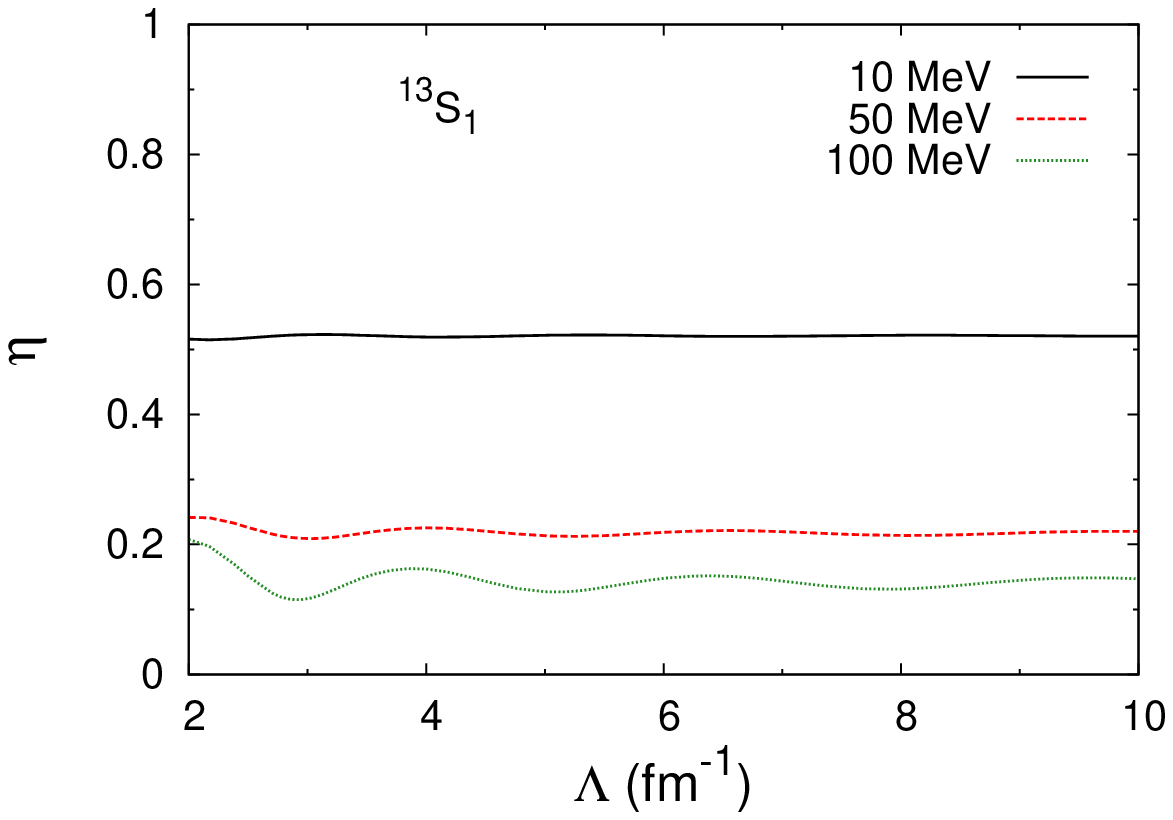}\\
	\includegraphics[width=0.45\textwidth]{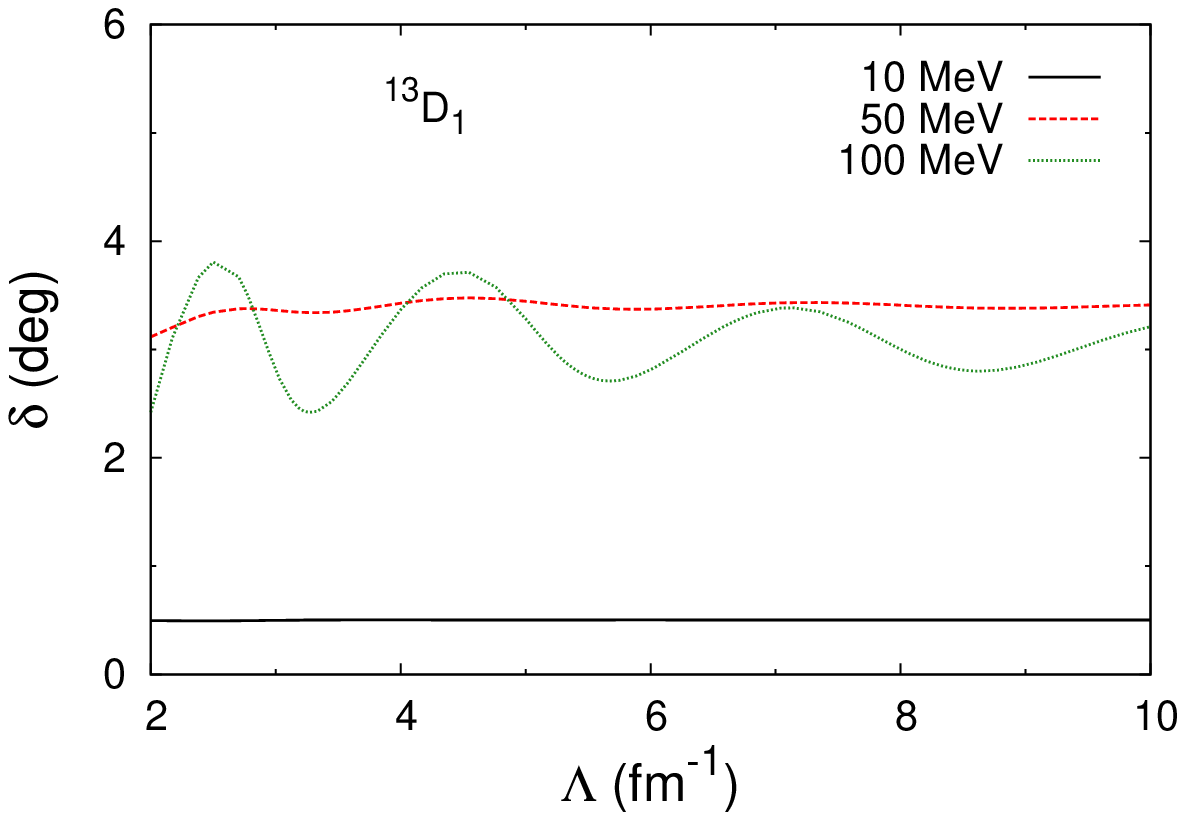} \hspace{2em}
	\includegraphics[width=0.45\textwidth]{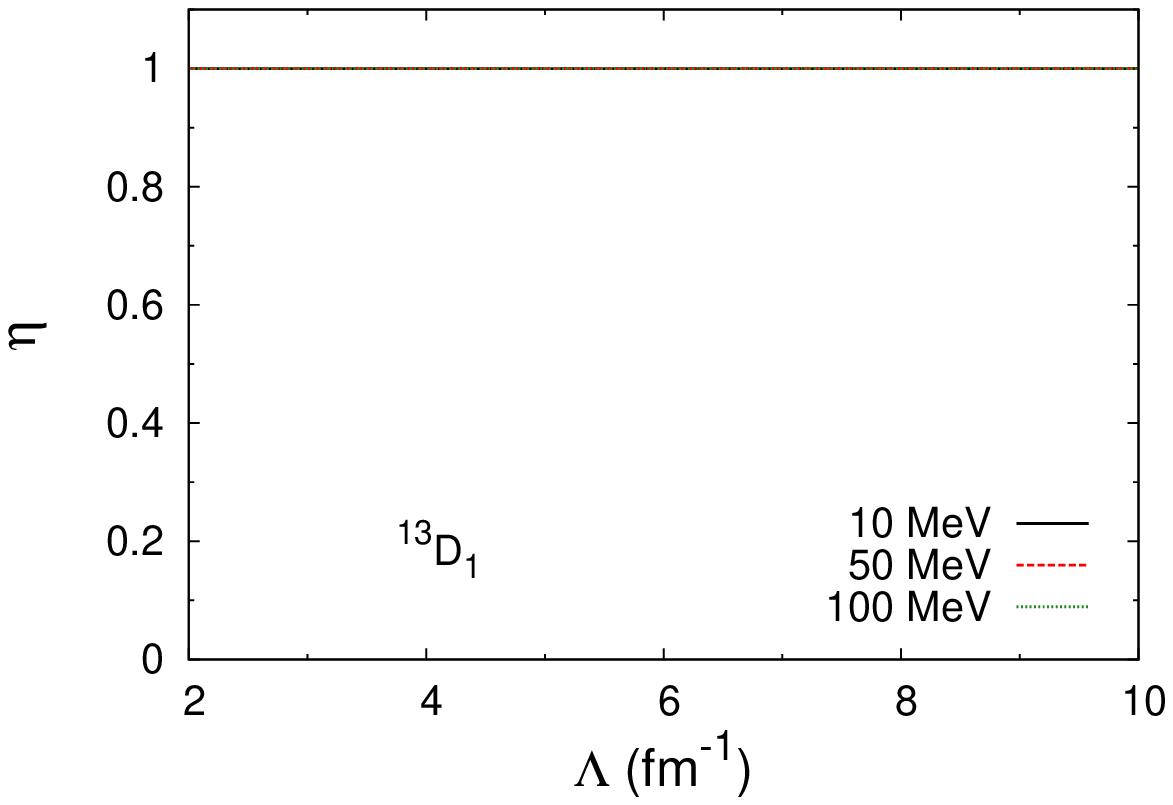} \\   
	\includegraphics[width=0.45\textwidth]{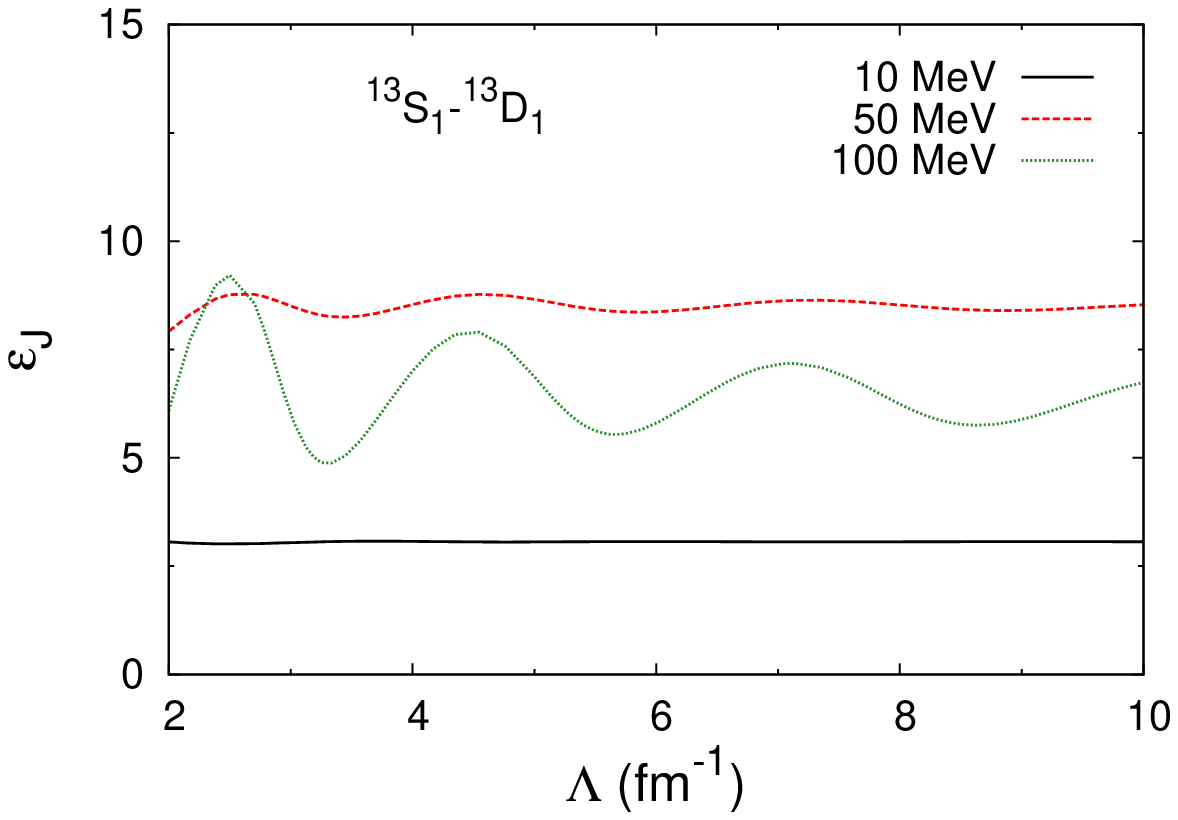} \hspace{2em}
	\includegraphics[width=0.45\textwidth]{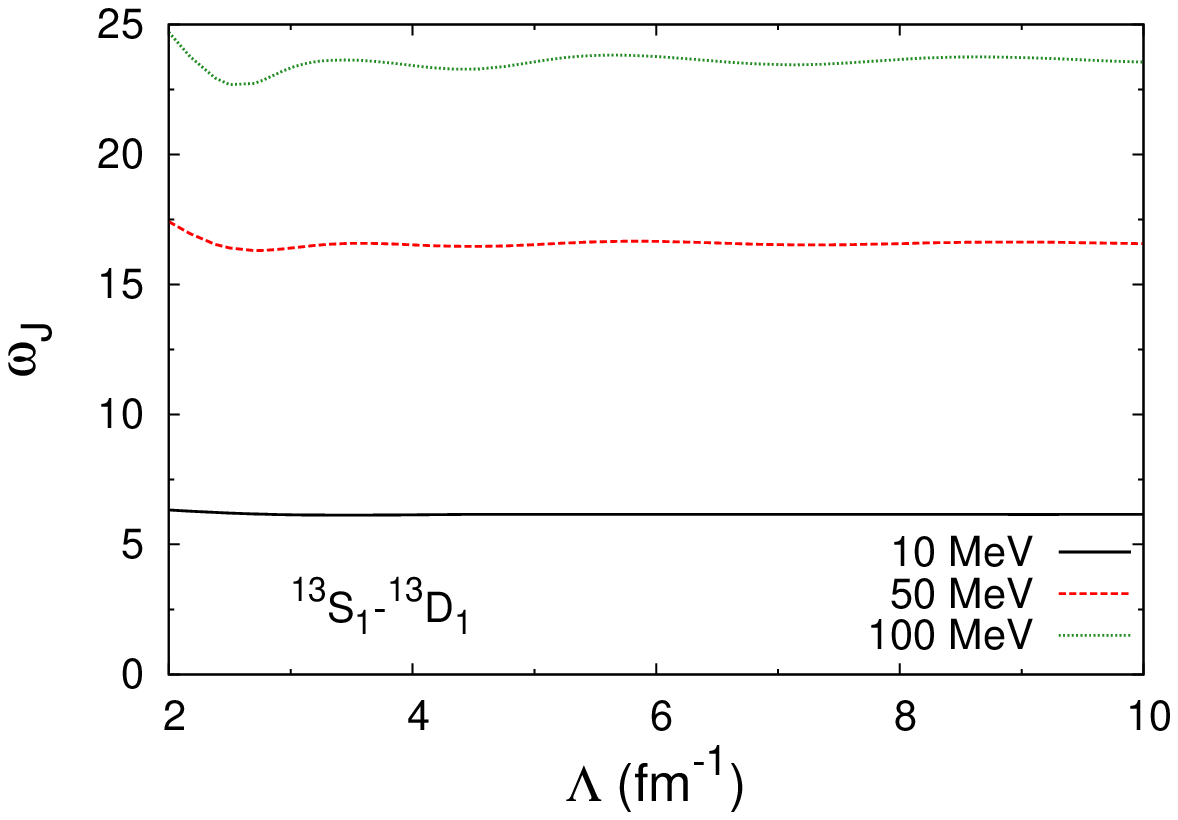}
	\caption{\label{Phase_LC21}{(Color online) Residual cutoff dependence of the phase shifts, inelasticities, and mixing angles in the $^{13}S_1$-$^{13}D_1$ waves at the laboratory energies of 10 MeV (black solid line), 50 MeV (red dashed line), and 100 MeV (green dotted line), for $V_c$ and $W_c$ in Fig. \ref{Counter_cutTripletS}.}}
\end{figure}

\begin{figure}[tb]
	\centering
	\includegraphics[width=0.45\textwidth]{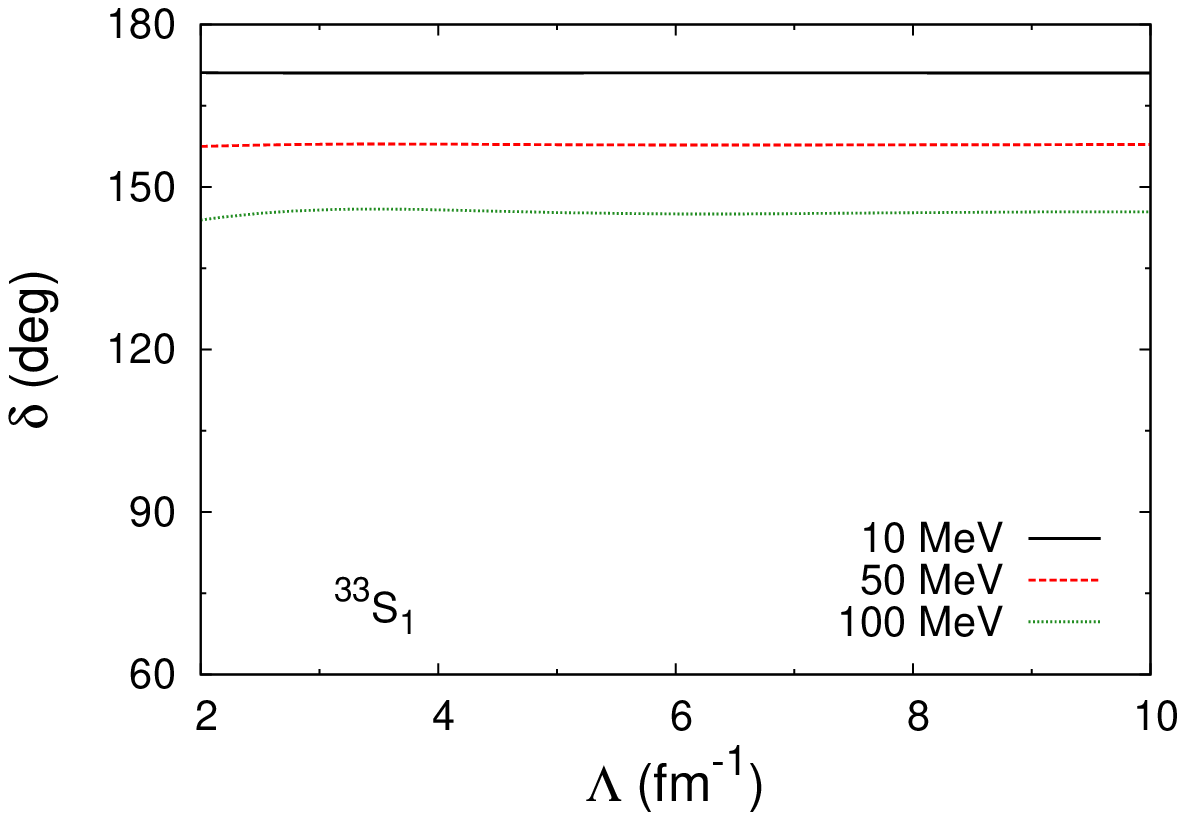} \hspace{2em}
	\includegraphics[width=0.45\textwidth]{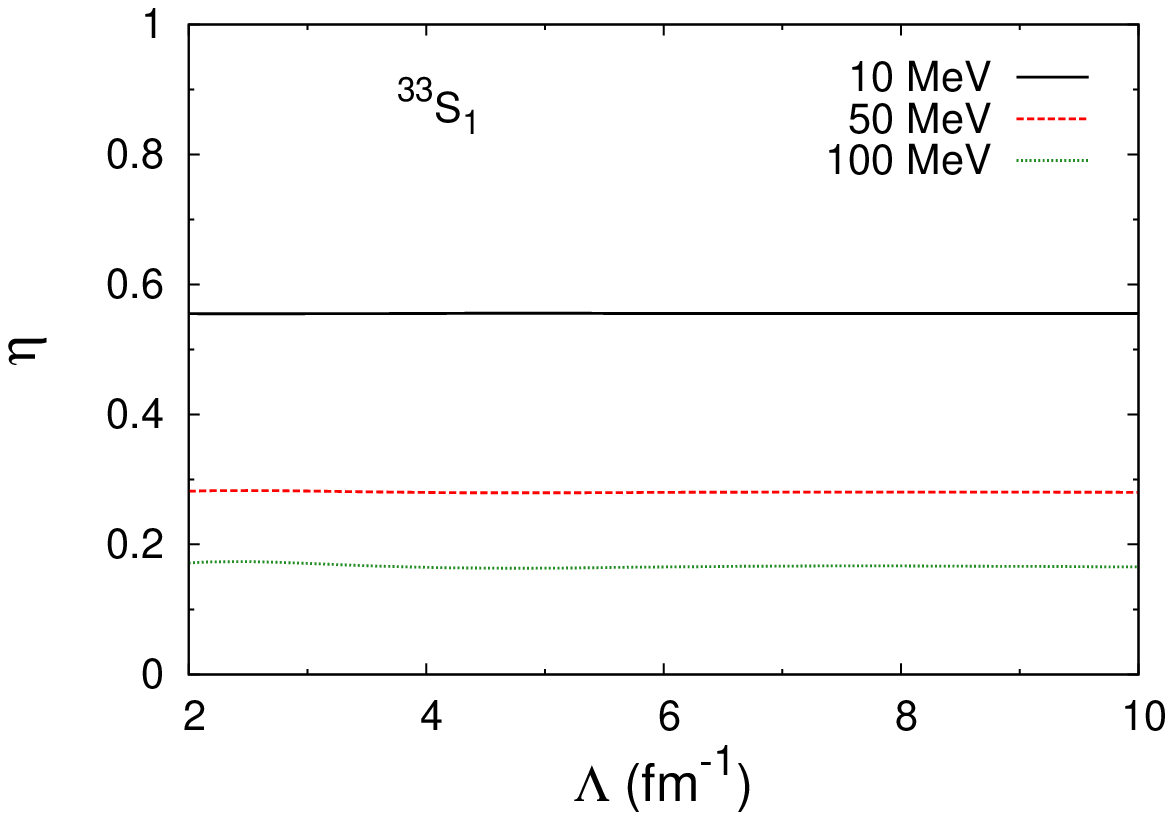}\\
	\includegraphics[width=0.45\textwidth]{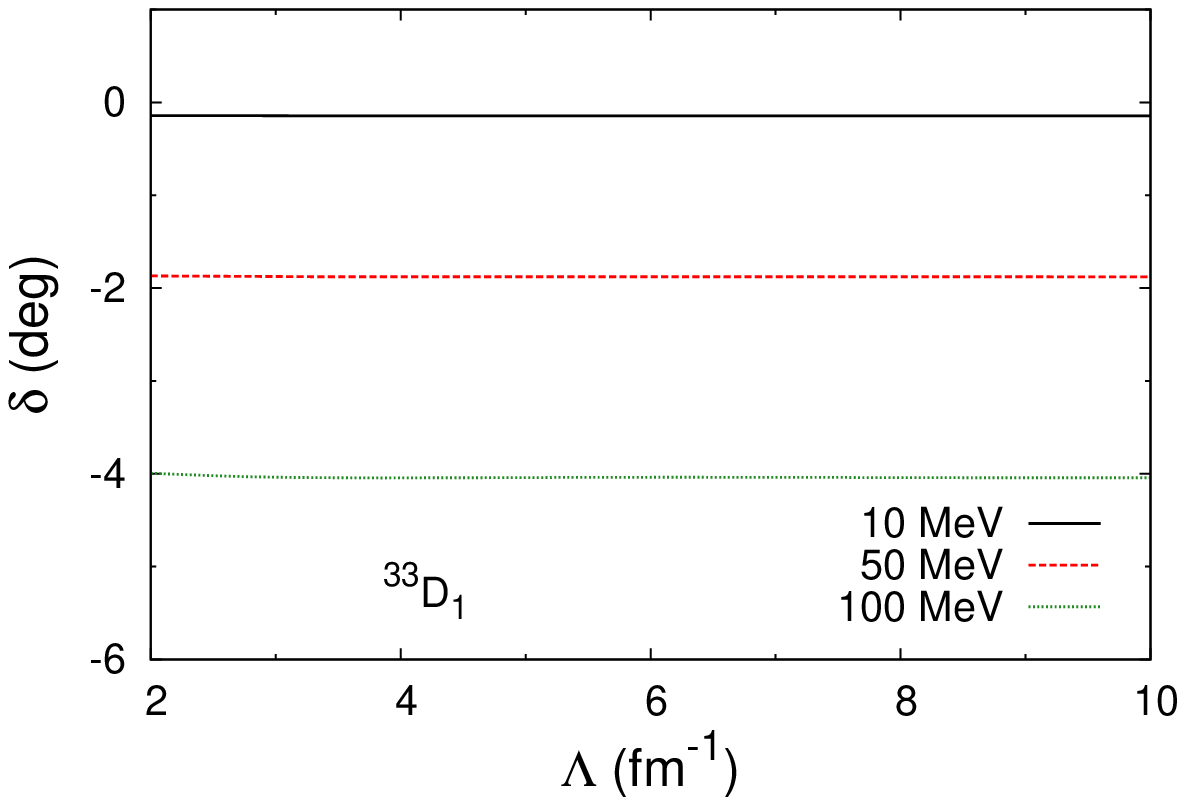} \hspace{2em}
	\includegraphics[width=0.45\textwidth]{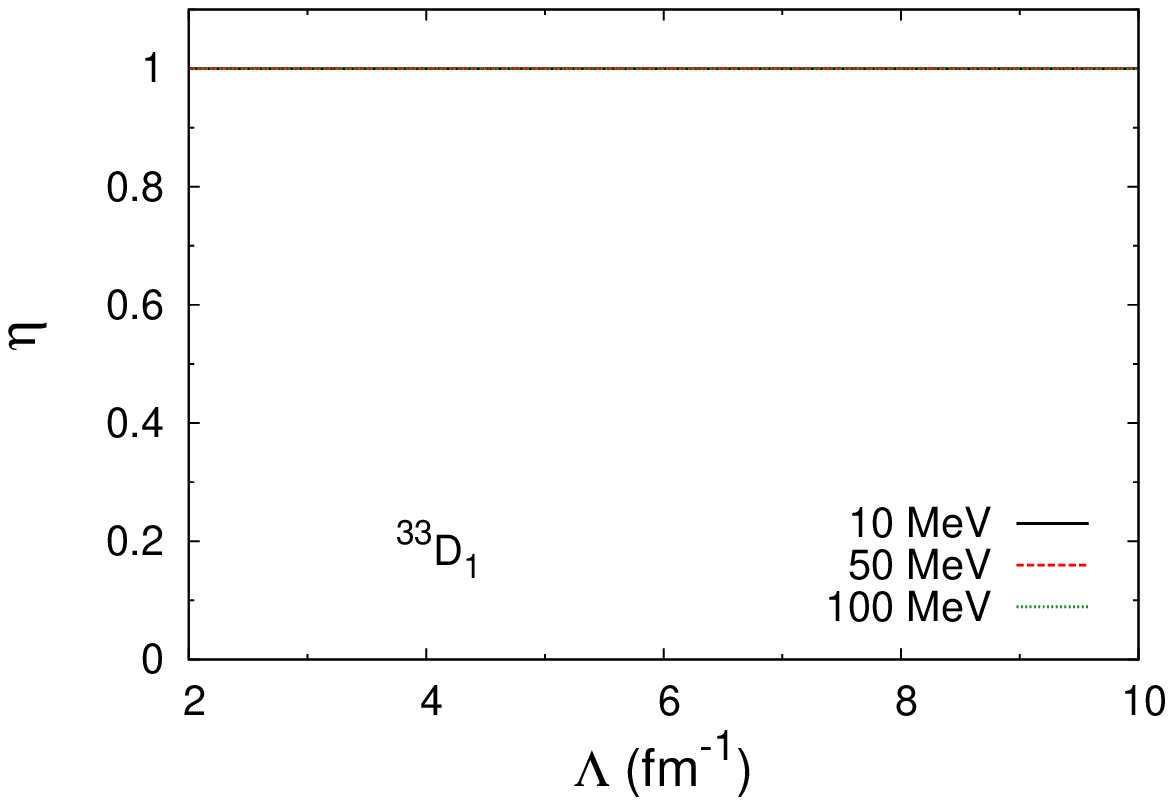} \\   
	\includegraphics[width=0.45\textwidth]{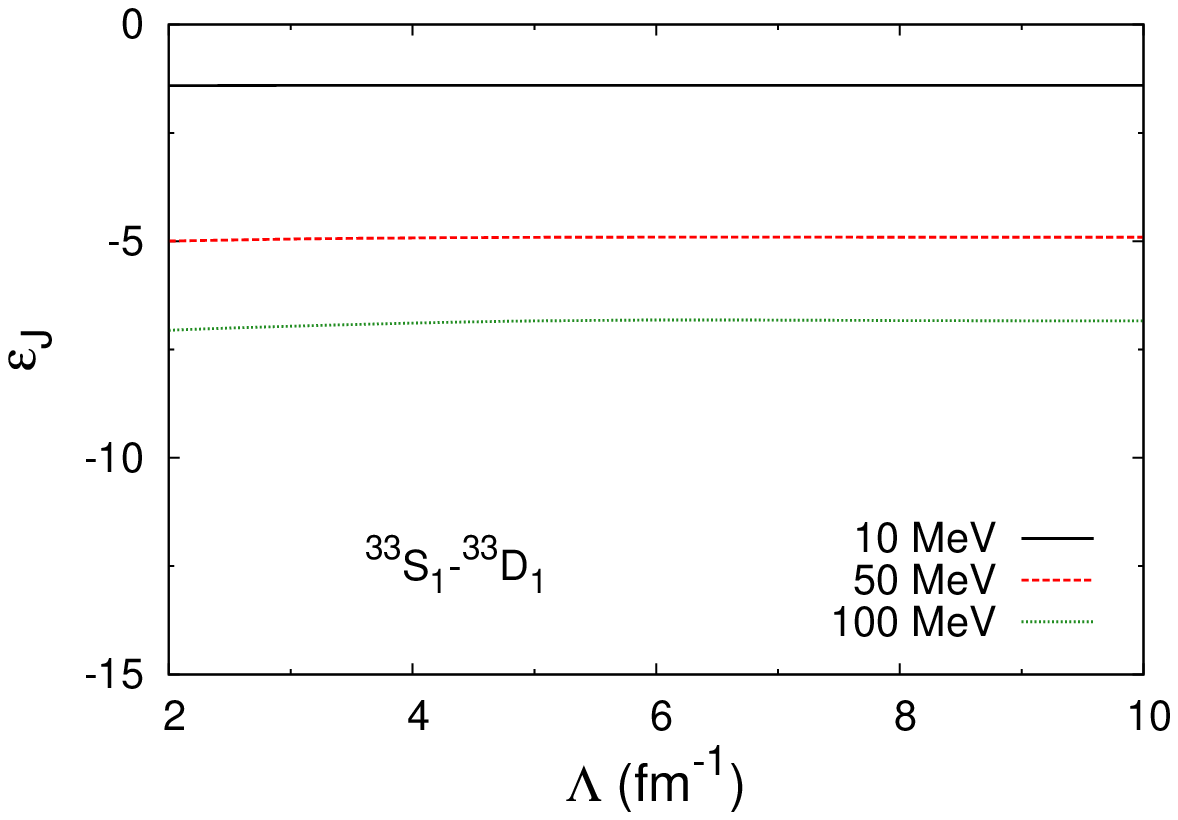} \hspace{2em}
	\includegraphics[width=0.45\textwidth]{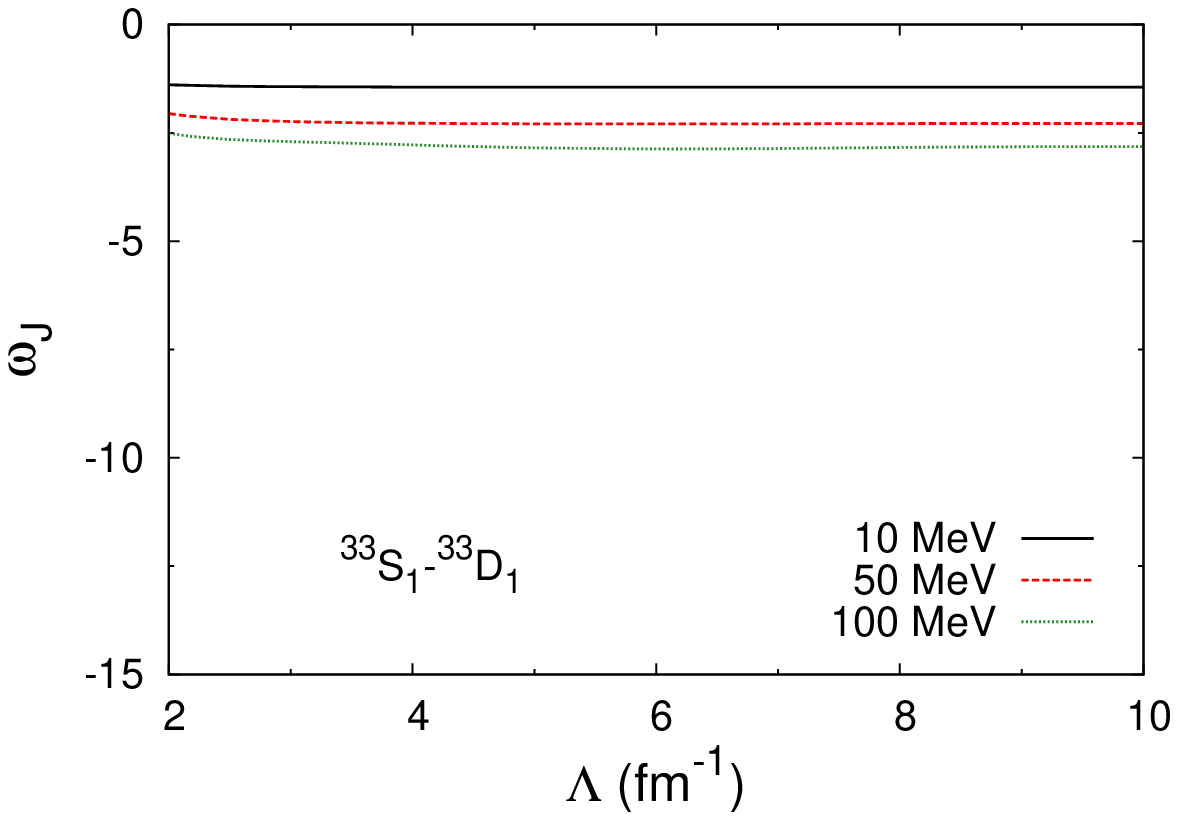}
	\caption{\label{Phase_LC22}{(Color online) Residual cutoff dependence of the phase shifts, inelasticities, and mixing angles in the $^{33}S_1$-$^{33}D_1$ waves at the laboratory energies of 10 MeV (black solid line), 50 MeV (red dashed line), and 100 MeV (green dotted line), for $V_c$ and $W_c$ in Fig. \ref{Counter_cutTripletS}.}}
\end{figure}

\begin{figure}[tb]
	\centering
	\includegraphics[width=0.45\textwidth]{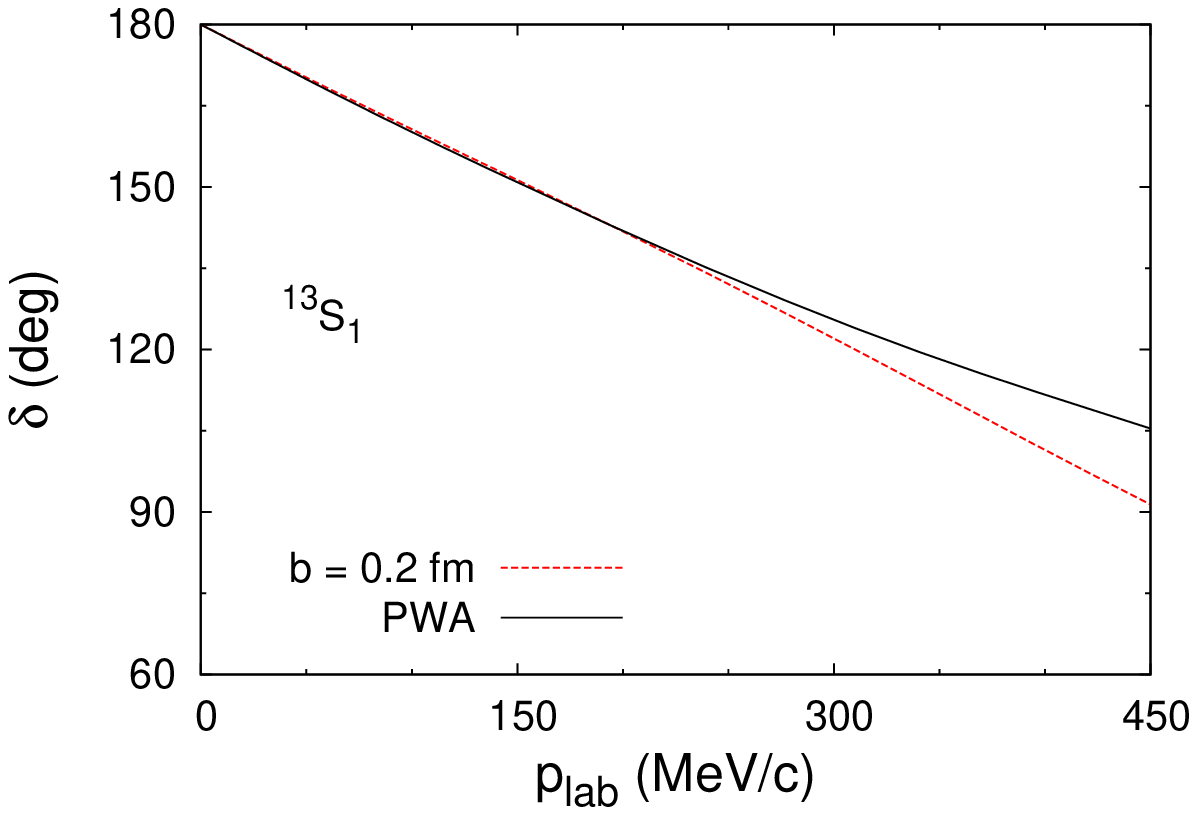} \hspace{2em}
	\includegraphics[width=0.45\textwidth]{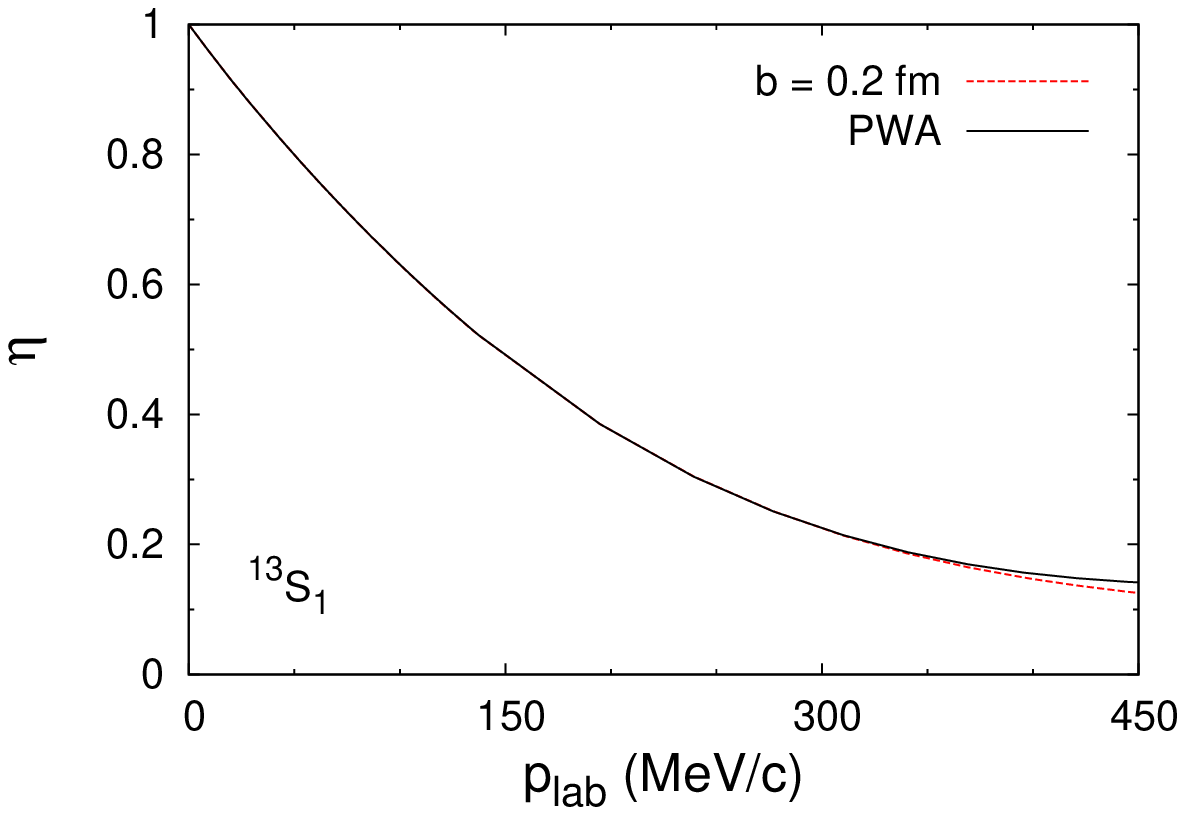} \\   
	\includegraphics[width=0.45\textwidth]{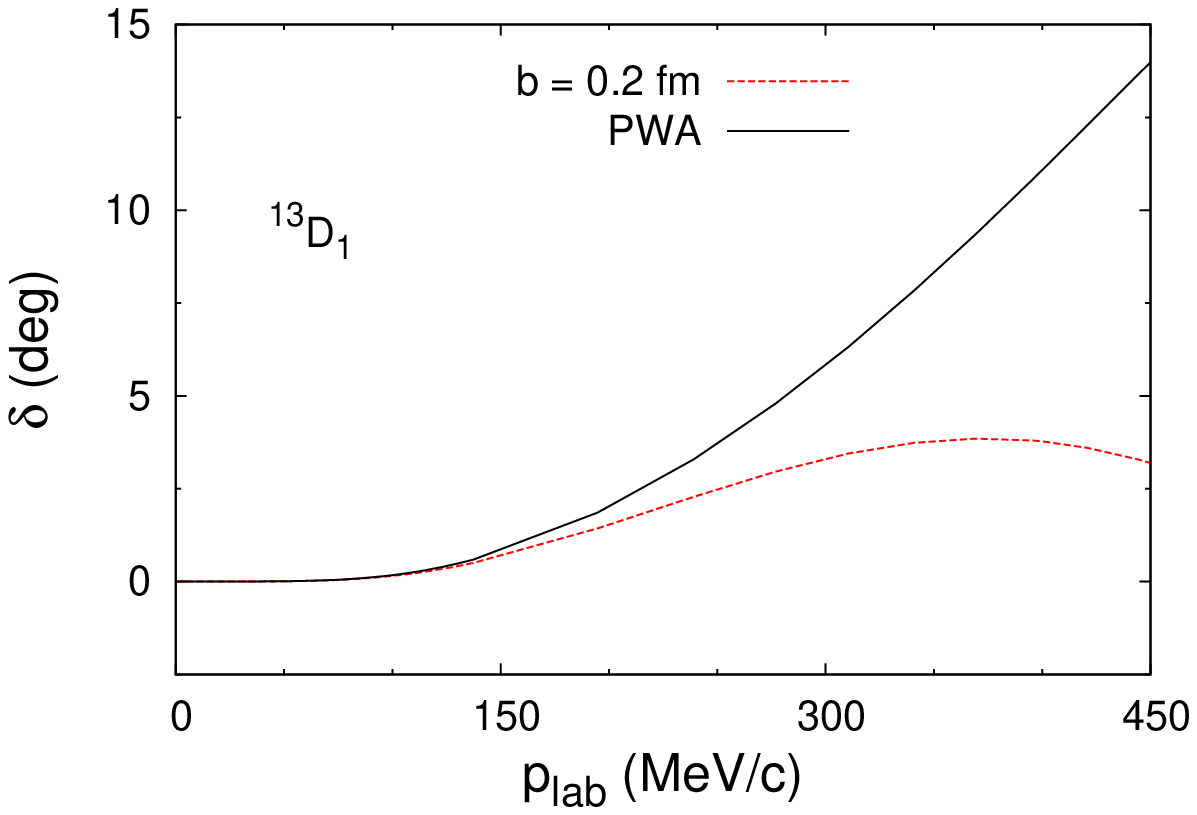} \hspace{2em}
	\includegraphics[width=0.45\textwidth]{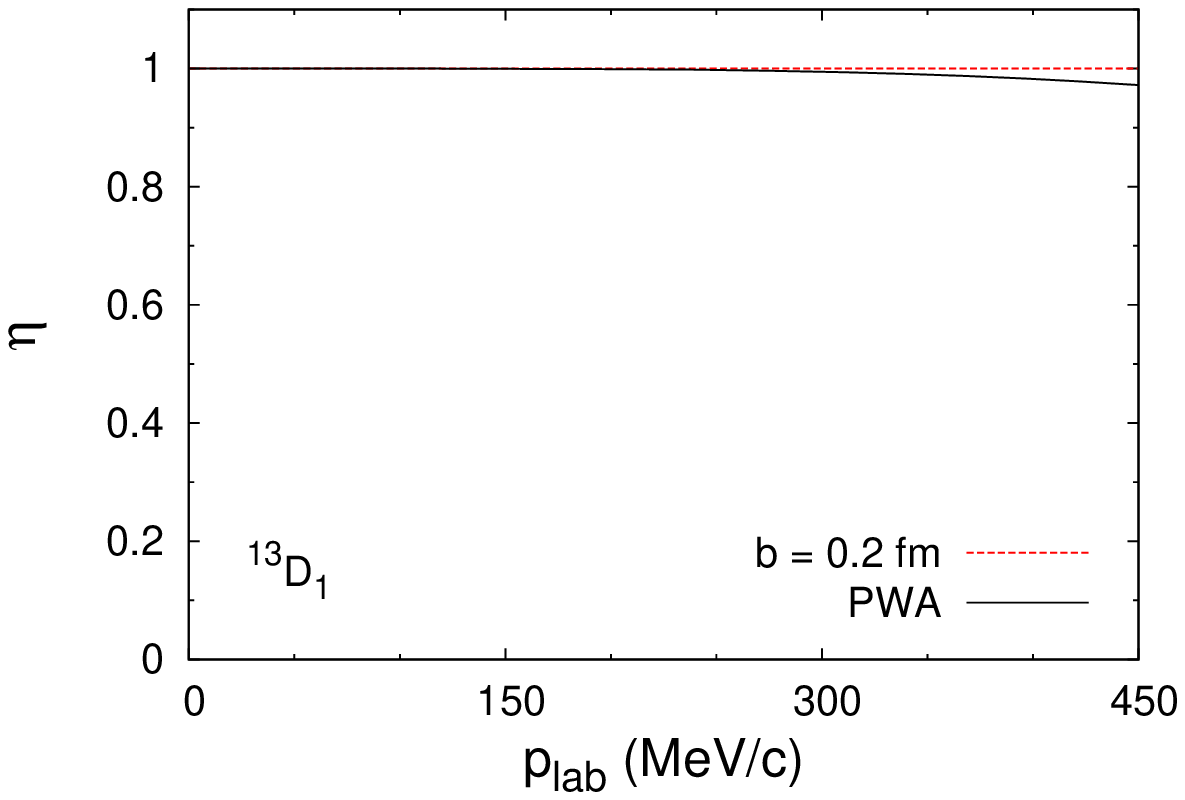}\\
	\includegraphics[width=0.45\textwidth]{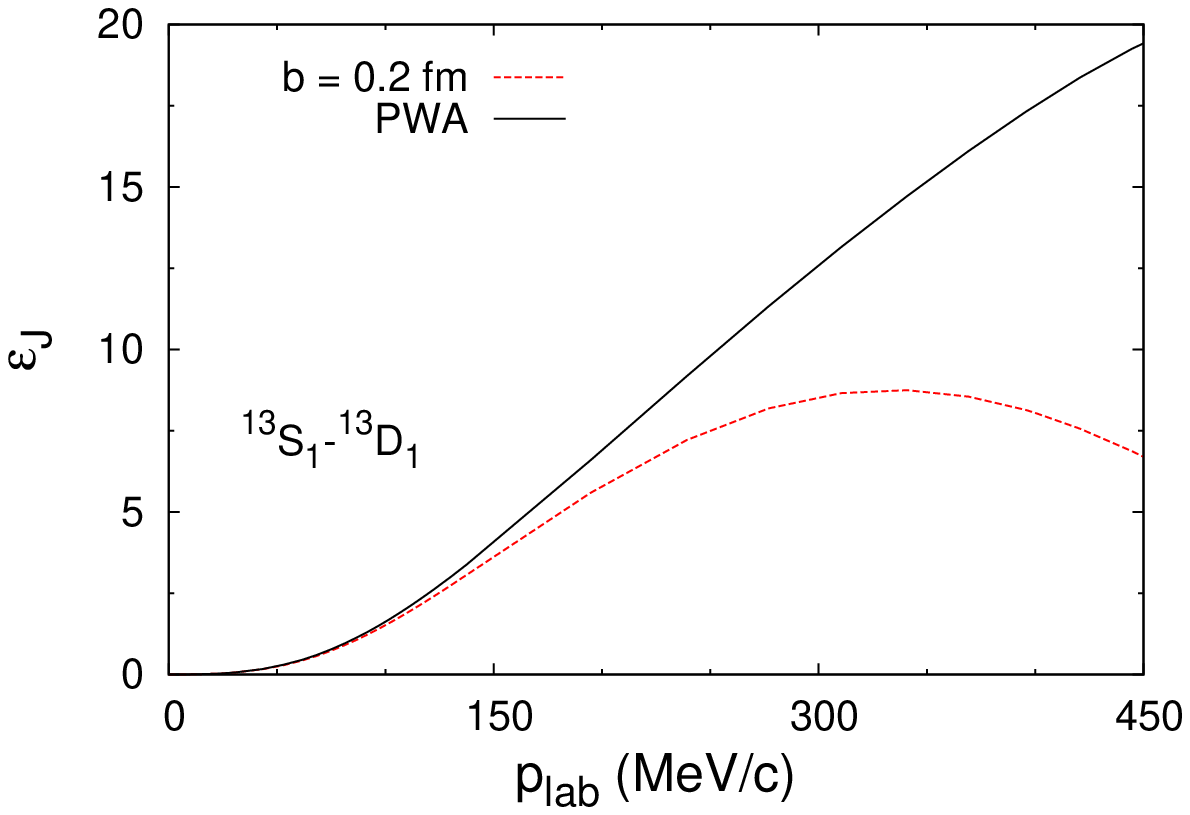} \hspace{2em}
	\includegraphics[width=0.45\textwidth]{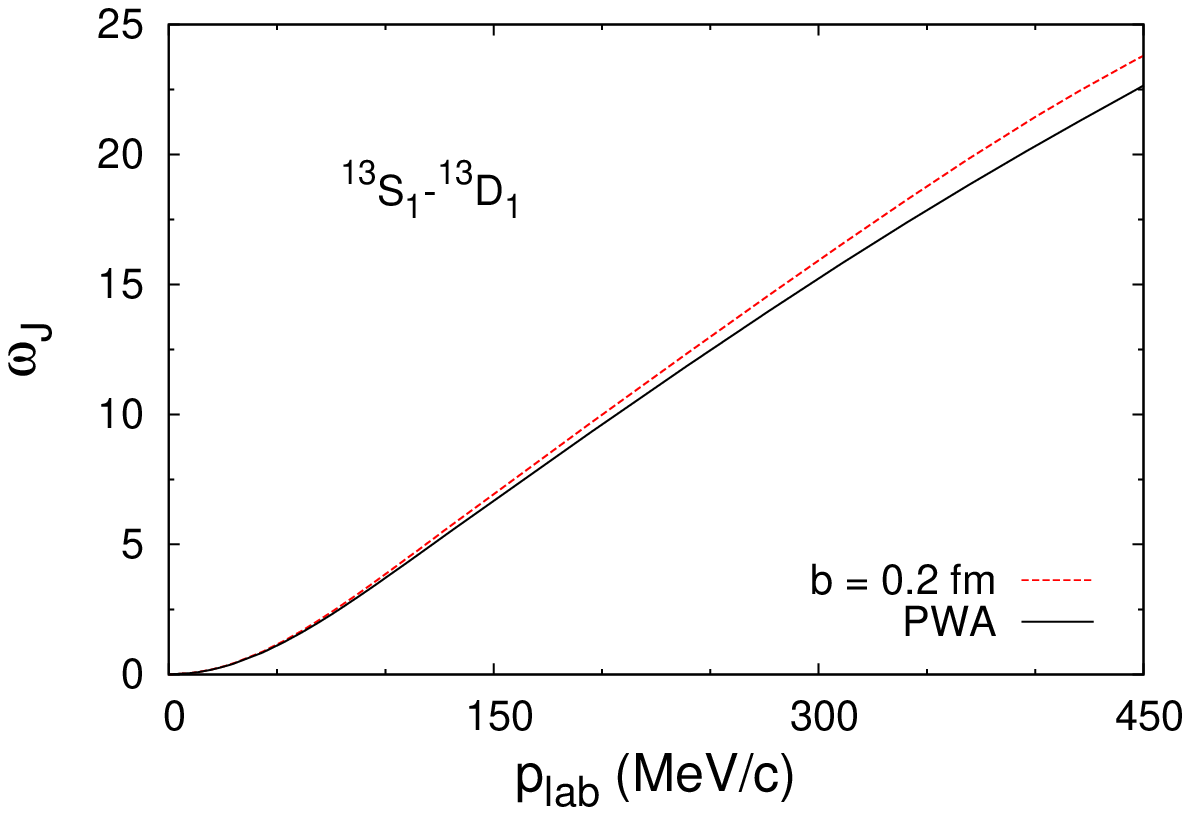}
	\caption{\label{Phase_plab21}{(Color online) 
	Phase shifts and mixing angle (left panels) and inelasticities and their mixing angle (right panels) for the coupled $^{13}S_1$-$^{13}D_1$ waves against laboratory momentum.
	The (red) dashed lines are from iterated one-pion exchange for $b=0.2$ fm and $V_c$, $W_c$ from Table \ref{tab:potentials2}, while (black) solid lines are the results of the PWA~\cite{Zhou:2012ui,Zhou:2013}.}}
\end{figure}

\begin{figure}[tb]
	\centering
	\includegraphics[width=0.45\textwidth]{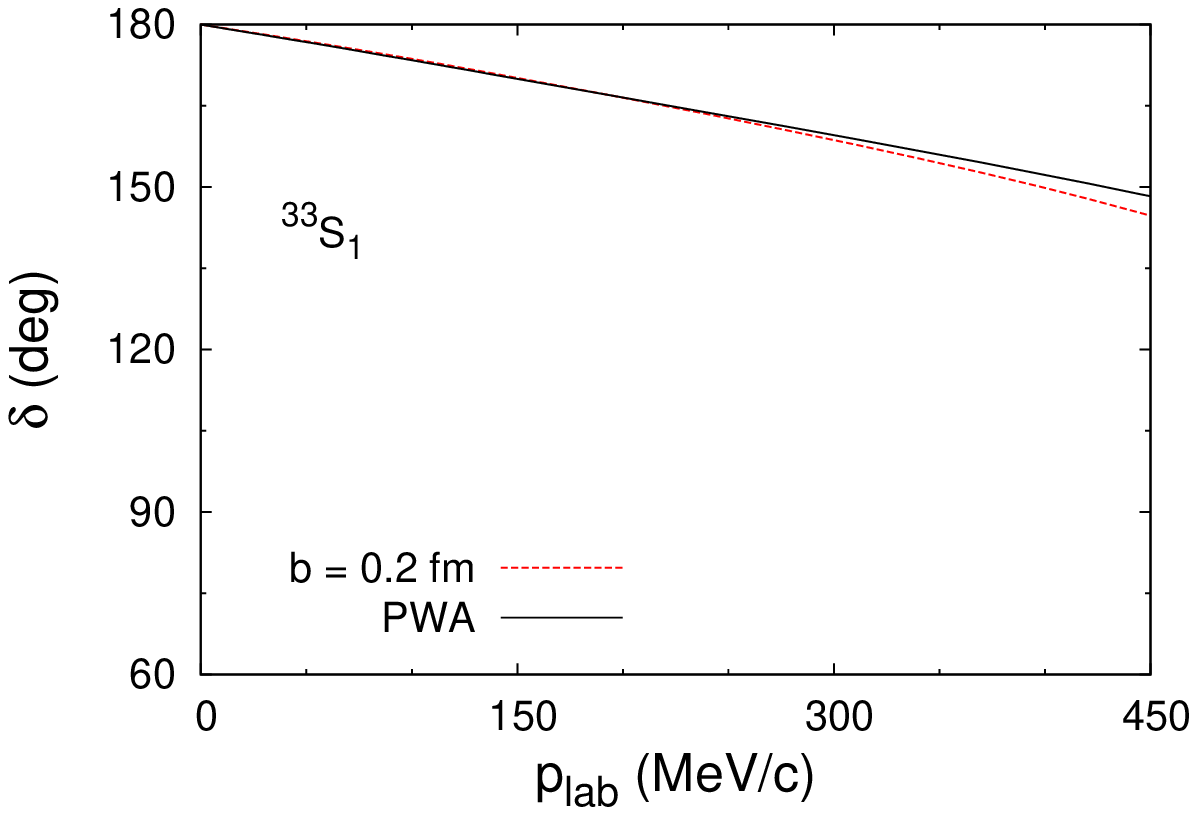} \hspace{2em}
	\includegraphics[width=0.45\textwidth]{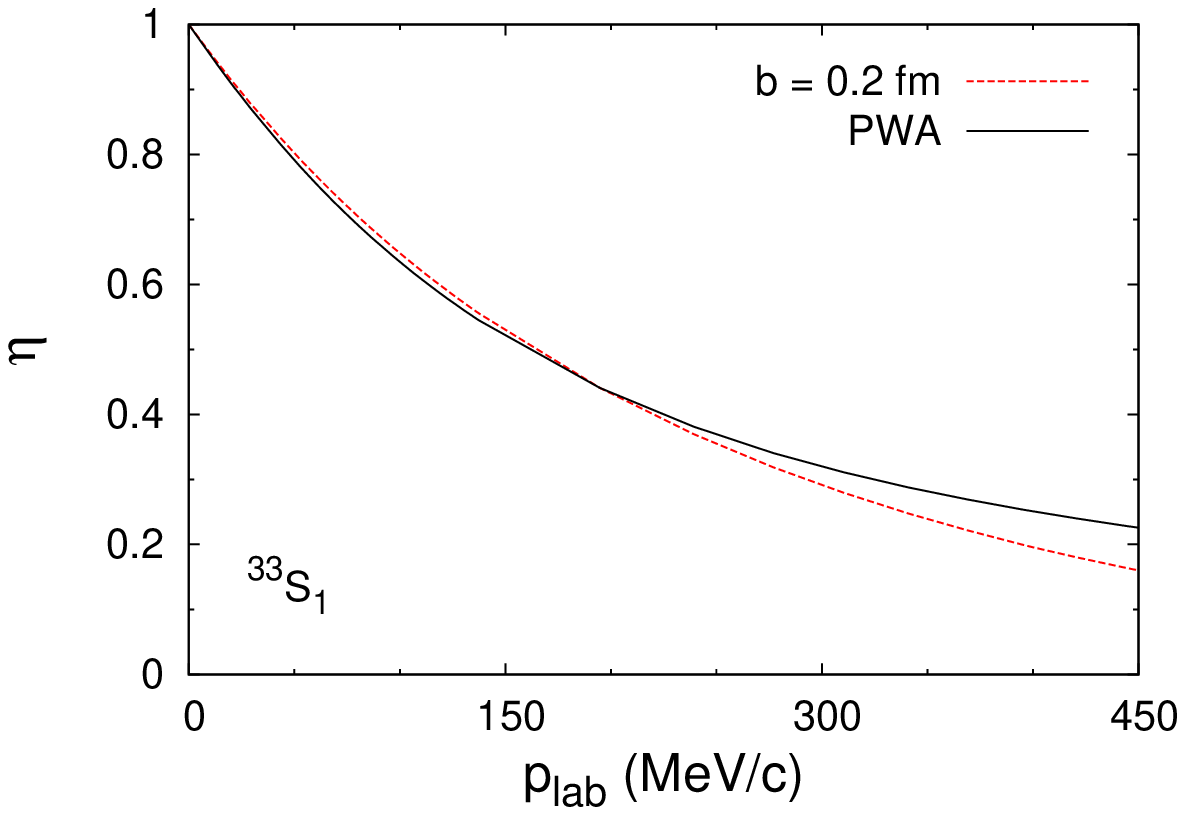} \\   
	\includegraphics[width=0.45\textwidth]{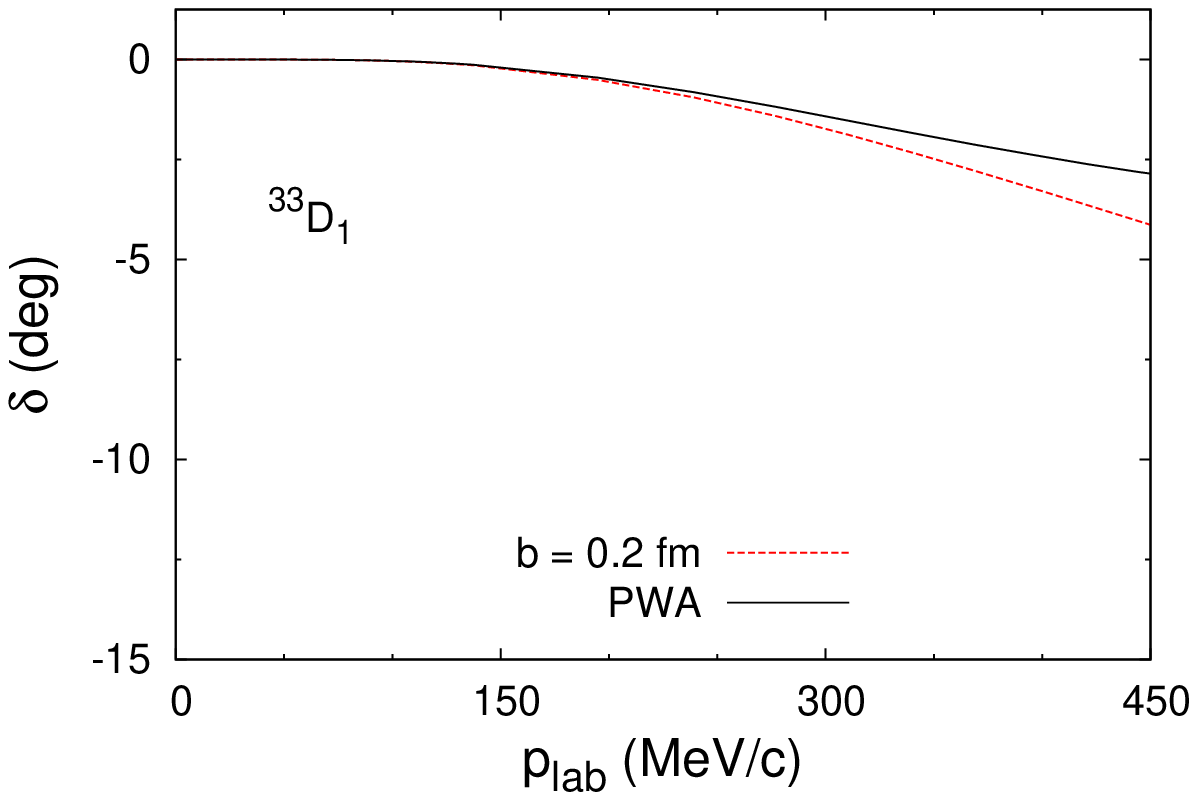} \hspace{2em}
	\includegraphics[width=0.45\textwidth]{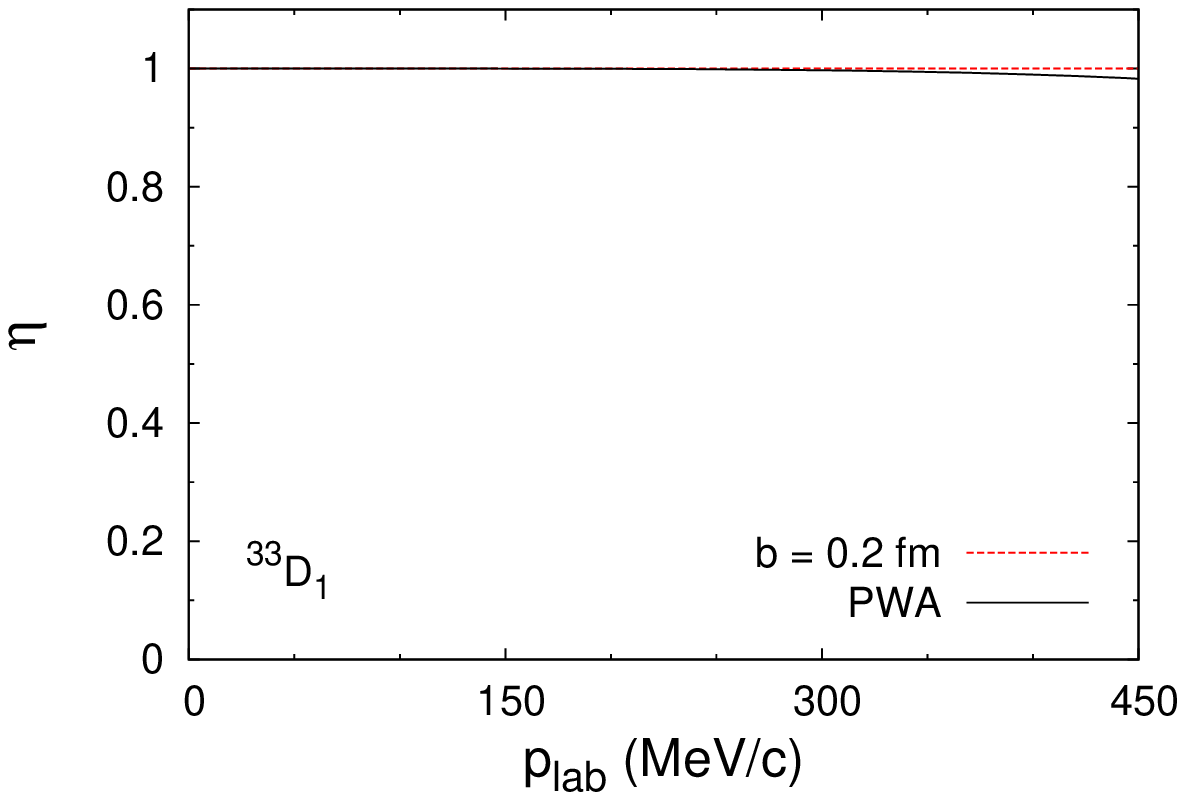}\\
	\includegraphics[width=0.45\textwidth]{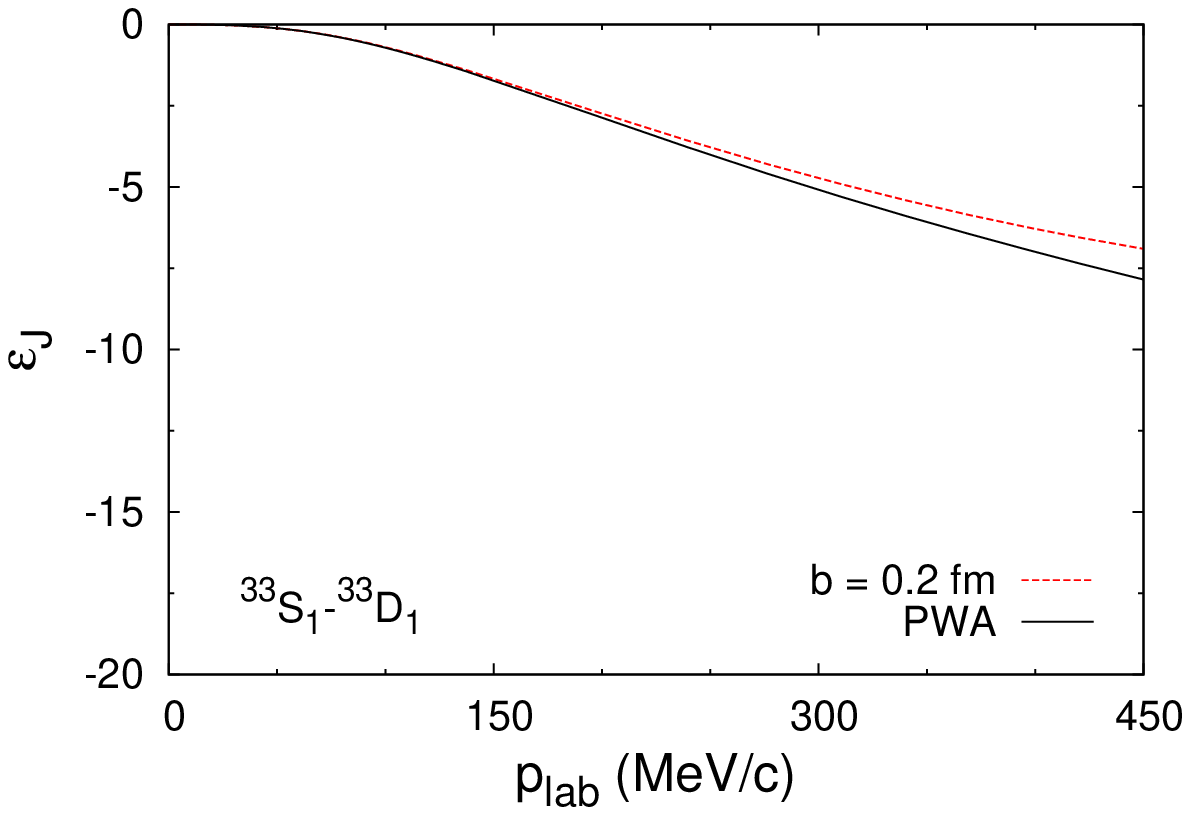} \hspace{2em}
	\includegraphics[width=0.45\textwidth]{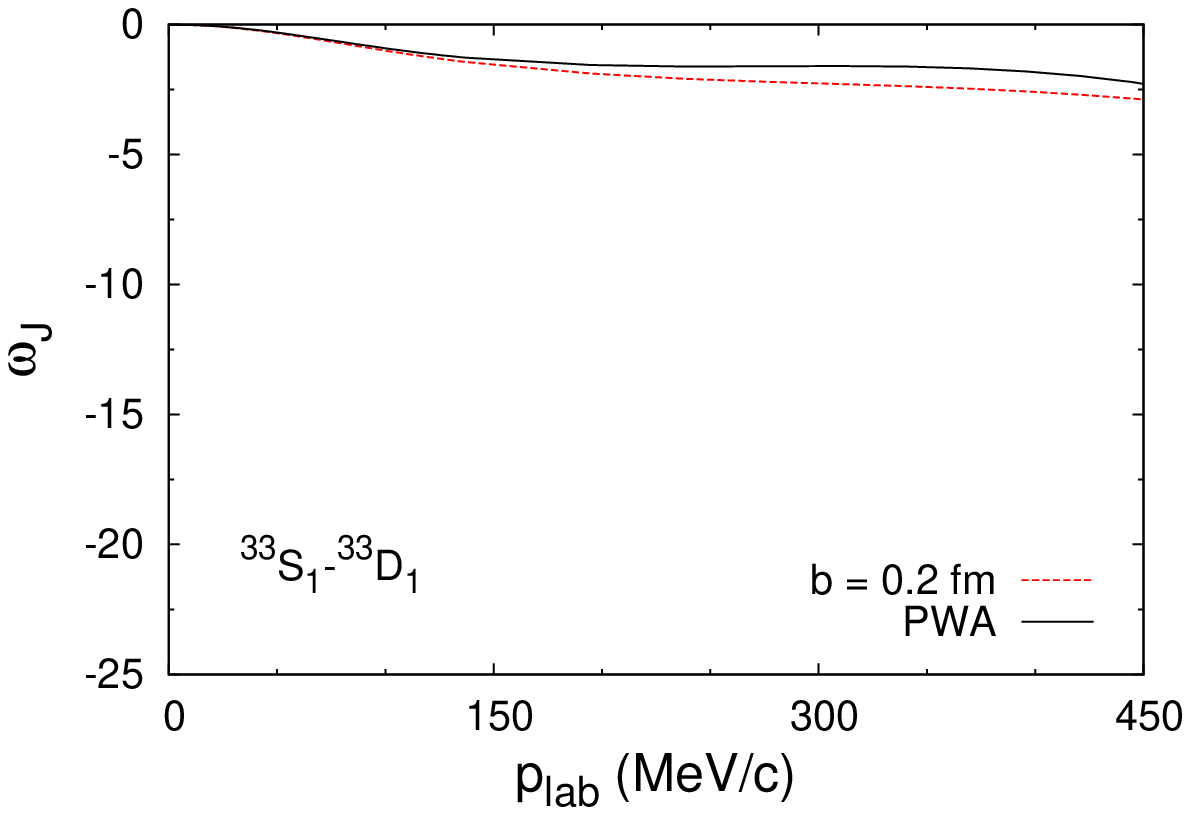}
	\caption{\label{Phase_plab22}{(Color online) 
	Phase shifts and mixing angle (left panels) and inelasticities and their mixing angle (right panels) for the coupled $^{33}S_1$-$^{33}D_1$ waves against laboratory momentum.
	The (red) dashed lines are from iterated one-pion exchange for $b=0.2$ fm and $V_c$, $W_c$ from Table \ref{tab:potentials2}, while (black) solid lines are the results of the PWA~\cite{Zhou:2012ui,Zhou:2013}.}}
\end{figure}

The cutoff dependence in the coupled waves exhibits the same oscillatory patterns, although it is weaker in the channels with higher angular momentum, particularly $^{33}D_{1}$ and $^{33}F_2$ where tensor OPE is weak. NDA prescribes short-range interactions in $^{13}S_1$ and $^{33}S_1$. As before, we adjust the corresponding LECs as functions of the cutoff in order to keep the $S$-wave phase shifts and inelasticites at $T_{\rm lab}=20$ MeV fixed. The corresponding $V_c$ and $W_c$ are shown in Fig. \ref{Counter_cutTripletS}.
The periodic behavior of $V_c$ in $^{13}S_1$ is similar to that seen with the same regulator in Ref. \cite{Beane:2000wh}, and it is accompanied by an oscillatory behavior in $W_c$. In the $^{33}S_1$ channel, where the tensor OPE is weaker, we find an ``ankle" structure in $V_c$, as before. We suspect that this is the beginning of a periodic pattern similar to $^{13}S_1$, just with a larger amplitude and lower frequency. These features are more visible in $W_c$.
With the counterterms, the oscillatory behavior of observables is damped, and they all now approach definite values asymptotically, as seen for $^{13}S_1$-$^{13}D_1$ in Fig. \ref{Phase_LC21} and $^{33}S_1$-$^{33}D_1$ in Fig. \ref{Phase_LC22}. As expected in an EFT, the residual cutoff dependence increases with energy. For the channel with stronger OPE, $I=0$, oscillations are still visible, but their amplitude decreases as $\Lambda$ increases. The residual cutoff dependence is particularly small for $I=1$. For both isospin values, $\eta=1$ in the $D$ wave where $W_c=0$. The renormalization with an $S$-wave counterterm is analogous to the coupled $^3S_1$-$^3D_1$ channel in the $N\!N$ system \cite{Frederico:1999ps,Beane:2001bc}. Using the values of the LECs at $b=0.2$ fm, given in Table \ref{tab:potentials2}, we compare observables with the PWA in Figs. \ref{Phase_plab21} and \ref{Phase_plab22} for $^{13}S_1$-$^{13}D_1$ and $^{33}S_1$-$^{33}D_1$, respectively. 
The best agreement might have been expected in the $S$-wave phase shifts and inelasticities, where LECs were fitted. The higher-energy discrepancies will presumably be reduced at higher orders. Empirical inelasticities in the $D$ waves are very close to 1, consistent with the absence of an imaginary counterterm in those channels. The inelasticity mixing is relatively large for $I=0$ but well described for both isospin values. For $I=1$, also the $D$-wave phase shift and the $D$-$F$ mixing angle are postdicted well.
Overall, the agreement is similar to other channels that contain a counterterm, except for the $I=0$ $D$ phase shift and $S$-$D$ mixing angle $\varepsilon_J$, where the discrepancies resemble more those in channels without counterterms.

\begin{table}[tb]
	\centering
	\caption{Values of the real ($V_c$) and imaginary ($W_c$) components of the short-range potential at $b=0.2$ fm for the coupled $S$-$D$ and $P$-$F$ partial waves. These of $S$ and $P$ waves are obtained by fitting to the PWA ``data" ~\cite{Zhou:2012ui,Zhou:2013} at $T_{\rm lab}=20$ MeV, while these of $D$ and $F$ waves are set to be $0$ by hand.}
	\tabcolsep=0.9em
	\renewcommand{\arraystretch}{0.9}
	\begin{tabular}{cd{4.2}d{2.0}d{4.2}d{2.0}d{4.2}d{2.0}d{4.2}d{2.0}}
		\hline
		\hline
		Partial wave &\multicolumn{1}{c}{$^{13}S_1$}&\multicolumn{1}{c}{$^{13}D_1$}
		            &\multicolumn{1}{c}{$^{33}S_1$}&\multicolumn{1}{c}{$^{33}D_1$}
		            &\multicolumn{1}{c}{$^{13}P_2$}& \multicolumn{1}{c}{$^{13}F_2$}
		            &\multicolumn{1}{c}{$^{33}P_2$}&\multicolumn{1}{c}{$^{33}F_2$}\\ 
		\hline
		$V$(fm$^{-1}$) & -37.3 & 0 & -18.8 & 0 & -83.7 & 0 & -44.5 & 0 \\
		$W$(fm$^{-1}$) & -7.4 & 0 & -43.4 & 0 &  -9.3 & 0 & -0.4 & 0 \\
		\hline
		\hline
	\end{tabular}
	\label{tab:potentials2}
\end{table}

\begin{figure}[tb]
   \centering
   \includegraphics[width=0.45\textwidth]{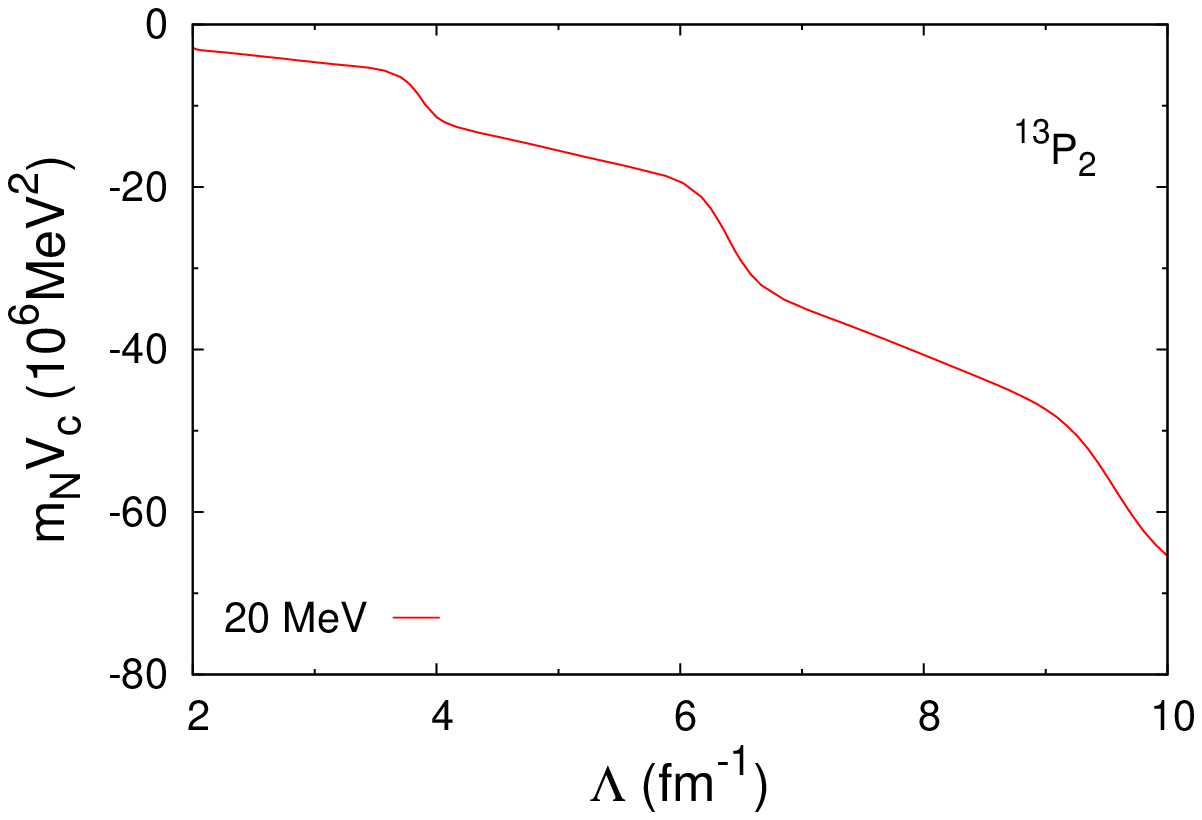} \hspace{2em}
   \includegraphics[width=0.45\textwidth]{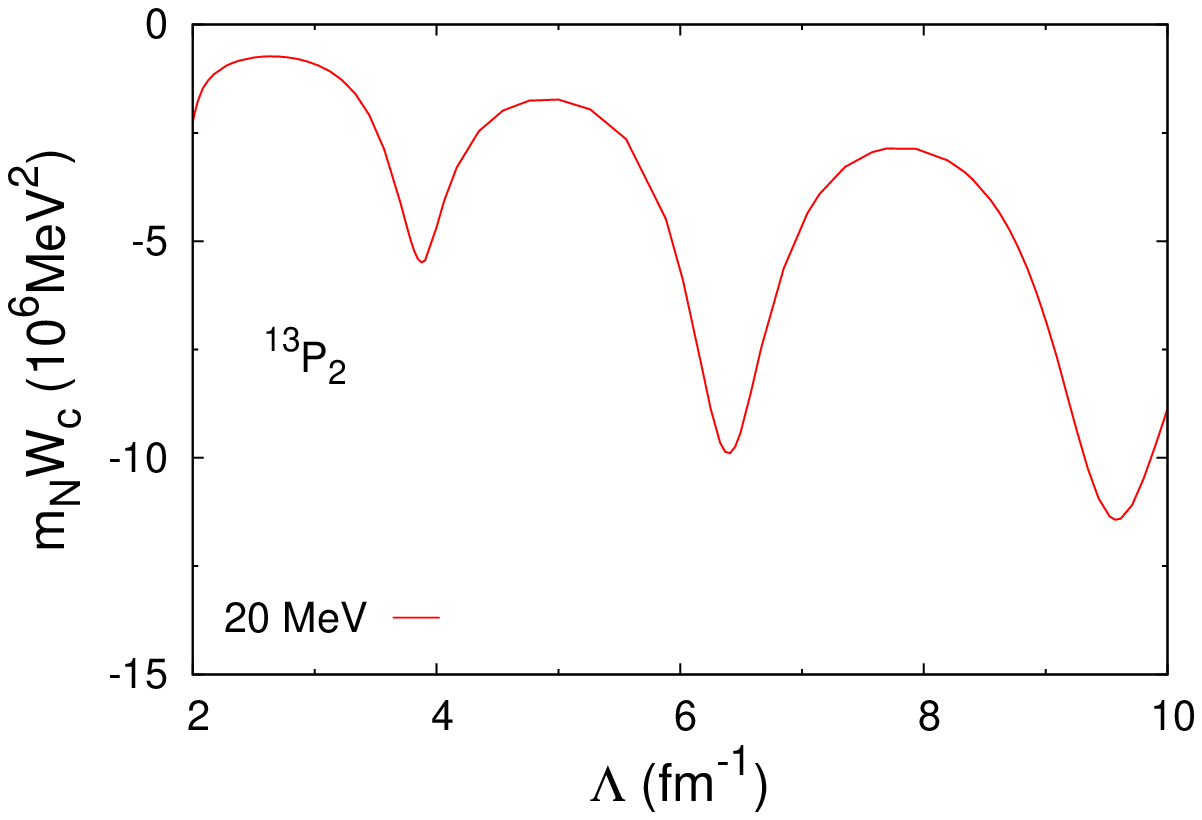} \\   
   \includegraphics[width=0.45\textwidth]{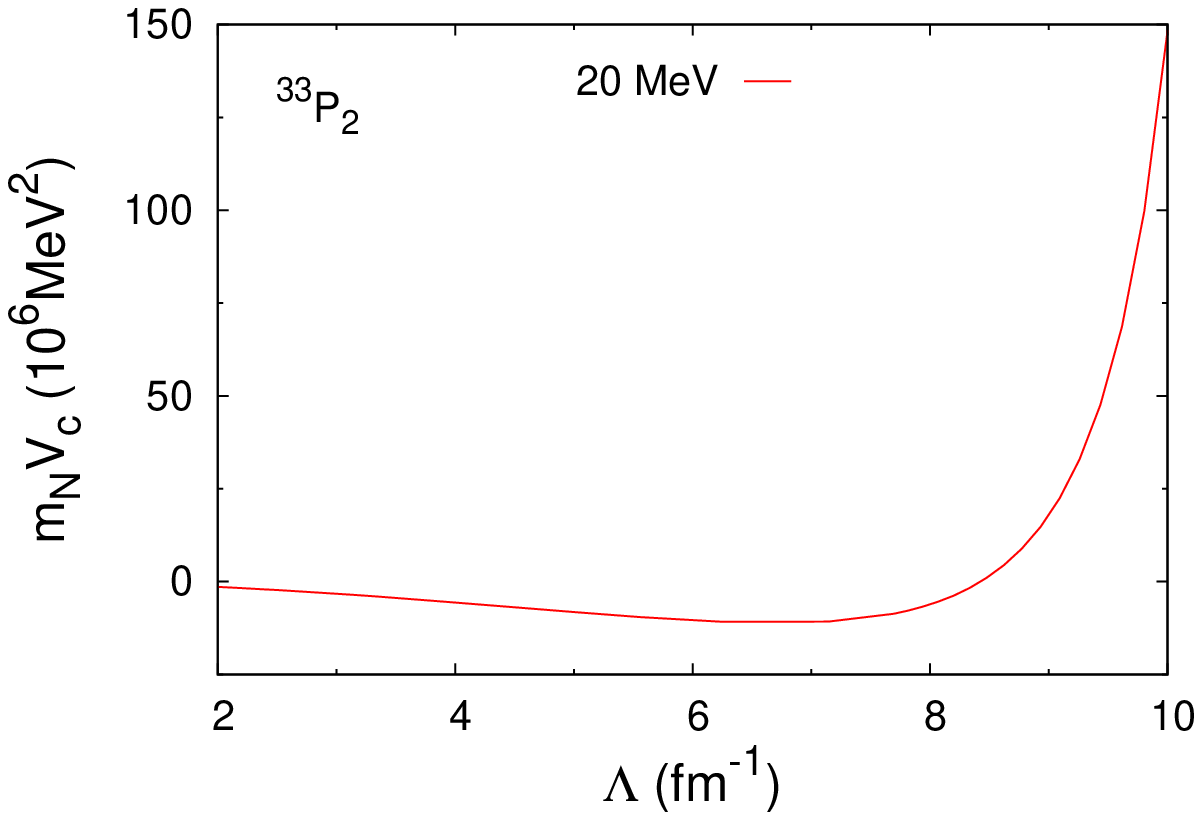} \hspace{2em}
   \includegraphics[width=0.45\textwidth]{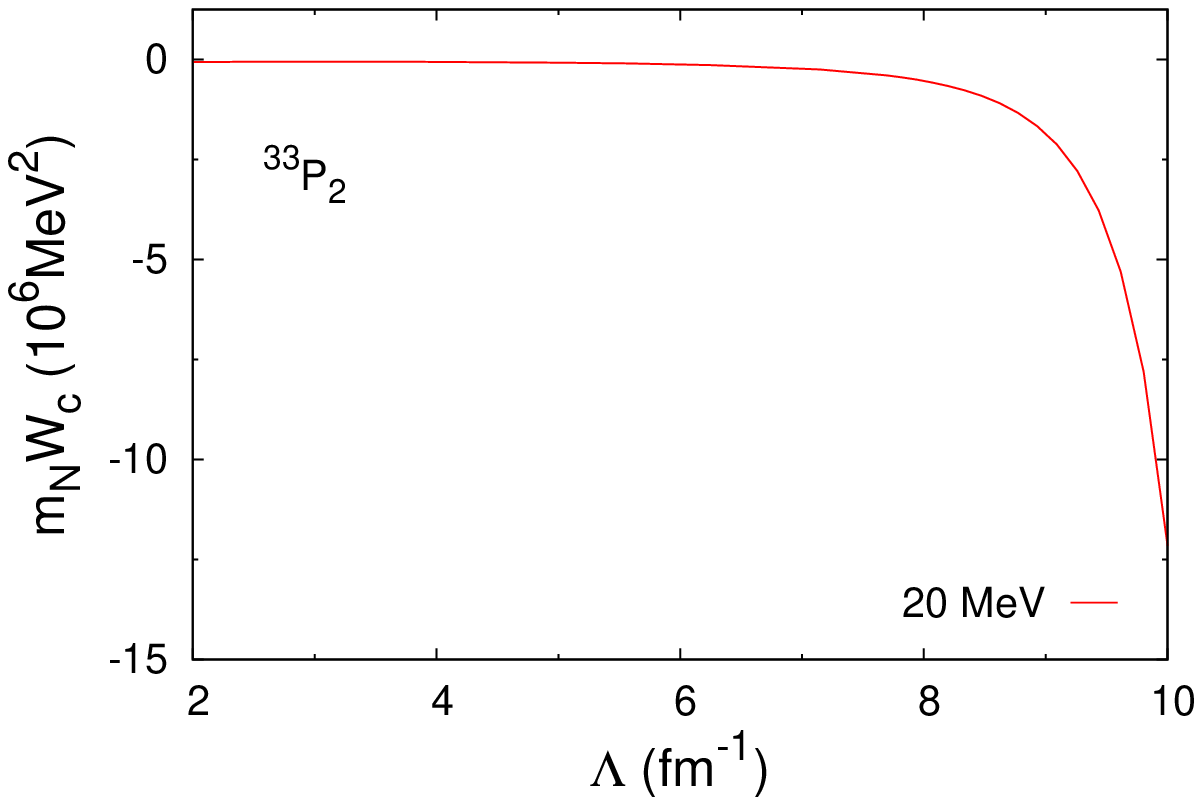} 
   \caption{\label{Counter_cutTripletP}{(Color online) 
   Cutoff dependence of $m_N V_c$ (left panels) and $m_N W_c$ (right panels) for the coupled spin-triplet $P$ waves. The PWA phase shifts and inelasticities are fitted at $T_{\rm lab}=20$ MeV.}}
\end{figure}

\begin{figure}[tb]
	\centering
	\includegraphics[width=0.45\textwidth]{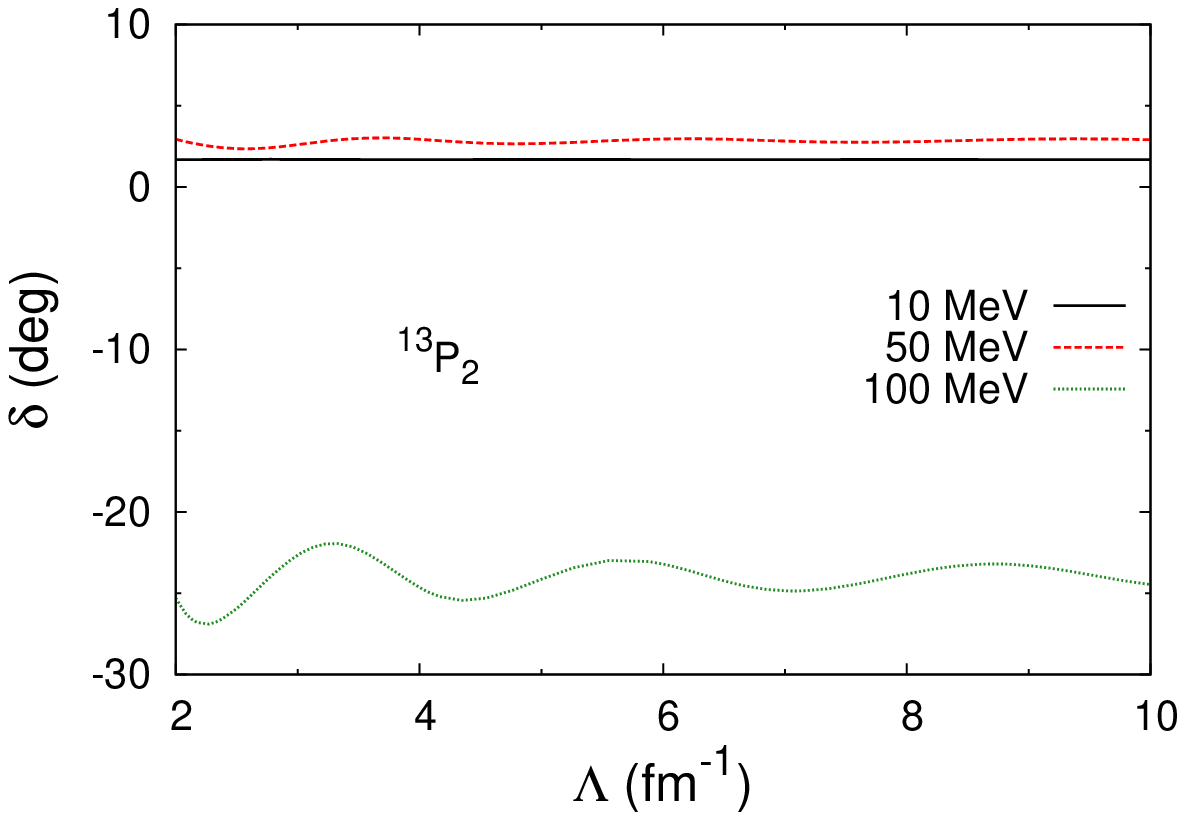} \hspace{2em}
	\includegraphics[width=0.45\textwidth]{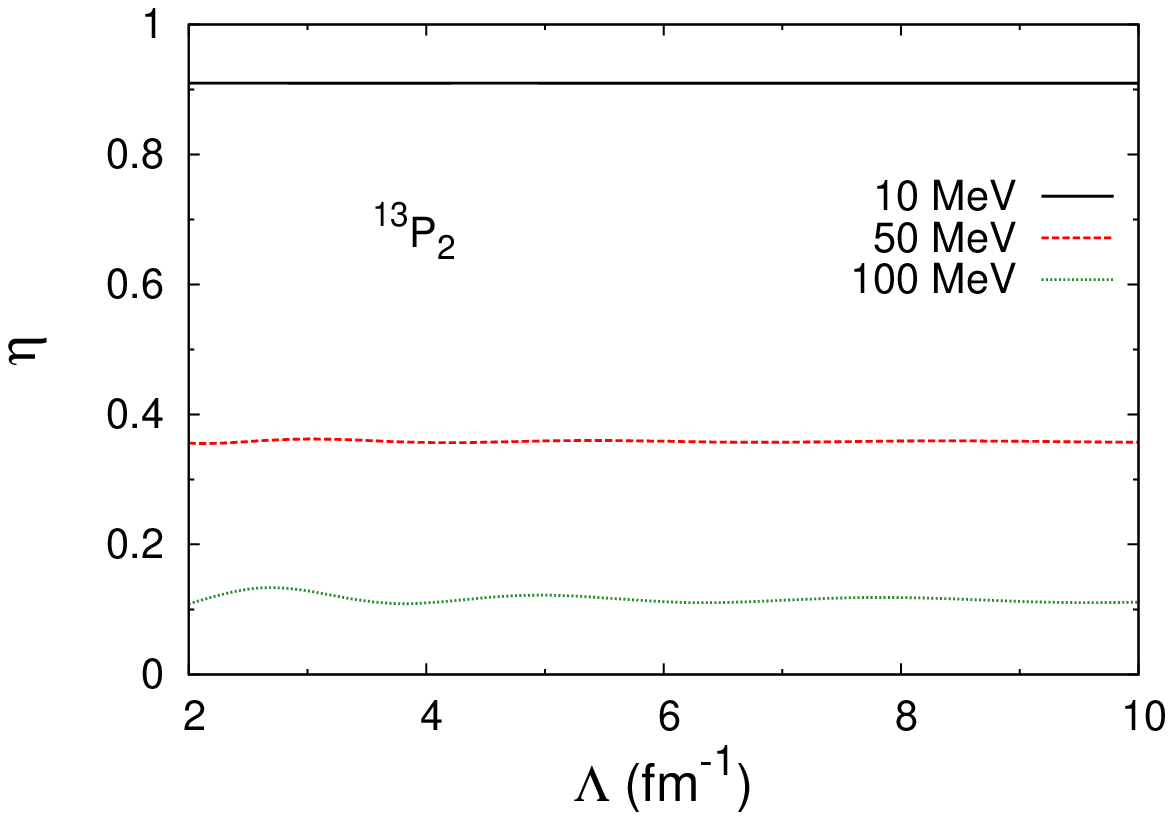}\\
	\includegraphics[width=0.45\textwidth]{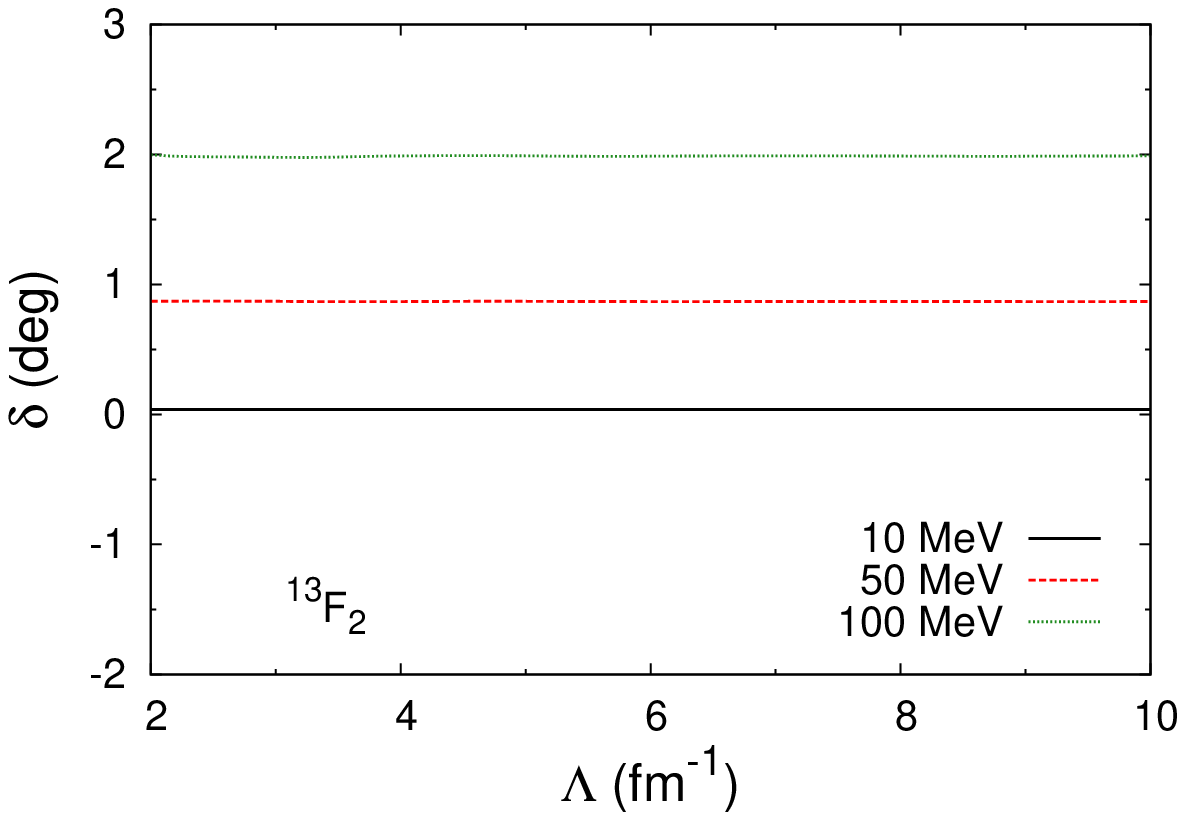} \hspace{2em}
	\includegraphics[width=0.45\textwidth]{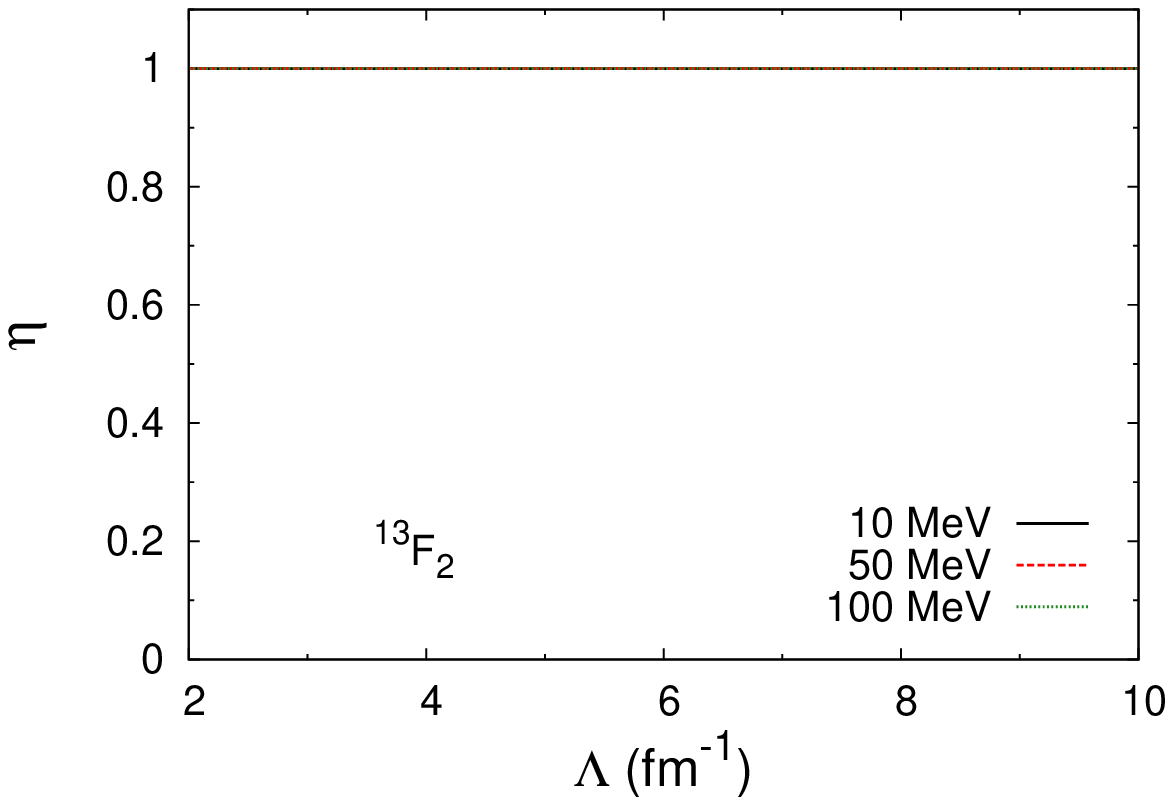} \\   
	\includegraphics[width=0.45\textwidth]{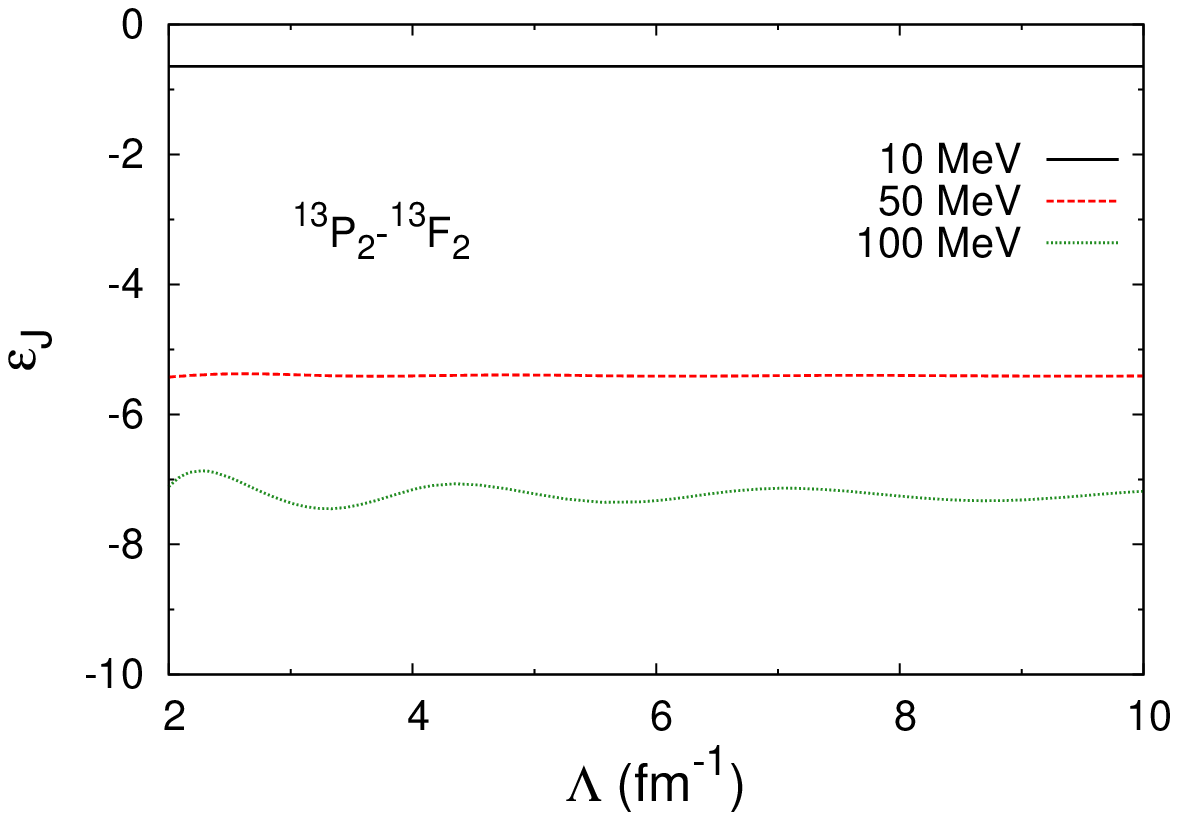} \hspace{2em}
	\includegraphics[width=0.45\textwidth]{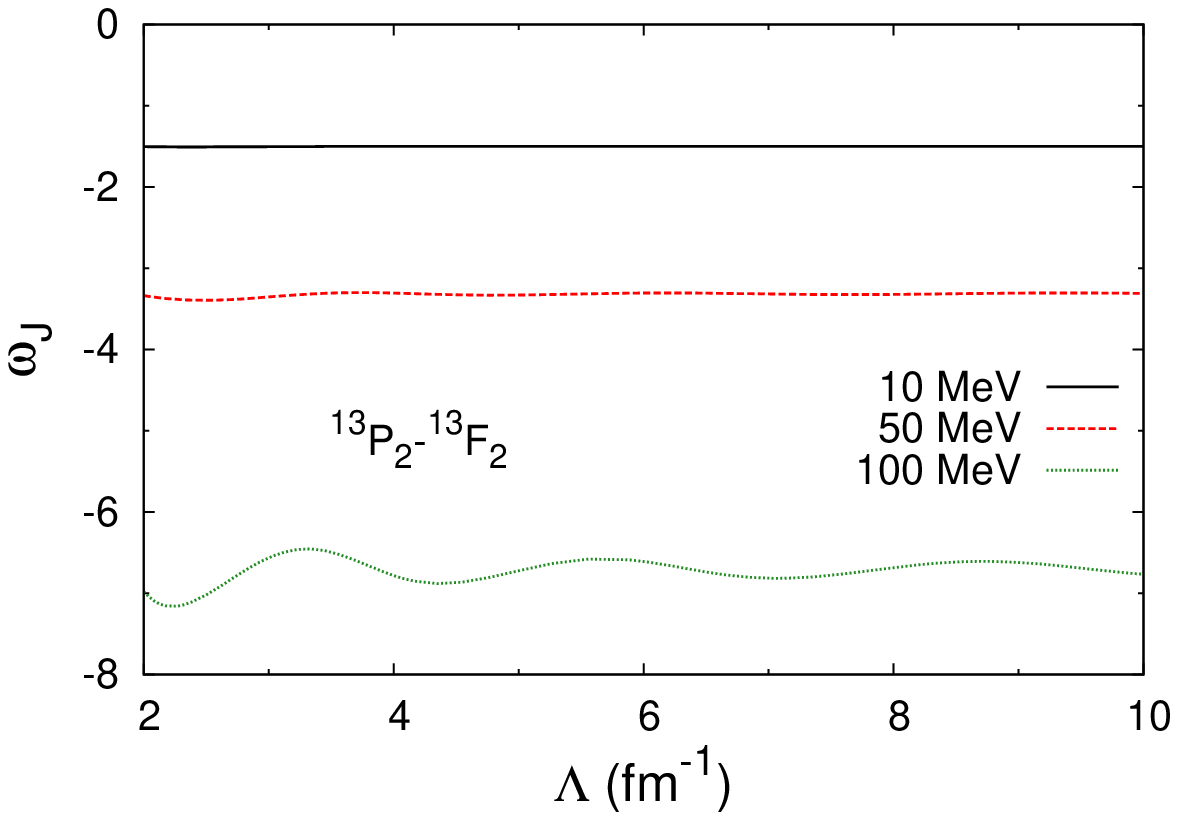}
	\caption{\label{Phase_LC31}{(Color online) Residual cutoff dependence of the phase shifts, inelasticities, and mixing angles in the $^{13}P_2$-$^{13}F_2$ waves at the laboratory energies of 10 MeV (black solid line), 50 MeV (red dashed line), and 100 MeV (green dotted line), for $V_c$ and $W_c$ in Fig. \ref{Counter_cutTripletP}.}}
\end{figure}

\begin{figure}[tb]
	\centering
	\includegraphics[width=0.45\textwidth]{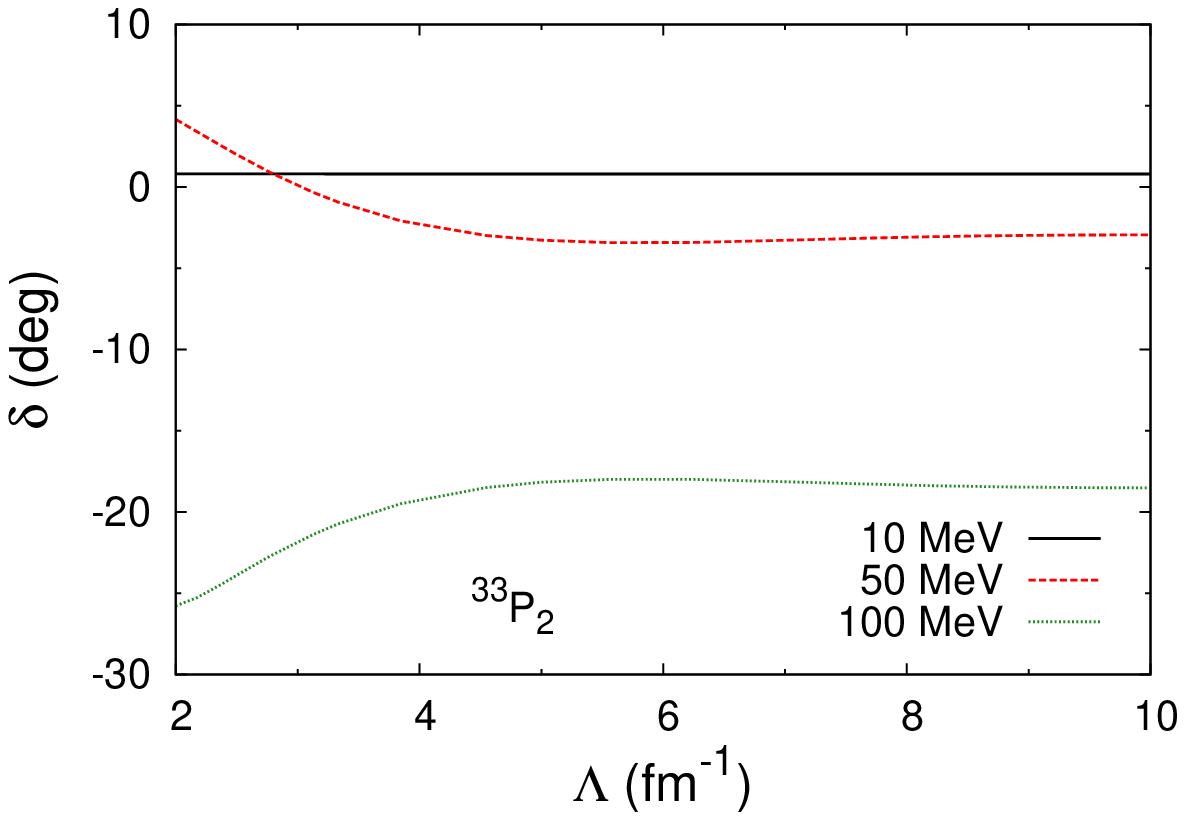} \hspace{2em}
	\includegraphics[width=0.45\textwidth]{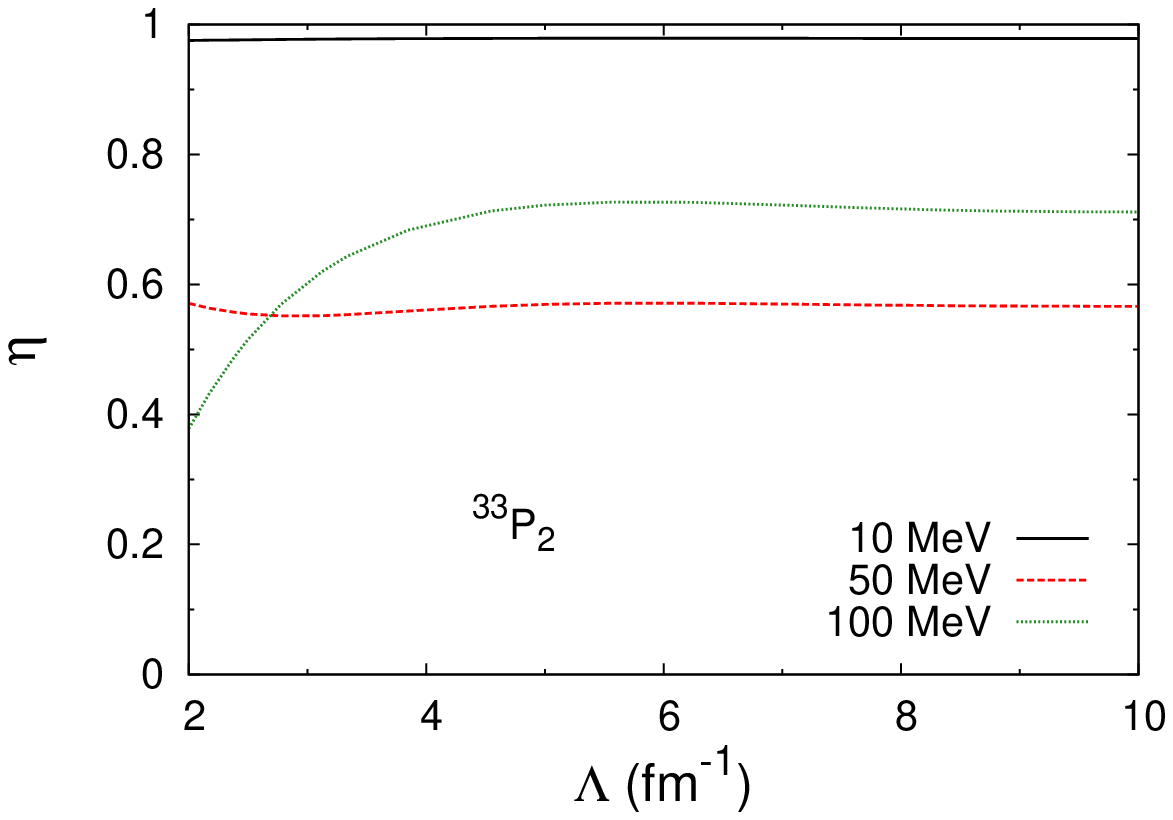}\\
	\includegraphics[width=0.45\textwidth]{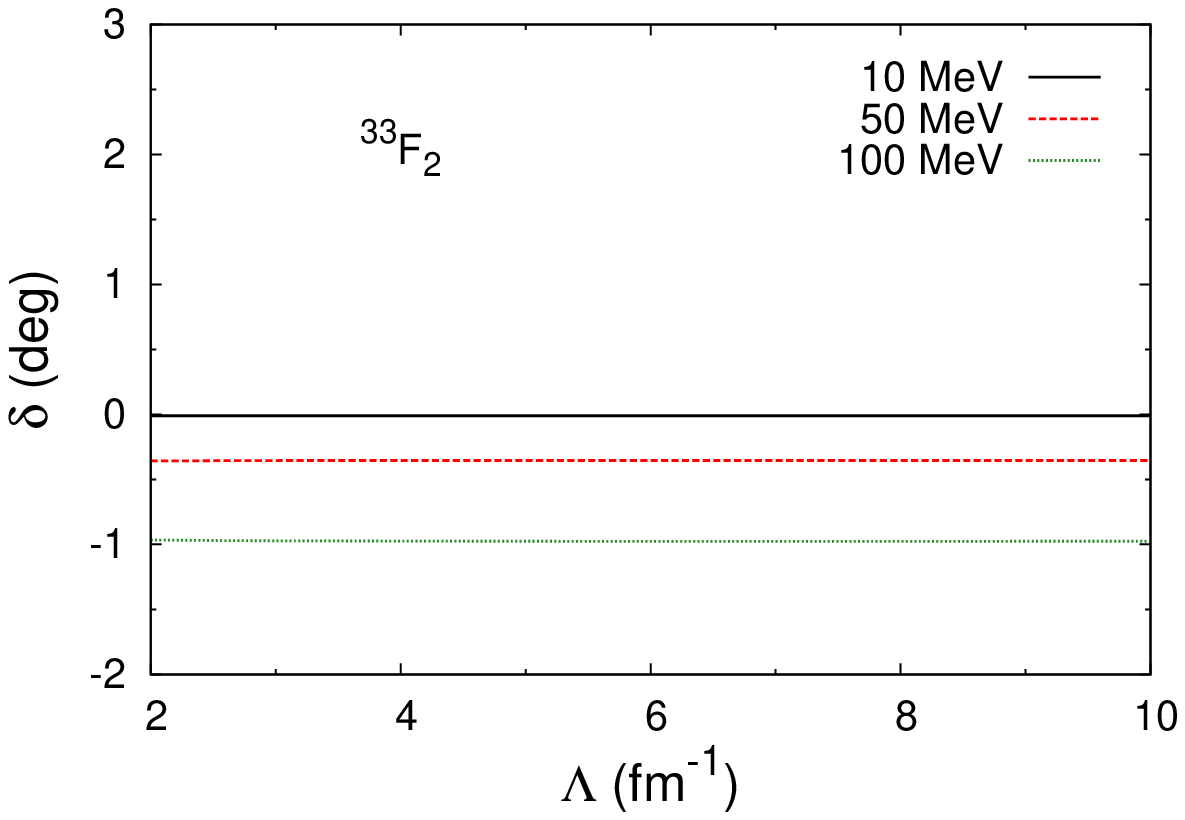} \hspace{2em}
	\includegraphics[width=0.45\textwidth]{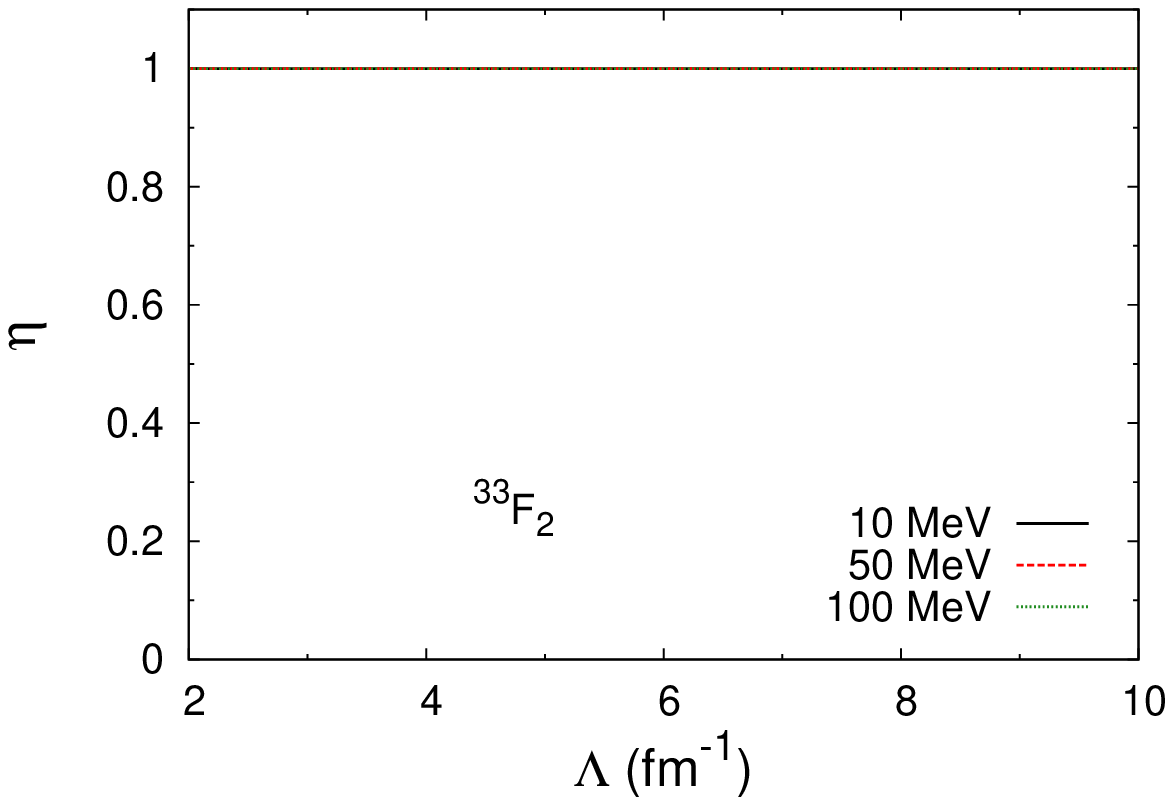} \\   
	\includegraphics[width=0.45\textwidth]{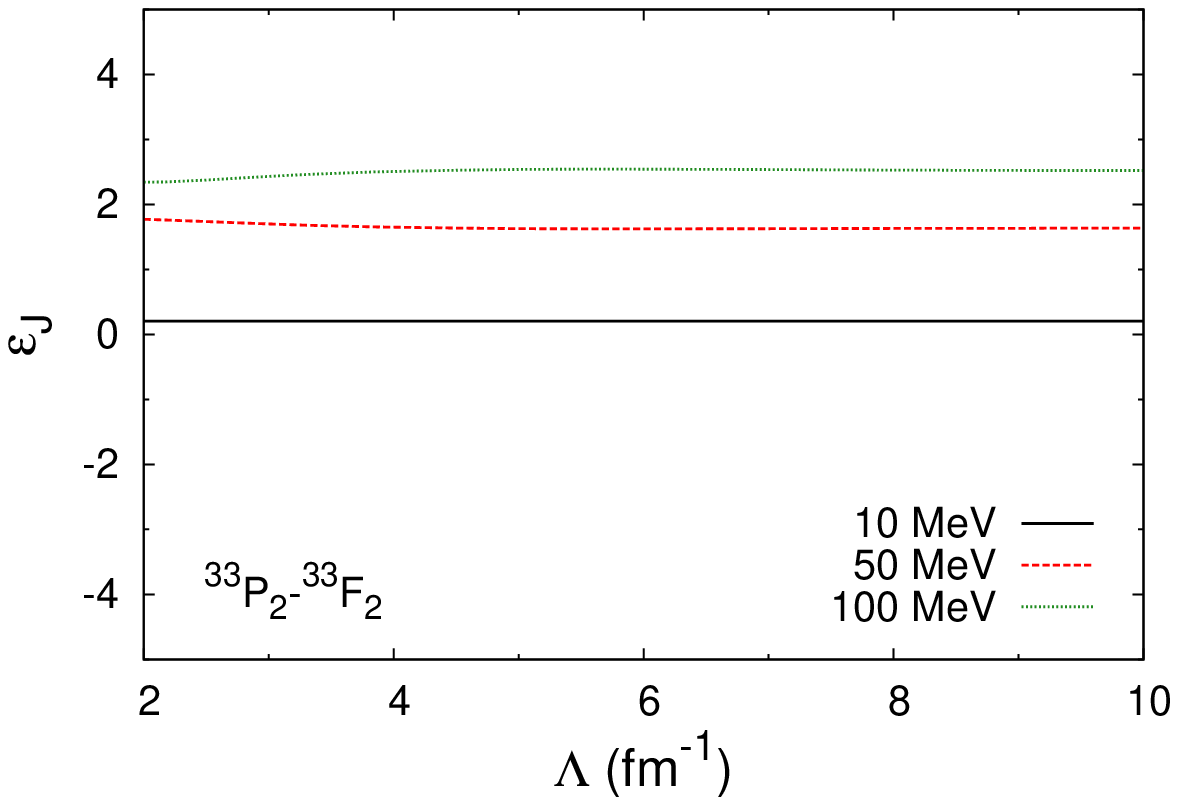} \hspace{2em}
	\includegraphics[width=0.45\textwidth]{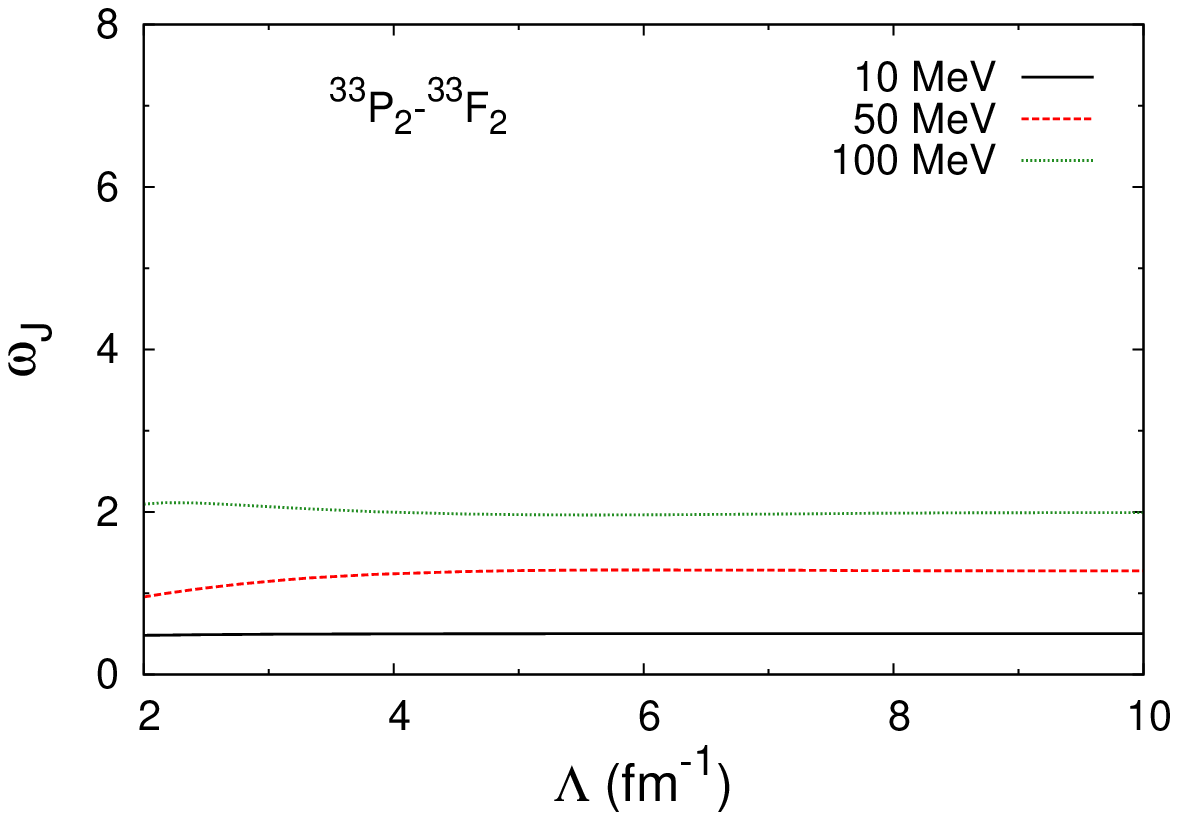}
	\caption{\label{Phase_LC32}{(Color online) Residual cutoff dependence of the phase shifts, inelasticities, and mixing angles in the $^{33}P_2$-$^{33}F_2$ waves at the laboratory energies of 10 MeV (black solid line), 50 MeV (red dashed line), and 100 MeV (green dotted line), for $V_c$ and $W_c$ in Fig. \ref{Counter_cutTripletP}.}}
\end{figure}

\begin{figure}[tb]
	\centering
	\includegraphics[width=0.45\textwidth]{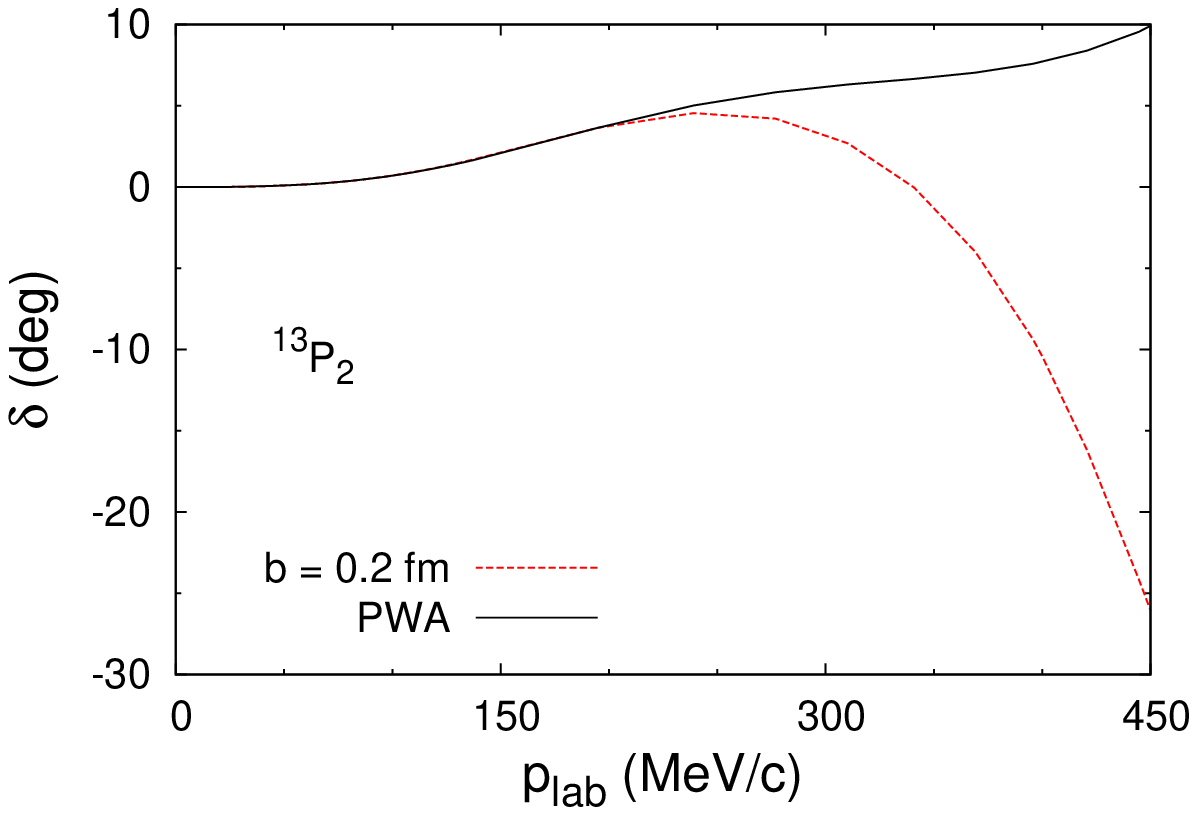} \hspace{2em}
	\includegraphics[width=0.45\textwidth]{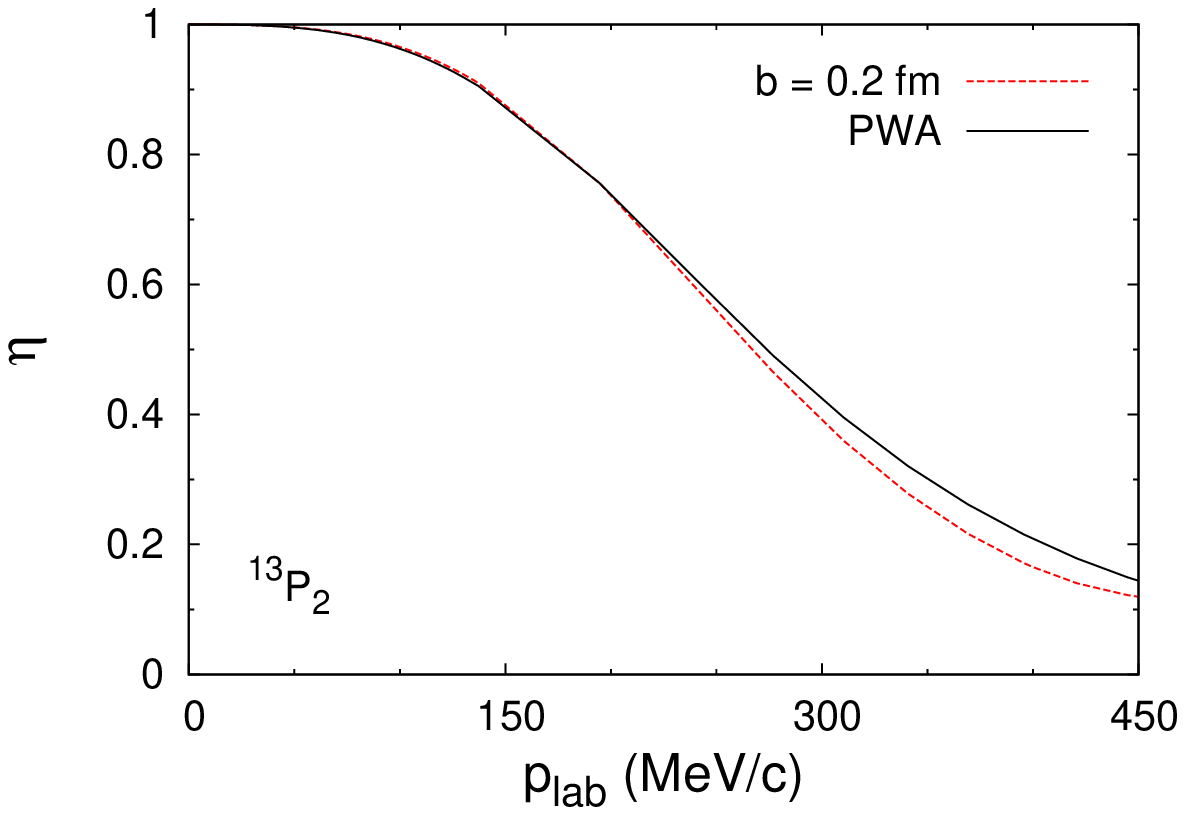} \\   
	\includegraphics[width=0.45\textwidth]{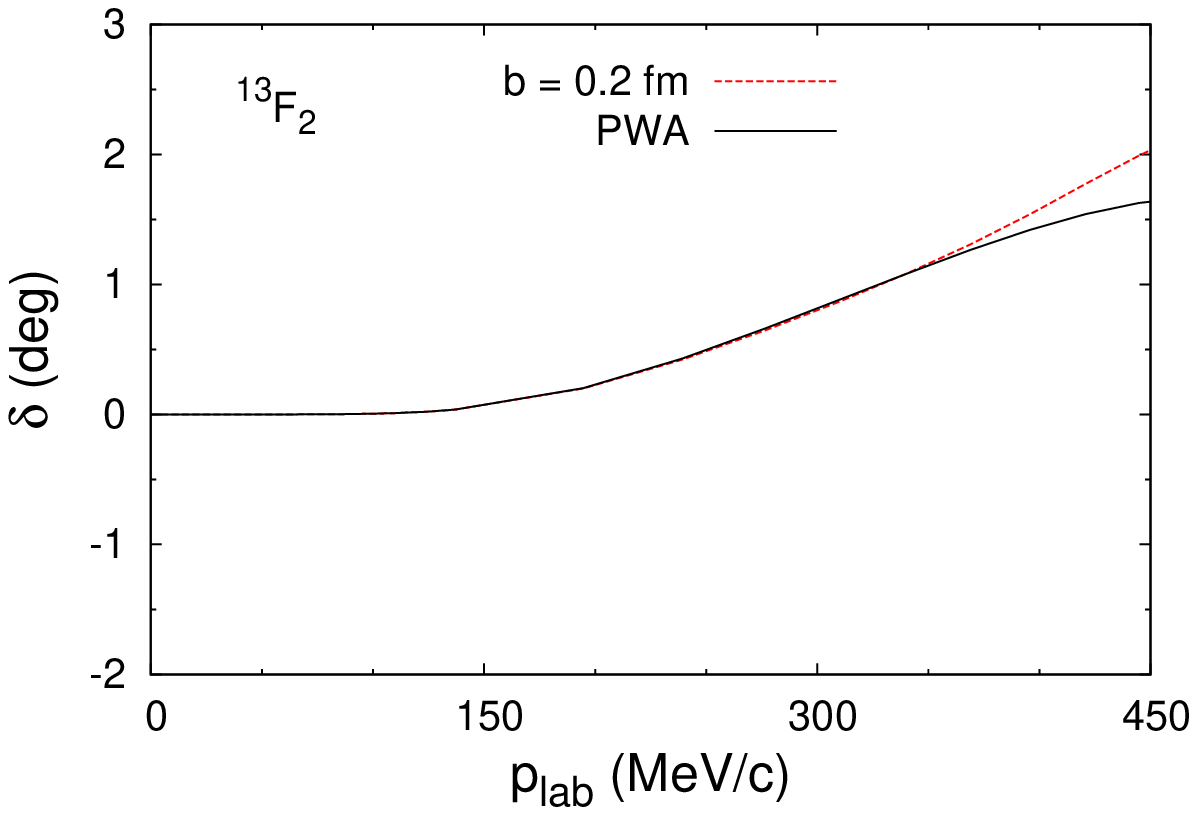} \hspace{2em}
	\includegraphics[width=0.45\textwidth]{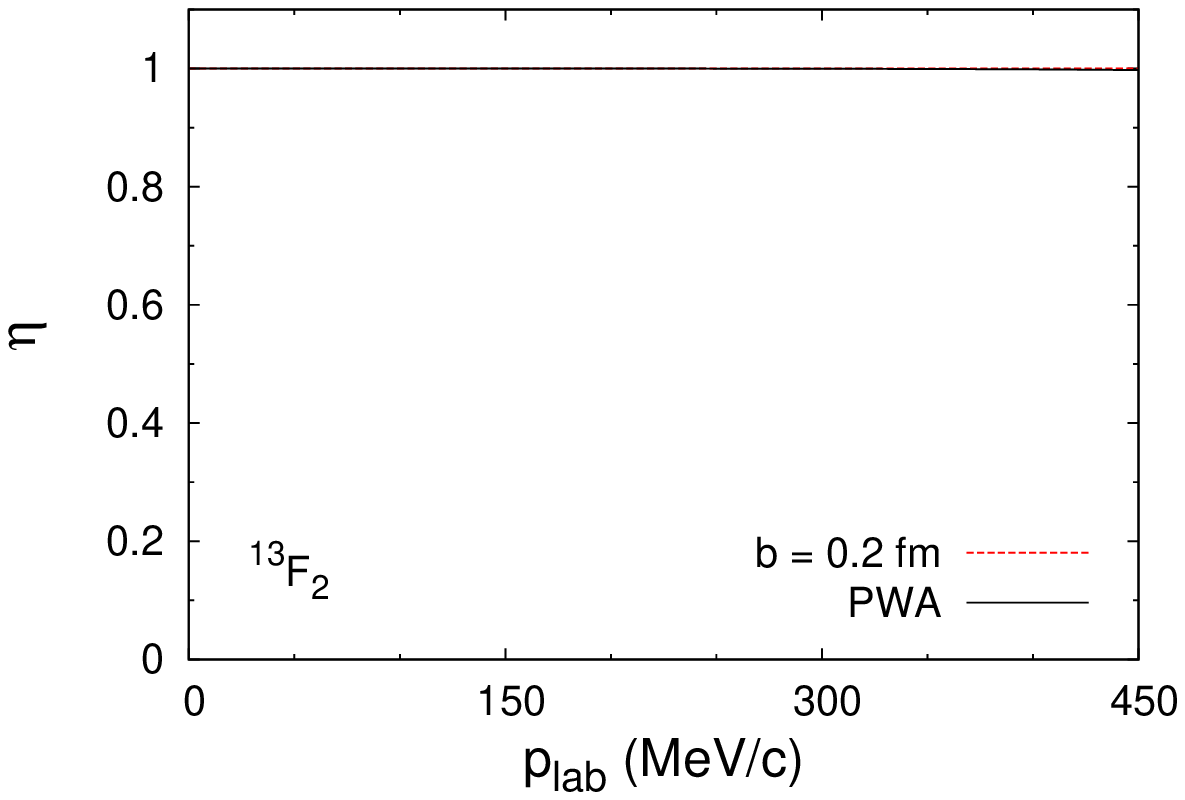}\\
	\includegraphics[width=0.45\textwidth]{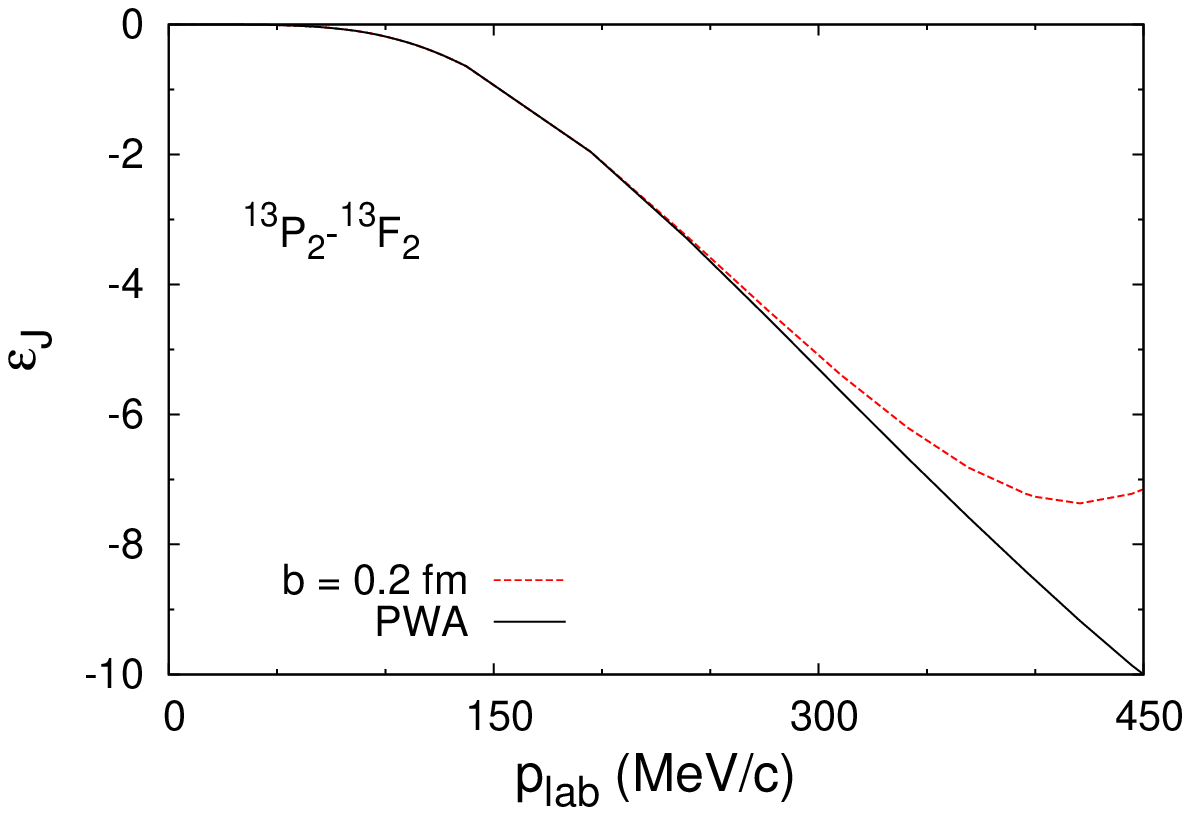} \hspace{2em}
	\includegraphics[width=0.45\textwidth]{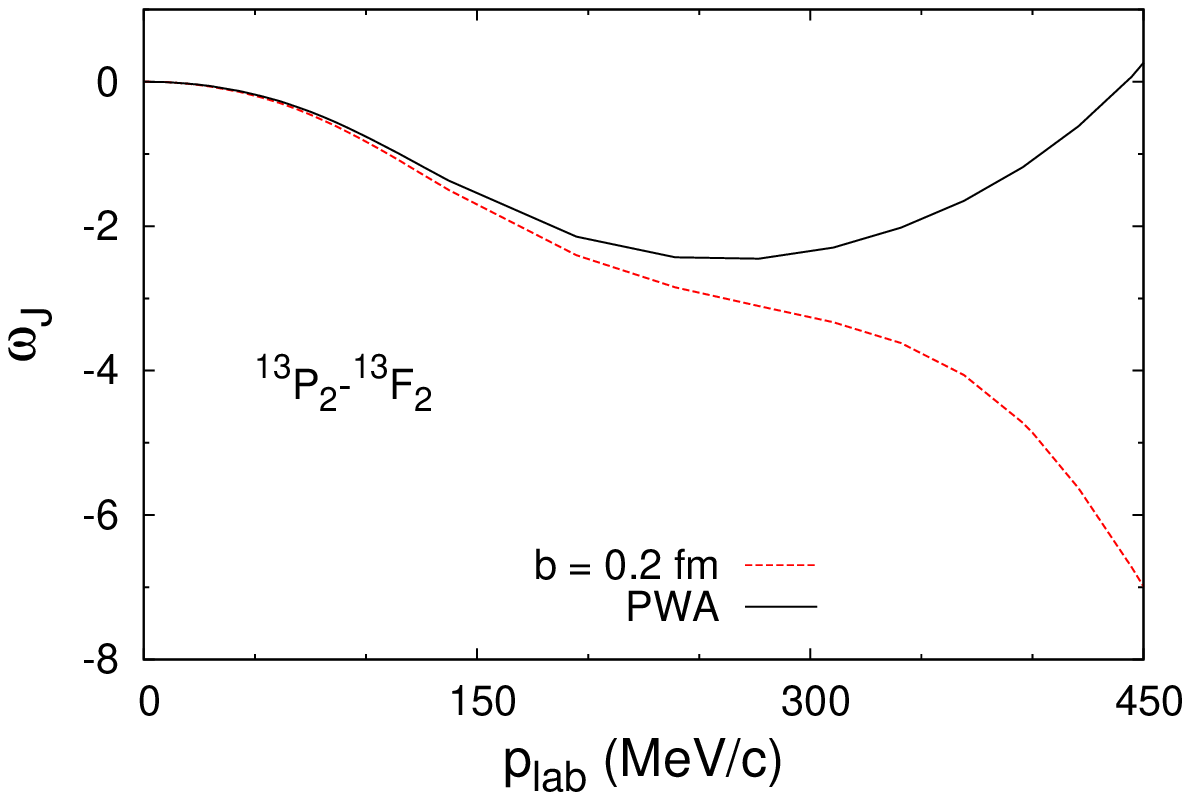}
	\caption{\label{Phase_plab31}{(Color online) 
	Phase shifts and mixing angle (left panels) and inelasticities and their mixing angle (right panels) for the coupled $^{13}P_2$-$^{13}F_2$ waves against laboratory momentum.
	The (red) dashed lines are from iterated one-pion exchange for $b=0.2$ fm and $V_c$, $W_c$ from Table \ref{tab:potentials2}, while (black) solid lines are the results of the PWA~\cite{Zhou:2012ui,Zhou:2013}.}}
\end{figure}

\begin{figure}[tb]
	\centering
	\includegraphics[width=0.45\textwidth]{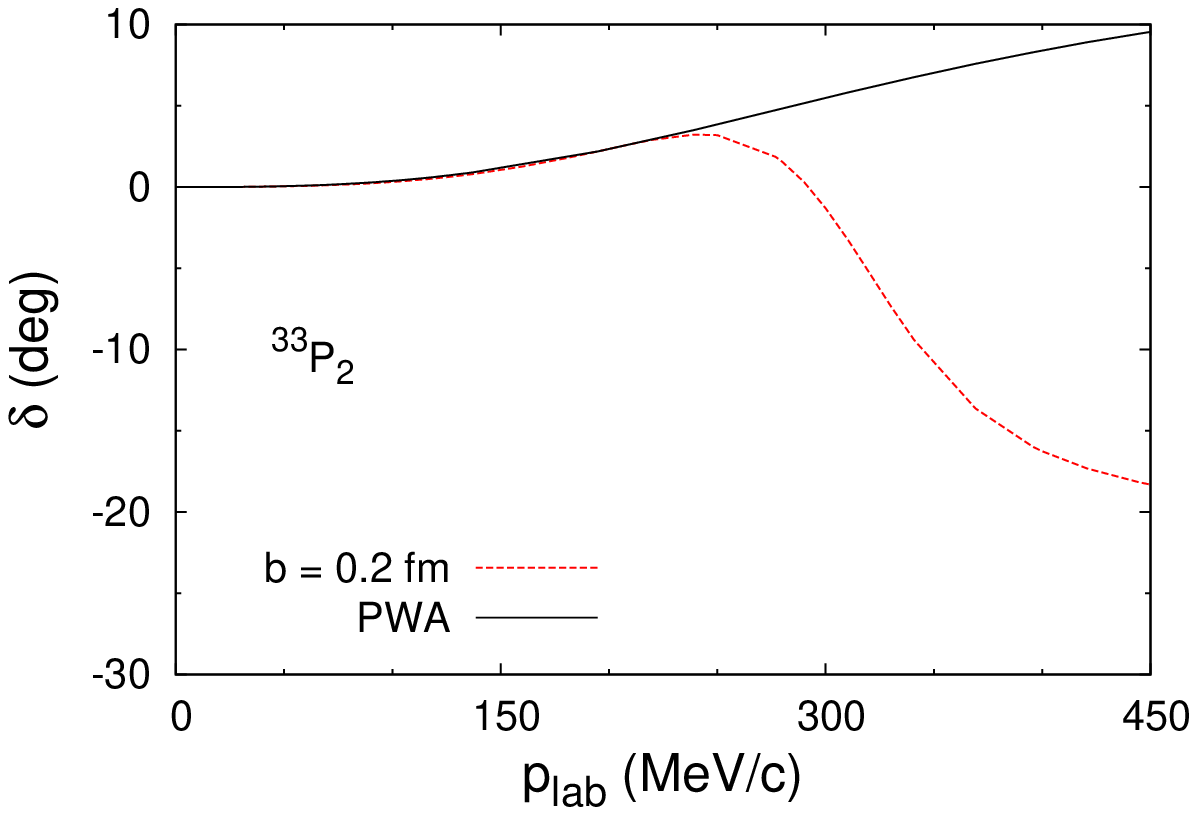} \hspace{2em}
	\includegraphics[width=0.45\textwidth]{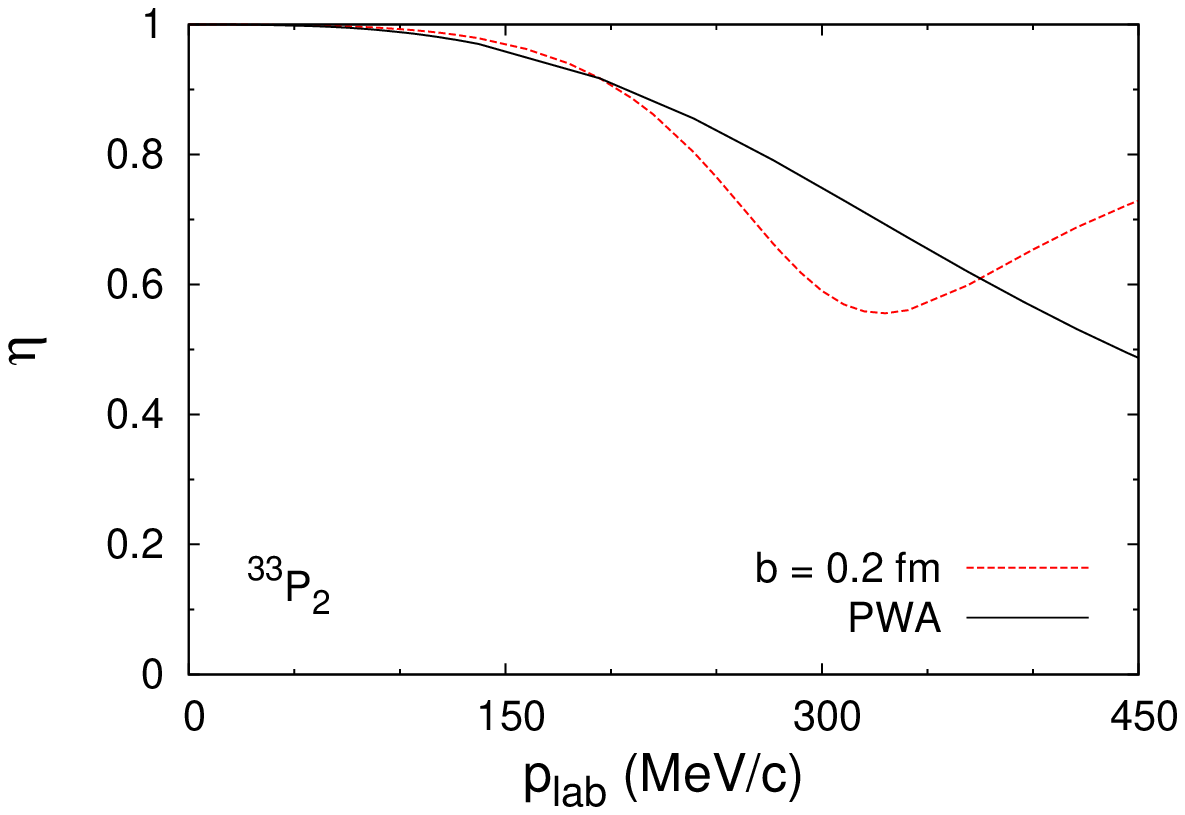} \\   
	\includegraphics[width=0.45\textwidth]{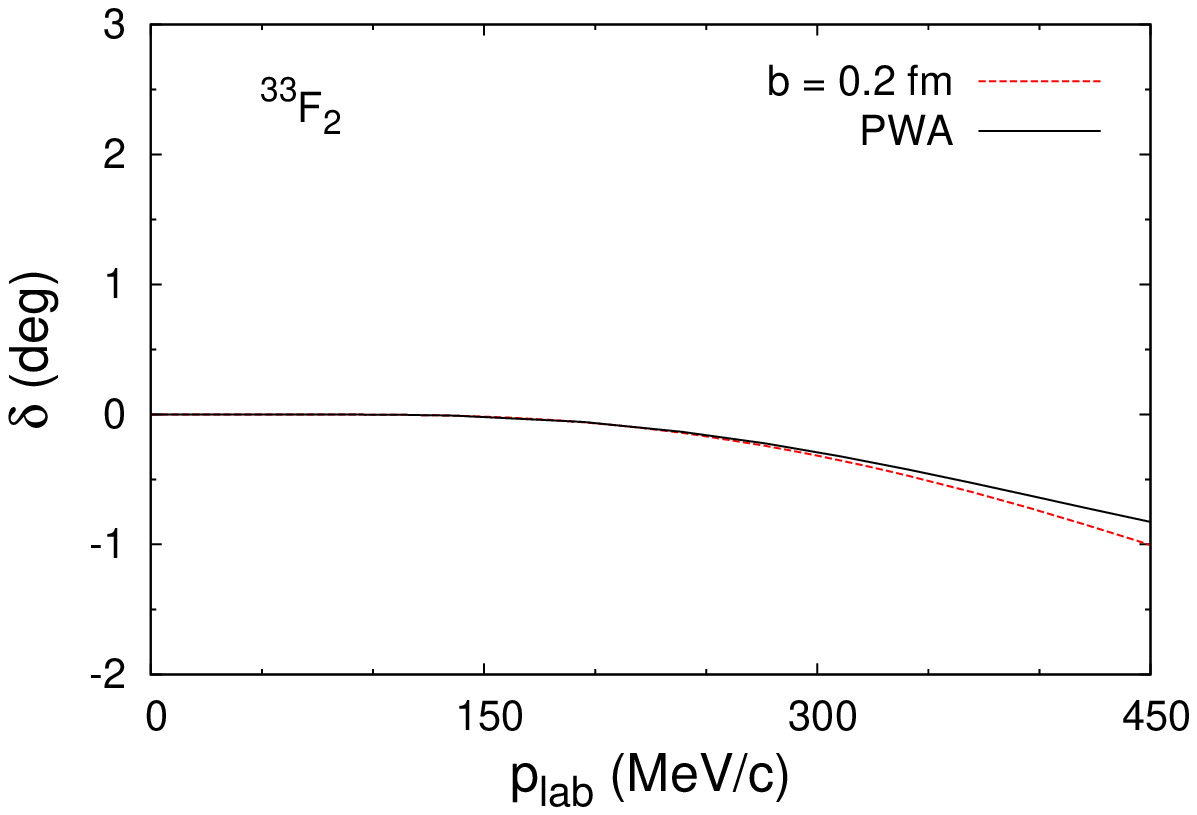} \hspace{2em}
	\includegraphics[width=0.45\textwidth]{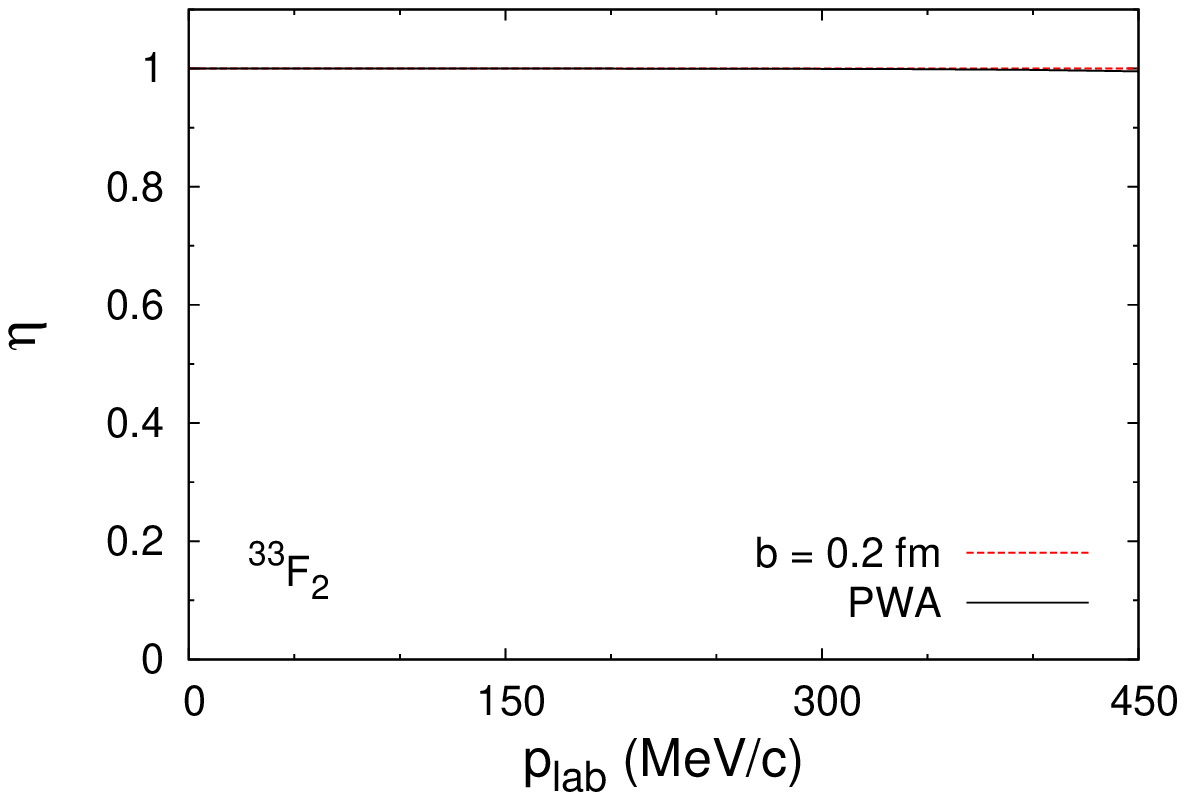}\\
	\includegraphics[width=0.45\textwidth]{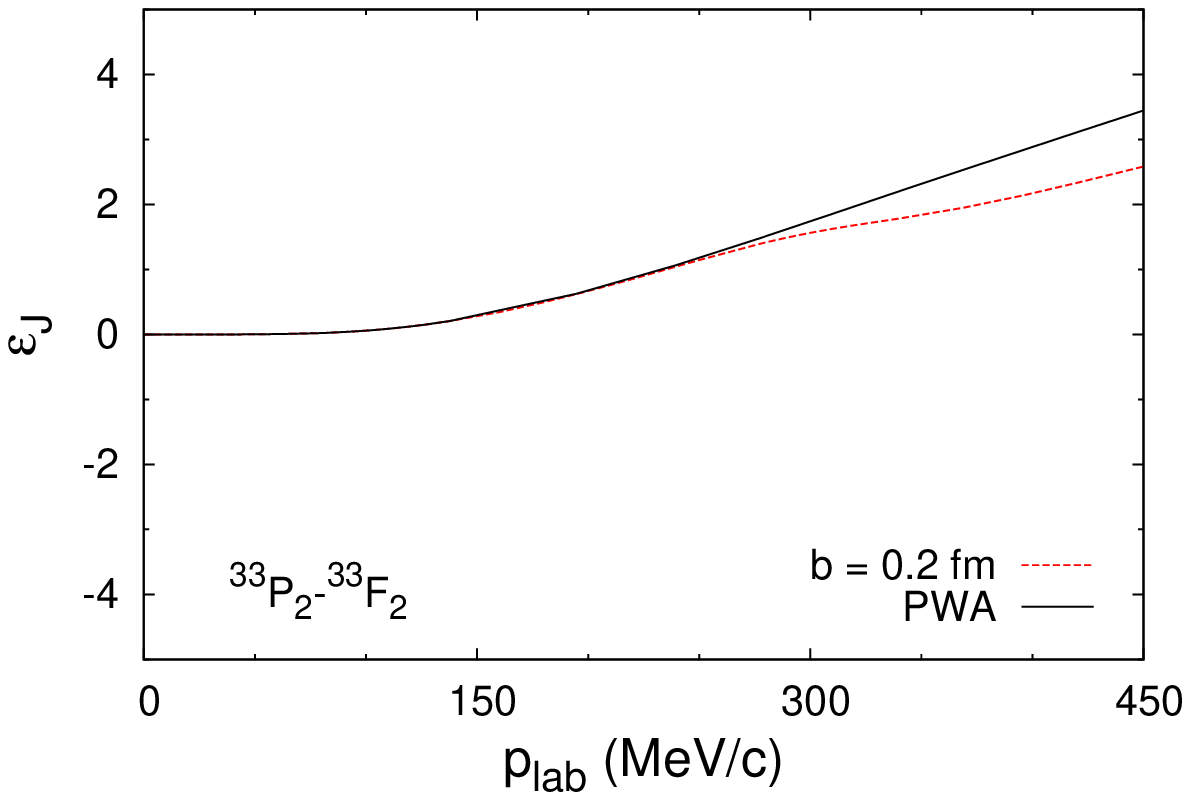} \hspace{2em}
	\includegraphics[width=0.45\textwidth]{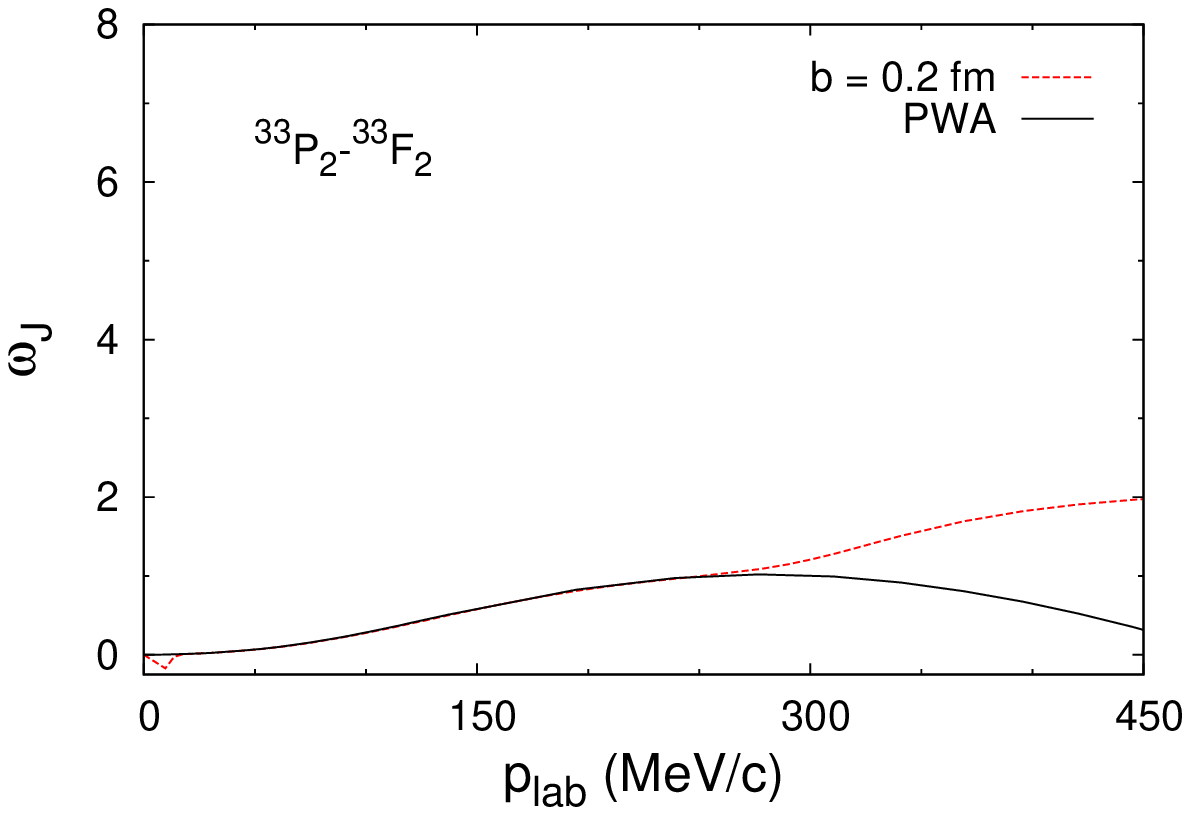}
	\caption{\label{Phase_plab32}{(Color online) 
	Phase shifts and mixing angle (left panels) and inelasticities and their mixing angle (right panels) for the coupled $^{33}P_2$-$^{33}F_2$ waves against laboratory momentum.
	The (red) dashed lines are from iterated one-pion exchange for $b=0.2$ fm and $V_c$, $W_c$ from Table \ref{tab:potentials2}, while (black) solid lines are the results of the PWA~\cite{Zhou:2012ui,Zhou:2013}.}}
\end{figure}

The situation is similar in the coupled $P$-$F$ waves, except that NDA prescribes no short-range interactions to accompany OPE. In fact, OPE is weaker than in $S$-$D$ waves and it is questionable whether it should be iterated. This is already the case for $^{33}P_2$-$^{33}F_2$ in $N\!N$~\cite{Wu:2018lai} and exacerbated here, 
where phase shifts and mixing angles are small and the cutoff dependence of iterated OPE is mild in the cutoff range we examine. However, if OPE is iterated the cutoff dependence needs to be ameliorated, particularly in $^{13}P_2$ where strong cutoff dependence is seen at all but the very lowest energies. Experience with $N\!N$ \cite{Nogga:2005hy,PavonValderrama:2005uj}
tells us that it is sufficient to add counterterms in the $P$ waves. The result from fitting them to the phase shifts and inelasticities at $T_{\rm lab}=20$ MeV is displayed in Fig. \ref{Counter_cutTripletP}. The behavior in the channel where OPE is larger, $^{13}P_2$, is similar to that in $^{13}S_1$ (Fig. \ref{Counter_cutTripletS}): oscillations and $V_c$ much larger in magnitude than $W_c$. 
In contrast, there is no obvious oscillation in $^{33}P_2$, where $V_c$ is positive when $\Lambda$ is large, which is different from other channels. The residual cutoff dependence is plotted in Figs. 
\ref{Phase_LC31} and \ref{Phase_LC32} for $^{13}P_2$-$^{13}F_2$ and $^{33}P_2$-$^{33}F_2$, respectively. Mirroring the similarity in counterterm behavior, $^{13}P_2$-$^{13}F_2$ resembles $^{13}S_1$-$^{13}D_1$ (Fig. \ref{Phase_LC21}), while qualitatively $^{33}P_2$-$^{33}F_2$ is similar to $^{33}S_1$-$^{33}D_1$ (Fig. \ref{Phase_LC22}). With the values for $V_c$ and $W_c$ in Table~\ref{tab:potentials2}, the momentum dependence of the resulting phase shifts, inelasticities, and mixing angles at $b=0.2$ fm are compared with the PWA in Fig.~\ref{Phase_plab31} for $^{13}P_2$-$^{13}F_2$ and in Fig.~\ref{Phase_plab32} for $^{33}P_2$-$^{33}F_2$. With some exceptions, the EFT results start to differ markedly from empirical values for $p_{\rm lab}\simge 300$ MeV/c, when structures appear where the PWA is smooth. (There is a small dip-bump structure at very low energies in the plot of the $^{33}P_2$-$^{33}F_2$ $\omega_J$, which might be due to the numerical implementation of Eq.~(\ref{Eq:Klarsfeld}) in a region where one $\eta$ equals $1$ and the other is very close to $1$.)

Therefore, we have confirmed that the renormalization of iterated OPE in spin-triplet $\overline{N}\!N$ scattering requires (in general complex) counterterms in the $^{13}S_1$, $^{33}S_1$, $^{13}P_0$, $^{33}P_1$, $^{13}P_2$, and $^{33}P_2$ channels. This was what was expected from the singular attractive nature of OPE. Higher spin-triplet waves will experience the same phenomenon, although the centrifugal barrier must reduce the need to iterate OPE in these waves. This is the case for the $N\!N$ system \cite{Nogga:2005hy,Birse:2005um}, and should be more so here for $I=1$, where OPE is weaker.

\section{Conclusions and outlook} \label{sec:Conclusions}

In summary, we have adapted the framework of the 
partial-wave analysis for $\overline{N}\!N$ scattering~\cite{Zhou:2012ui,Zhou:2013} to study the renormalization of the iterated static one-pion-exchange potential in low partial waves.
Both the PWA and the Weinberg approach followed by the J\"ulich group \cite{Kang:2013uia,Dai:2017ont} are based on the exact solution of the Schr\"odinger (or the equivalent Lippmann-Schwinger) equation with the $G$-parity-transformed version of 
$N\!N$ potentials obtained from Chiral EFT at large distances. However, they differ in their accounting of short-range physics: the PWA uses an energy-independent spherical well with two parameters for each wave, while the J\"ulich potential employs a separable regulator with short-range parameters mostly dictated by naive dimensional analysis (for the imaginary parts, some higher powers of momenta are included). 
We have implemented here the requirement that the short-range interaction provide order-by-order renormalizability. Power counting indicates that the 
leading-order long-range potential is OPE, which should be iterated for momenta comparable with the pion decay constant, at least in the lower waves \cite{Hammer:2019poc}. In a first approach to renormalizability in the $\overline{N}\!N$ system, we have examined iterated OPE together with
a spherical
well in the $S$ and $P$ waves. 

In the $S$ waves, coupled or not, NDA prescribes short-range interactions at the same order as one-pion exchange~\cite{Weinberg:1990rz,Weinberg:1991um}.
The 
$^{1}S_0$ waves converge with respect to the cutoff even without the short-range LECs, and once the latter are included a good description of the PWA is seen.
In contrast, in the coupled $^{3}S_1$-$^{3}D_1$ waves the LECs are needed just as in $N\!N$ \cite{Frederico:1999ps,Beane:2001bc} to remove the strong cutoff dependence, and a PWA description of more or less the same quality as the singlet channels results after renormalization. In the $S$ waves inelasticities are relatively high, in line with the NDA assumption that real and imaginary parts of the LECs are of the same order.

In contrast, NDA, based on perturbation theory, assigns $P$-wave short-range interactions to next-to-next-to-leading order, that is, two orders down in the $Q/M_{\rm QCD}$ expansion \cite{Ordonez:1992xp,Ordonez:1995rz}. By construction of NDA, if this estimate were correct there would be no significant cutoff dependence at leading order. We have verified explicitly that this is the case for $^{1}P_1$ waves where tensor OPE vanishes, as well as $^{33}P_0$ and $^{13}P_1$ where it is repulsive. The PWA phase shifts are not large and are well described at low momenta, but agreement deteriorates quickly with lab momenta above 250 MeV/$c$. Without short-range interactions, there are no inelasticities, in line with the small PWA values of $1-\eta$, except for $^{33}P_0$. 

The story is different for the remaining $P$ waves: one finds strong cutoff dependence in those waves where OPE has a $-r^{-3}$ singularity in an eigenchannel of the tensor operator. This is particularly true of the uncoupled $^{13}P_0$ and the coupled $^{13}P_2$-$^{13}F_2$ waves, where the tensor force is the strongest and three phase-shift cycles are seen in the cutoff range 2 to 10 fm$^{-1}$. In $^{13}P_0$, the spikes occur at cutoffs similar to those in $^{13}S_1$, starting at a $\Lambda$ below 500 MeV. From the naturalness perspective, there is no justification for the absence of counterterms in these $P$ waves. $^{13}P_0$ and $^{13}P_2$ are also the $P$-wave channels where inelasticities are relatively large. 
Just as for $^{3}S_1$-$^{3}D_1$, one complex counterterm per wave is sufficient for renormalization and a good description of the PWA results at low momenta. 

Our work raises several questions. We have considered only LO here, to highlight the problems of NDA when OPE is iterated in waves beyond $S$. Our results are fully consistent with the $N\!N$ case studied in Ref. \cite{Nogga:2005hy}. The different running of the LECs is likely due to the different regulator, since it resembles the running of $S$-wave LECs in the uncoupled, annihilation-free problem tackled in Ref. \cite{Beane:2000wh} with the same regulator.
For $N\!N$, the renormalized description of the empirical phase shifts improves at higher orders \cite{Valderrama:2009ei,PavonValderrama:2011fcz,Long:2011qx,Long:2011xw,Long:2012ve,Long:2013cya,PavonValderrama:2016lqn,SanchezSanchez:2017tws,Wu:2018lai}. One may hope that, likewise, the description of the $\overline{N}\!N$ PWA \cite{Zhou:2012ui,Zhou:2013} will improve at higher orders, where the chiral 
two-pion-exchange potential used in the PWA appears. The question is whether TPE accounted for perturbatively, as required by renormalization, will produce the desired effects. 

A related issue is the relative importance of the real and imaginary parts of the LECs. There seems to be a correlation between the magnitude of the inelasticity and the strength of the real part of the potential: the channels with higher inelasticity have a short-range potential at LO, either because of NDA or OPE renormalization, or both. It is not clear whether this means the imaginary part should be LO, as assumed here, or merely an enhancement of a subleading correction. A resolution probably requires the study of convergence in a calculation where imaginary parts are treated in perturbation theory.

Even more fundamental is the question of whether the power counting we discuss here can be revised more drastically. 
It is straightforward to apply our procedure to higher partial waves, but 
the centrifugal barrier 
should render OPE perturbative at the low energies of interest. When OPE is iterated, there are problems already in the $P$ waves. Despite the proper low-momentum behavior, at momenta above 300 MeV/$c$ structures appear in $^{3}P_2$-$^{3}F_2$ waves that are not seen in the PWA, while the converse holds for $^{11}P_1$. These discrepancies could be due to an unnecessary iteration of OPE \cite{Wu:2018lai}, which results from not accounting for suppression from angular-momentum factors. The convergence of perturbative pions in the $P$ and higher waves should be investigated.

Moreover, for almost all phase shifts we assigned values at zero energy found in the Groningen PWA by extrapolation from higher energies.
With one exception, we had no trouble fitting the LECs, whether they were required by renormalization or by the power counting used in $N\!N$. For $^{31}S_0$, we could find a solution only with attractive LECs that overcome the OPE repulsion, which suggests that the phase shift should start at $180$ degrees, as is the case in the PWA for $^{11}S_0$, $^{13}S_1$, $^{33}S_1$, and $^{13}P_0$. The issue of the value of the zero-energy phase shift is tied to the existence of shallow bound states, which for $\bar{N}\!N$ have a long and uncertain history --- see for example the compilation of recent results in Ref. \cite{Haidenbauer:2018wso} or the broader review \cite{Richard:2019dic}. 
Shallow bound states certainly demand at least some interaction to be treated nonperturbatively at LO, but if they are absent perturbation theory might be sufficient. We plan to return to this issue in the future.

\section*{Acknowledgments}
RGET and UvK acknowledge stimulating discussions with participants of the ESNT Workshop ``Nuclear Physics with Antiprotons" (CEA-Saclay, November 2021) organized within the framework of the ANR-21-CE31-0020 project.
This work was supported in part
by the Doctoral Fund Project under grant No. 2020BQ03 of Nanfang College, Guangzhou (DZ), by the National Natural Science Foundation of China (NSFC) under Grant No. 11735003 (BL),
and 
by the U.S. Department of Energy, Office of Science, Office of Nuclear Physics, under award DE-FG02-04ER41338 (UvK).

\end{document}